\documentclass[twocolumn,floats,amsmath,amssymb,prb]{revtex4}

\usepackage{graphicx}
\usepackage{textcomp}
\usepackage{amsmath}

\begin{document}

\title{Path-integral simulation of solids}
\author{Carlos P. Herrero}
\author{Rafael Ram\'{\i}rez}
\affiliation{Instituto de Ciencia de Materiales de Madrid,
         Consejo Superior de Investigaciones Cient\'{\i}ficas (CSIC),
         Campus de Cantoblanco, 28049 Madrid, Spain }
\date{\today}

\begin{abstract}
The path-integral formulation of the statistical mechanics of quantum
many-body systems is described, with the purpose of introducing practical 
techniques for the simulation of solids. 
Monte Carlo and molecular dynamics methods for distinguishable quantum
particles are presented, with particular attention to the
isothermal-isobaric ensemble.
Applications of these computational techniques to different types of
solids are reviewed, including 
noble-gas solids (helium and heavier elements), group-IV materials
(diamond and elemental semiconductors), and
molecular solids (with emphasis on hydrogen and ice). 
Structural, vibrational, and thermodynamic properties of these
materials are discussed. 
Applications also include point defects in solids (structure and
diffusion), as well as nuclear quantum effects in solid surfaces and
adsorbates.   Different phenomena are discussed, as
solid-to-solid and orientational phase transitions, 
rates of quantum processes, classical-to-quantum crossover, 
and various finite-temperature anharmonic effects
(thermal expansion, isotopic effects, electron-phonon interactions).
Nuclear quantum effects are most remarkable in the presence of light 
atoms, so that especial emphasis is laid on solids containing hydrogen
as a constituent element or as an impurity.   \\
\end{abstract}

\pacs{67.10.Fj, 71.15.Pd, 05.10.Ln, 61.43.Bn} 


\maketitle

\section{Introduction}

Theoretical methods in condensed-matter science can nowadays 
accurately predict many observable properties of different kinds of 
materials.  A large number of these properties depend on the atom 
vibrations around their potential minima. The well-known harmonic 
approximation
for lattice vibrations in crystals may predict rather accurately some
of their properties, but omits other basic phenomena such as thermal
expansion and some isotopic effects, which are due to the anharmonicity 
of the interatomic interactions.
From a theoretical point of view, consideration of realistic interatomic 
potentials is a difficult subject, since in actual cases one has to deal
with quantum many-body systems at finite temperatures.

A variety of theoretical techniques exist to handle these
problems in an approximate way, such as mean-field, infinite-dimension, 
reduced dimensionality, quasi-harmonic, and several kinds of perturbative 
methods. Another approach to tackle these problems consists in considering
simplified models for which exact solutions can be found, and that
will hopefully capture the basic physics of the problem.
An alternative route is provided by quantum simulations, which 
in principle can yield `numerically exact' solutions \cite{ce86,ma99}.
This means that observable quantities can be obtained for a given
many-body Hamiltonian without uncontrolled approximations, taking into
account statistical error bars and adequate corrections for the finite
size of the simulation cells and the discretized steps in the 
integration of continuous variables.
This is the case of path-integral (PI) simulations of many-body
systems, which have been increasingly used in recent years, in parallel
with the growing power of available computing facilities. 

In his original description of the PI approach, Feynman
acknowledged that it was a third formulation of non-relativistic
quantum theory, mathematically equivalent to the previous ones by
Schr\"odinger and Heisenberg, and therefore it did not contain
fundamentally new results \cite{fe48}.  He stressed, however, 
the satisfaction of recognizing old things from a new point of view.
At that time it was difficult to appreciate that this new perspective
of quantum mechanics was suitable for its implementation as a
simulation method in a digital computer, in particular in the field of
statistical mechanics. Nowadays, the discretized PI expression
of the statistical partition function, interpreted as a quantum-classical
isomorphism, allows one to apply powerful methods such as Monte Carlo (MC)
and molecular dynamics (MD) to quantum statistical mechanics.

The PI simulation of condensed phases experienced a rapid
development in the 1980s and 1990s.  Some excellent
reviews were published at that time. Chandler and Wolynes presented
a deep analysis of the isomorphism between quantum many-body theory
and classical statistical mechanics \cite{ch81}, building upon
the previous proposals of using MC calculations for the quantum
partition function by Morita \cite{mo73} and Baker \cite{ba79}.
Berne and Thirumalai \cite{be86} discussed the numerical problems
associated to the study of time correlation functions by PI simulations.
A didactic introduction to PI simulations was presented by
Gillan \cite{gi90}, who applied quantum transition-state theory (QTST) 
of activated processes to study the diffusion
of H in metals. Ceperley's review is an outstanding introduction to
path integrals with a focus on the MC techniques required to study
collective effects in boson systems such as condensed $^4$He \cite{ce95}.
Also Chakravarty's review of the MC and MD techniques for systems
of distinguishable particles, bosons and fermions offers an insightful
introduction to PI simulation techniques \cite{ch97}. Quantum algorithms
based on MD methods have been extensively treated by Tuckerman and
Hughes \cite{tu98}. 
In addition to the mentioned reviews, Feynman's books \cite{fe65,fe72} 
contain the grounds of his PI formalism, and remain as a source of basic 
knowledge for present-day students and researchers.
More recently, the extensive monograph by Kleinert \cite{kl90}
addressed the application of path integrals to several scientific
areas, while a book by Tuckerman has focused on both PI algorithms
and applications in statistical mechanics \cite{tu10}.

For condensed-matter systems, {\em ab-initio} electronic-structure
calculations based on density-functional theory (DFT) currently provide 
good approximations to the total energy and stable atomic configurations.
These methods, however, usually treat atomic nuclei as classical particles, 
and typical quantum effects like zero-point motion are not 
directly accessible.
These effects can be taken into account by approaches such as the 
quasi-harmonic approximation (QHA), but are difficult to include accurately 
when large anharmonicities are present, as may happen for light atoms like
hydrogen.  Thus, to consider the quantum nature of atomic nuclei in 
condensed-matter systems, path-integral simulations (MC or MD) 
have proved to be well-suited.
A significant advantage of this method is that nuclear degrees of
freedom can be quantized in an efficient way, thus allowing to include
quantum and thermal fluctuations in many-body systems at finite 
temperatures.  In this manner, MC or MD 
sampling applied to evaluate finite-temperature path integrals are very
useful for performing quantitative and nonperturbative calculations of
anharmonic effects in condensed matter \cite{gi88,ce95,ma09b}.

For a given solid, quantum effects related to the atomic vibrational
motion are relevant at temperatures lower than the Debye temperature
of the material, $\Theta_D$.
At $T \lesssim \Theta_D$ the vibrational amplitudes differ from the
classical expectancy, so that anharmonic effects on the properties of
the material are enhanced with respect to a classical calculation.
This includes structural, thermodynamic, and also electronic
properties. For example, the low-temperature ($T \to 0$) crystal volume
of a (quantum) solid is usually larger than that giving the minimum
potential energy in a classical calculation (zero-point expansion).
Moreover, the coexistence lines in phase diagrams of different
substances are known to be displaced by quantum nuclear motion, as
has been revealed by their dependence on isotopic mass.
Also, the electronic gap in semiconductors has been found to be
renormalized due to nuclear quantum effects on the electron-phonon
interaction.
For properties that depend on a particular vibrational mode with
frequency $\omega$ (e.g., optical or acoustic modes), 
quantum effects are expected to be relevant at temperatures
$T \lesssim \hbar \, \omega / k_B$.

A typical quantum effect is the isotopic dependence of several
properties of a crystal, which do not vary with the atomic masses in
a classical approach.  Thus, it is well known that
the actual molar volumes of chemically identical crystals 
with different isotopic composition are not equal, as a consequence
of the dependence of the atomic vibrational amplitudes
on the atomic mass \cite{bu88,ho91,no97}.
Lighter isotopes have larger vibrational amplitudes (as expected in 
a harmonic approximation) and in general they display larger
molar volumes as a consequence of anharmonicity.
(There are important exceptions to this trend, as happens for systems
including hydrogen bonds; see below.)
This effect is most noticeable at low temperatures,
since the atoms in a solid feel the anharmonicity
of the interatomic potential due to zero-point motion.
At higher temperatures, the vibrational amplitudes are larger, but the 
isotope effect on the crystal volume is less prominent, because
those amplitudes become less mass-dependent. In the high-temperature
(classical) limit this isotope effect disappears.
Path-integral simulations have turned out to be
sensitive enough to quantify the dependence of crystal volume on the
isotopic mass of the constituent atoms \cite{no97,he99,he09c}.

In this topical review we present an introduction to MC and MD 
algorithms used in PI 
simulations of condensed-matter systems, i.e., the so-called 
path-integral Monte Carlo (PIMC) and path-integral molecular dynamics 
(PIMD).  We restrict ourselves to the formalism where quantum effects 
related to atomic indistinguishability can be neglected. 
Nuclear exchange effects become important when the de Broglie 
wavelength, $\lambda = h / \sqrt{2 \pi m k_B T}$, 
is comparable to or larger than the average distance between atoms 
in a solid ($m$: nuclear mass).
This is not the case for most of the solids and temperatures studied 
here, and therefore the PI implementation of exchange for bosonic and
fermionic degrees of freedom is not considered. The PI simulation
of time dependent correlation functions is another aspect that is
not discussed in this review (see Refs.~\cite{be86b,ho06,pe09,ha13}). 
We summarize the computational requirements
to simulate solids in the isothermal-isobaric ($NPT$) ensemble, either 
using PIMC or PIMD sampling ($N$ stands for the number of particles, 
while $P$ and $T$ denote pressure and temperature, respectively). 
We do not present an exhaustive description of existing alternative 
algorithms for performing PI simulations, rather we prefer to present 
a minimum of technical information relevant to understand the machinery 
of a PI computer code, following our own experience in this subject. 

In the context of PIMC simulations, there has been a large
amount of work for lattice models with the so-called `world-line'
method. Applications of this procedure have been from lattice gauge
theory of particle physics to models of high-temperature
superconductors.
Here we will discuss only continuum models, as the techniques
employed for lattice models, although related in principle,
turn out to be in practice rather different
\cite{hi82,ra85,li92b,ba92}.
Other quantum simulation methods, such as variational, diffusion,
and continuous-time quantum MC, have been employed for the
simulation of condensed-matter systems and were reviewed
elsewhere \cite{fo01,ne10,gu11}.

 The review is organized as follows. In Sec.\,II, we describe the
theoretical background necessary for the implementation of PI simulations 
of solids. Especial emphasis is laid upon the formulation of the
discretization procedure for path integrals at finite temperatures, and
its application to calculate average observable quantities in the
isothermal-isobaric ensemble. 
In the remainder of the review we concentrate on the application of these
computational methods to different types of solids:
noble-gas solids (Sect.~III), group-IV materials (Sect.~IV),
and molecular solids (mainly hydrogen and ice; Sect.~V).
In Sect.~VI we present simulations of the structure and diffusion of
point defects in materials, and in Sect.~VII we concentrate on nuclear 
quantum effects in solid surfaces and adsorbates.
The review closes with a summary in Sect.~VIII.

\section{Theory
\label{sec:Theory}}

In this section, the definition of the partition function in the $NPT$
ensemble is presented. Then, the discrete PI representation of the
partition function is derived for a one-particle ensemble, explaining
the suggestive quantum-classical isomorphism. After dealing with the
many-body generalization, we summarize the basic algorithms required
for both PIMC and PIMD simulations. We comment also on comparative
advantages and drawbacks of both methods. Some technical points are 
illustrated by PI simulations of solid neon in the $NPT$ ensemble. 
Moreover, we explain the application of PI techniques to calculate 
rate constants by QTST,
vibrational frequencies of solids by a linear-response approach,
and free energies by thermodynamic integration.

\subsection{Isothermal-isobaric partition function
\label{sub:Isothermal-isobaric-partition-fu} }

We focus on the quantum $NPT$ simulation of a solid composed of atoms 
enclosed in a cell with volume $V$
and periodic boundary conditions. The partition function 
in the $NPT$ ensemble is given by
\begin{equation}
\triangle(N,P,T) = 
    \intop_{0}^{\infty}dVe^{-\beta PV}Z(N,V,T)\;,\label{eq:Z_NPT}
\end{equation}
where $\beta$ is the inverse temperature $\beta^{-1}=k_{B}T$ and
$k_{B}$ is the Boltzmann constant. The canonical partition function
is defined as
\begin{equation}
Z(N,V,T) = \mathrm{Tr}\left[e^{-\beta\widehat{H}}\right] = 
   \sum_{k}e^{-\beta E_{k}}.\label{eq:Z_tr}
\end{equation}
$\widehat{H}$ is the Hamiltonian operator, which represents the sum
of kinetic ($\widehat{K}$) and potential ($\widehat{U}$) energy terms.
$e^{-\beta\widehat{H}}$ is the canonical \textit{density operator}. 
This exponential operator is diagonal in the basis of Hamiltonian
eigenfunctions. Its trace, $Z$, is the sum, for the complete set
of energy eigenfunctions, of the Boltzmann factor associated to
each energy eigenvalue, $E_{k}$.

\subsection{Path-integral discretization
\label{sub:Path-integral-discretization} }

For didactic purposes, the discretization of the density operator for
a free particle is presented in Appendix \ref{sec:Appendix}. Specifically
it is shown that the temperature dependence of the density operator
may be approximated by following three steps: an exact factorization
of the exponential operator, a formulation of a \textit{high-temperature
approximation} (HTA), and a repeating loop calculation. Here these
steps are applied to a general case. For simplicity we consider first
the canonical partition function of an ensemble of only one particle
with mass $m$ moving in an external potential $U(\mathbf{R})$,
$\mathbf{R}$ being a position vector. 
Let us denote a position eigenfunction as the ket 
$|\mathrm{\mathbf{R}}_{1}\rangle$.
The trace of the density operator over this complete basis set is 
\begin{equation}
Z = \int d\mathbf{R}_{1} \,
   \langle\mathbf{R}_{1}|e^{-\beta\widehat{H}}|\mathbf{R}_{1}\rangle \; .
\label{eq:Z}
\end{equation}
The temperature dependence of the partition function can be approximated
by considering first the exact factorization
\begin{equation}
e^{-\beta\widehat{H}} = 
   \left(e^{-\beta\widehat{H}/L}\right)^{L}   \; ,
\label{eq:factorization}
\end{equation}
and formulating a convenient HTA for the operator $e^{-\beta\widehat{H}/L}$.
A computationally feasible HTA is the Trotter or primitive 
approximation \cite{sh81}
\begin{equation}
e^{-\beta\widehat{H}/L} \approx  e^{-\beta\widehat{U}/2L} \,
       e^{-\beta\widehat{K}/L} \, e^{-\beta\widehat{U}/2L}   \;.
\label{eq:Trotter_expansion}
\end{equation}
The error of the HTA becomes vanishingly small as $\beta/L\rightarrow0$,
i.e., for Trotter number $L\rightarrow\infty$. We refer to the
literature for improved approximations as higher order 
expansions \cite{ta84},
cumulant expansions, or pair-product approximations \cite{ce95}.
Let us apply the primitive HTA in Eq.~(\ref{eq:Trotter_expansion})
to the initial state vector $|\mathbf{R}_{1}\rangle$. 
Note that $|\mathbf{R}_{1}\rangle$ is an eigenstate of $\widehat{U},$ 
\begin{equation}
\widehat{U}|\mathbf{R}_{1}\rangle = 
   U(\mathbf{R}_{1}) \, |\mathbf{R}_{1}\rangle   \;,
\end{equation}
but not of $\widehat{H}$. Therefore, the effect of $e^{-\beta\widehat{H}/L}$
on $|\mathbf{R}_{1}\rangle$ will be to project the state 
$|\mathbf{R}_{1}\rangle$ into a set of different position eigenvectors. 
One has
\begin{equation}
  e^{-\beta\widehat{H}/L} \, |\mathbf{R}_{1}\rangle  \approx 
  e^{-\beta\widehat{U}/2L} \, e^{-\beta\widehat{K}/L} \, 
   |\mathbf{R}_{1}\rangle \, e^{-\beta U(\mathbf{R}_{1})/2L}  \; .
\label{eq:HTA}
\end{equation}
By inserting the resolution of the identity in momentum space
\begin{equation}
 \widehat{I} = \intop d\mathbf{P}\:|\mathbf{P}\rangle\langle\mathbf{P}| \; ,
\end{equation}
and with the eigenvalue equation
\begin{equation}
  \widehat{K} \, |\mathbf{P}\rangle = 
       \frac{P^{2}}{2m} \, |\mathbf{P}\rangle  \; ,
\end{equation}
one gets 
\begin{multline}
  e^{-\beta\widehat{H}/L} \, |\mathbf{R}_1\rangle \approx   \\
  \int d\mathbf{P} \, e^{-\beta\widehat{U}/2L} \, |\mathbf{P}\rangle \, 
  e^{-\beta P^2/2Lm} \, \langle\mathbf{P}|\mathbf{R}_{1}\rangle  \, 
  e^{-\beta U(\mathbf{R}_{1})/2L}  \; .     
\end{multline}
Now, by inserting a second resolution of the identity, this time in
real space
\begin{equation}
\widehat{I} = \intop d\mathbf{R}_{2} \:
       |\mathbf{R}_{2}\rangle\langle\mathbf{R}_{2}|  \; ,
\end{equation}
one obtains

\begin{widetext}

\begin{equation}
  e^{-\beta\widehat{H}/L}| \, \mathbf{R}_{1}\rangle \approx 
 \int d\mathbf{R}_{2} \, 
 e^{-\beta\left[U(\mathbf{R}_{2})+U(\mathbf{R}_{1})\right]/2L} \, 
  |\mathbf{R}_{2}\rangle  \int d\mathbf{P} \, 
  \langle\mathbf{R}_{2}|\mathbf{P}\rangle \, 
  e^{-\beta P^2/2Lm} \, \langle\mathbf{P}|\mathbf{R}_{1}\rangle  \; .
\label{eq:SIA_zw}
\end{equation}
We recall that the momentum representation of a position function is
\begin{equation}
 \langle\mathbf{P}|\mathbf{R}\rangle =
 \left(\frac{1}{2\pi\hbar}\right)^{3/2}e^{i\mathbf{P\cdot R}/\hbar}  \;,
\end{equation}
where $\mathbf{P}\cdot\mathbf{R}$ is the scalar product. By substitution
of this expression and its complex conjugate, 
$\langle\mathbf{R}|\mathbf{P}\rangle,$
into Eq.~(\ref{eq:SIA_zw}) one gets 
\begin{equation}
 e^{-\beta\widehat{H}/L}| \, \mathbf{R}_{1}\rangle \approx
 \left(\frac{1}{2\pi\hbar}\right)^3  \int d\mathbf{R}_{2} \, 
 e^{-\beta\left[U(\mathbf{R}_{2})+U(\mathbf{R}_{1})\right]/2L} \,
 |\mathbf{R}_{2}\rangle  \int d\mathbf{P} \, 
 e^{-\beta P^{2}/2Lm} \, 
 e^{i\mathbf{P}\cdot(\mathbf{R}_{1}-\mathbf{R}_{2})/\hbar}  \;.
\end{equation}
The momentum integral is Gaussian, so that it can be done analytically
with the result
\begin{equation}
 e^{-\beta\widehat{H}/L}|\mathbf{R}_{1}\rangle  \approx
 \left(\frac{mL}{2\pi\beta\hbar\text{\texttwosuperior}}\right)^{3/2}
 \int d\mathbf{R}_{2} \, \exp \left\{ -\beta
 \left[\frac{mL}{2\beta^{2} \hbar^{2}} 
 \left(\mathbf{R}_{1}-\mathbf{R}_{2}\right)^2 + 
  \frac{U(\mathbf{R}_{2}) + U(\mathbf{R}_{1})}{2L}\right] \right\} 
 |\mathbf{R}_{2} \rangle  \; .
\label{eq:SIA}
\end{equation}

\end{widetext}

This approximation for the effect of 
the exponential operator on $|\mathbf{R}_{1}\rangle$ has a clear 
physical meaning. It translates this state into a superposition
of different positions $|\mathbf{R}_{2}\rangle$ that extend over
the whole space. The weight of the new state $|\mathbf{R}_{2}\rangle$
is a \textit{positive} \textit{real} number (that one with exponential
notation in the integrand) that depends on both the squared distance
$\left(\mathbf{R}_{1}-\mathbf{R}_{2}\right)^{2}$ and the potential
energy at the positions $\mathbf{R}_{1}$ and $\mathbf{R}_{2}$. Note
that the term depending on $\left(\mathbf{R}_{1}-\mathbf{R}_{2}\right)^{2}$
results from the kinetic energy operator of the quantum Hamiltonian.

Having the result of Eq.~(\ref{eq:SIA}) for the application of 
$e^{-\beta\widehat{H}/L}$
to the state $|\mathbf{R}_{1}\rangle$, the effect of a new factor
$e^{-\beta\widehat{H}/L}$ is a straightforward repeating loop. This
operator will project the state $|\mathbf{R}_{2}\rangle$ onto a new
position $|\mathbf{R}_{3}\rangle$ with identical weights to those
given in Eq.~(\ref{eq:SIA}), up to a trivial relabeling of the subindex
of the position eigenvectors. Note that the weight of the state
$|\mathbf{R}_{3}\rangle$
will be the product of the weights associated to the transitions
$|\mathbf{R}_{1}\rangle\rightarrow|\mathbf{R}_{2}\rangle$ and 
$|\mathbf{R}_{2}\rangle\rightarrow|\mathbf{R}_{3}\rangle$,
integrated over all $|\mathbf{R}_{2}\rangle$ positions.
Successive applications of the exponential operator 
$e^{-\beta\widehat{H}/L}$ will end after $L$ steps 
with the state $|\mathbf{R}_{L+1}\rangle$.
The weight of $|\mathbf{R}_{L+1}\rangle$ is then the product of the
$L$ weights associated to each step in the chain of transitions 
$|\mathbf{R}_{1}\rangle \rightarrow |\mathbf{R}_{2}\rangle 
  \ldots \rightarrow|\mathbf{R}_{L+1}\rangle$,
integrated over all intermediate positions.

To obtain the partition function of Eq.~(\ref{eq:Z}), the last operation
is to perform the dot product with the bra $\langle\mathbf{R}_{1}|$ and
to integrate over $\mathbf{R}_{1}.$ The orthogonality of the position
eigenstates implies that
\begin{equation}
  \langle \mathbf{R}_{1}|\mathbf{R}_{L+1} \rangle =
  \delta(\mathbf{R}_{1}-\mathbf{R}_{L+1})  \; ,
\end{equation}
where $\delta$ denotes the Dirac delta function. Thus the final state
$|\mathbf{R}_{L+1}\rangle$ is restricted to be identical to the initial
state $|\mathbf{R}_{1}\rangle$. With this condition, i.e., 
$\mathbf{R}_{L+1}=\mathbf{R}_{1}$, the discretized approximation 
of $Z$ is 

\begin{widetext}

\begin{equation}
 Z \approx \left(\frac{mL}{2\pi\beta\hbar^2}\right)^{\frac{3L}{2}}
 \int d\mathbf{R}_{1}\ldots d\mathbf{R}_{L}\, 
 \exp \left\{ -\beta \sum_{i=1}^L \left[ \frac{mL}{2\beta^{2}\hbar^{2}}
 \left(\mathbf{R}_{i}-\mathbf{R}_{i+1}\right)^2 +
 \frac{U(\mathbf{R}_i)}{L} \right] 
 \right\} _{\mathbf{R}_{L+1}=\mathbf{R}_{1}}  \; .
\label{eq:Z_PI}
\end{equation}

\end{widetext}

There are several points to stress. The first one is that the limit
$L\rightarrow\infty$ of this expression converges to the exact
path-integral formulation of $Z$. We refer to the literature for those
readers interested in the continuum limit of this 
expression \cite{fe72,kl90,tu10}.
Quantum PI simulations rely on the use of a discretized form such
as the one in Eq.~(\ref{eq:Z_PI}). A second aspect is that 
Eq.~(\ref{eq:Z_PI}) is formally equivalent to the partition function 
of a classical \textit{ring-polymer} composed of $L$ \textit{beads} or
\textit{replicas} $(\mathbf{R}_{1},\mathbf{R}_{2},\ldots,\mathbf{R}_{L})$
of the same quantum particle. The ring-polymer interacts through an
effective potential given by the sum of two contributions
\begin{equation}
U_{\rm spr} = \left\{ \sum_{i=1}^{L}\frac{mL}{2\beta^{2}\hbar^{2}}
       \left(\mathbf{R}_{i}-\mathbf{R}_{i+1}\right)^{2}
       \right\} _{\mathbf{R}_{L+1}=\mathbf{R}_{1}}    \; ,
\end{equation}

\begin{equation}
U_{\rm int} = \sum_{i=1}^{L}\frac{U(\mathbf{R}_{i})}{L}   \; .
\end{equation}
The first term, $U_{\rm spr}$, represents a harmonic interaction between
neighboring beads at $\mathbf{R}_{i}$ and $\mathbf{R}_{i+1}$. The coupling
constant increases along with the number of beads, the particle mass
and the squared temperature. The second term, $U_{\rm int}$, 
is the total potential energy of the beads with the true Hamiltonian
scaled by a factor $1/L$ (i.e., it corresponds to the mean potential
energy for the true Hamiltonian). 

This suggestive result is the so-called \textit{quantum-classical
isomorphism.} It establishes that the quantum partition function can
be approximated with arbitrary accuracy by means of a classical system
interacting through the effective potential 
\begin{equation}
  U_{\rm eff} = U_{\rm spr} + U_{\rm int}   \; .
\end{equation}
A particularity of the classical isomorph, 
when compared to a real classical system,
is that its potential energy $U_{\rm spr}$ depends explicitly on both
the temperature and particle mass. The classical limit for the potential
$U(\mathbf{R})$ is easily achieved within the discretized PI algorithm
by setting the Trotter number $L$ = 1, so that $U_{\rm spr} \equiv 0$
and $U_{\rm int} \equiv U$.

\subsection{Many-body extension
\label{sub:Many-body-extension}}

The generalization of the previous one-particle results is straightforward
for the case that identical quantum particles are treated by Boltzmann
statistics. The position vectors need now a second index to label
each particle. Thus $\mathbf{R}_{ij}$ denotes the position vector
associated to the $i'$th bead of the $j'$th particle. Assuming for
simplicity that there are $N$ identical particles with mass $m$,
then the harmonic interaction includes a sum over the $N$ particles
\begin{equation}
 U_{\rm spr} = \sum_{j=1}^{N} \left\{ \sum_{i=1}^{L}
  \frac{mL}{2\beta^{2}\hbar^{2}} 
  \left( \mathbf{R}_{ij}-\mathbf{R}_{(i+1)j} \right)^{2} 
  \right\} _{\mathbf{R}_{(L+1)j}=\mathbf{R}_{1j}}  \;.
\label{eq:U_spr}
\end{equation}
The interaction energy term is simply
\begin{equation}
  U_{\rm int} =  \frac{1}{L} \, \sum_{i=1}^{L} U_{i}  \; ,
\label{eq:U_int}
\end{equation}
where $U_{i}$ is the potential energy associated to the $i$'th replica
of the system, composed of $N$ classical particles at 
$(\mathbf{R}_{i1},\mathbf{R}_{i2},\ldots,\mathbf{R}_{iN})$: 
\begin{equation}
U_{i} \equiv  U\left(\mathbf{R}_{i1},\mathbf{R}_{i2},\ldots,
        \mathbf{R}_{iN}\right)  \; .
\label{eq:U_i}
\end{equation}
 $U_{\rm int}$ is then the average potential energy over the $L$ replicas
of the system.

The canonical partition function is approximated as 
\begin{multline}
 Z(NVT) \approx \\   \frac{1}{N!} 
  \left(\frac{mL}{2\pi\beta\hbar\text{\texttwosuperior}}\right)^{3NL/2}
  \int d\Gamma \exp\left[-\beta\left(U_{\rm spr}+U_{\rm int}\right)\right] 
   \; ,
\label{eq:Z_NVT}
\end{multline}
where $d\Gamma = \prod_{j=1}^{N} \prod_{i=1}^{L} d\mathbf{R}_{ij}$, 
and the particle indistinguishability has been considered by the $N!$
term. The thermal behavior of the quantum system can be explored by
applying conventional MC or MD simulation techniques on the classical
ring-polymer isomorph. 

For a series of PI simulations of the same model at different temperatures,
it is sensible to use the same value of $\epsilon$ in the HTA of
Eq.~(\ref{eq:HTA}), so that 
\begin{equation}
  \epsilon = \frac{\beta}{L}=\frac{1}{k_{B}TL}  
\end{equation}
remains independent of $T$. This condition implies that $L$ should
be varied as $T^{-1}$. Thus, the number of beads, $L$, as well as
the computational cost of the simulations, increases as the temperature
lowers. In general, different thermodynamic properties may display
different convergence with the Trotter number $L$ \cite{ce95}.
An empirical rule to estimate a reasonable value of $TL$ is the following.
For a solid where the largest vibrational frequency of its phonon
spectrum is $\omega_{max}$ one should take
\begin{equation}
  \epsilon^{-1} = k_{B} T L \geq 4 \hbar \, \omega_{max}  \; .
\end{equation}
In words this means that the thermal energy corresponding to the 
temperature $T L$ of the HTA must be at least about four times larger 
than the energy quantum of the system under consideration.

\subsection{Path-integral Monte Carlo
\label{sub:Path-integral-Monte}}

For simplicity we consider here a rigid simulation cell, so that volume
fluctuations in the $NPT$ ensemble are restricted to be isotropic.
The general case of a flexible simulation cell can be treated by techniques
that were originally introduced for MD simulations, but can be also
applied to MC methods \cite{pa80,pa81,so97}. In a possible setup of 
the PIMC method in the isothermal-isobaric ensemble, the coordinates
$\mathbf{R}_{ij}$ are updated according to three different types
of sampling schemes. 

The first one consists of trial moves of the individual coordinates
$\mathbf{R}_{ij}$. The trials may be performed sequentially for every
bead coordinate $i$ and every atom $j$. The trial configuration
is accepted with probability
\begin{equation}
  P_{\rm acc} = \mathrm{min \left[1,
   \exp\left(-\beta\triangle\mathit{U_{\rm eff}}\right)\right]}  \; ,
\label{eq:P_accept}
\end{equation}
where $\triangle U_{\rm eff}$ is the difference of $U_{\rm eff}$ 
between the new and old configurations. 

The second type of sampling corresponds to trial moves of the center
of mass (\textit{centroid}) of the ring-polymers
\begin{equation}
\mathbf{X}_{j} = \frac{1}{L} \sum_{i=1}^{L} \mathbf{R}_{ij}  \; ,
\label{eq:centroid}
\end{equation}
which are performed sequentially for every atom $j$ of the simulation
cell. This type of move maintains unaltered the shape of the individual
ring-polymers. This move is essential to explore efficiently 
the configurational space, specially 
at temperatures where thermal fluctuations are significant
for the spatial atomic delocalization \cite{be86}. The acceptance
criterion is the same as in Eq.~(\ref{eq:P_accept}), although for this 
type of move the harmonic interaction, $U_{\rm spr}$, remains unaltered
and needs not be recalculated.

The third type of sampling consists of trial changes in the volume
of the simulation cell. The volume change ($V\rightarrow V_{new}$)
implies that all particle coordinates change accordingly, so that they
are scaled by $s=(V_{new}/V)^{1/3}$. The acceptance probability of
the trial configuration is \cite{fr96}
\begin{multline}
  P_{\rm acc} = \\ \mathrm{min} \left\{ 1,\exp\left(-\beta \left[ 
    \triangle\mathit{U_{\rm eff}} + \mathit{P \, \triangle V - k_{B}T N L
    \ln} \frac{V_{new}}{V}  \right] \right) \right\}  \; .
\end{multline}
For given $T$ and $P$ the maximum change allowed for random moves
of the bead coordinates ($\mathbf{R}_{ij}$), ring-polymer centroids
($\mathbf{X}_{j}$), and volume is typically adjusted to yield an
acceptance ratio of attempted moves of about 50\% for each kind of
sampling. 

Alternative methods for PIMC simulations are formulated based on the
Fourier representation of the ring-polymer configurations
\cite{fr84,co86,ch98b},
and also on combined schemes using both real and Fourier space
representations of ring-polymers \cite{vo97}.

The trajectory generated by the MC algorithm is made of a set of 
configurations, each one defined by the volume and the bead positions 
$\left(V,\left\{ \mathbf{R}_{ij}\right\} \right)$.
For each configuration it is straightforward the estimation of several
important observables, as the kinetic energy, internal energy, and
pressure. The expression of these estimators in either $NPT$ or $NVT$
ensembles is the same. The average kinetic energy, 
$\langle K \rangle$, is derived from the thermodynamic relation
\begin{equation}
\langle K \rangle = \frac{m}{\beta} \, \frac{\partial\ln Z}{\partial m} \; .
\label{meank}
\end{equation}
Applying this expression to Eq.~(\ref{eq:Z_PI}) results on the primitive
estimator of the kinetic energy 
\begin{equation}
K_{p} = \frac{3NL}{2\beta} - U_{\rm spr}    \; ,
\end{equation}
where the $U_{\rm spr}$ is calculated with Eq.~(\ref{eq:U_spr}). Note
that we have written the result in terms of $K_{p}$ instead of 
$\left\langle K_{p}\right\rangle $, since the former refers to the 
kinetic energy estimator of a single configuration 
$\left(V,\left\{ \mathbf{R}_{ij}\right\} \right)$,
while the latter denotes the ensemble average, i.e., an average over
the whole MC trajectory. The virial estimator of the kinetic energy
is physically equivalent, but it displays an improved numerical convergence
\cite{he82,pa84} 
\begin{equation}
 K_{v} = \frac{3N}{2\beta} + \frac{1}{2L} \sum_{j=1}^{N} \sum_{i=1}^{L}
  \left(\mathbf{R}_{ij}-\mathbf{X}_{j}\right) \cdot 
  \nabla_{\mathbf{R}_{ij}} U_{i}  \; ,
\label{eq:kin_vir}
\end{equation}
where $-\nabla_{\mathbf{R}_{ij}}U_{i}$ is the force on the bead located
at $\mathbf{R}_{ij}$, i.e., the force on the atom $j$ of the $i$'th
replica of system, as derived from the potential function 
$U(\mathbf{R}_{i1},\mathbf{R}_{i2},\ldots,\mathbf{R}_{iN})$
in Eq.~(\ref{eq:U_i}). The position vector $\mathbf{X}_{j}$ is the
centroid of the atom $j$ [see Eq.~(\ref{eq:centroid})]. Note
that while the virial estimator requires the calculation of the atomic
forces at all bead positions, the primitive estimator needs only to
know the harmonic energy $U_{\rm spr}$. The estimator of the internal
energy is derived analogously from
\begin{equation}
  \langle E \rangle = - \frac{\partial \ln Z}{\partial \beta}
\end{equation}
as
\begin{equation}
E = K_{v} + U_{\rm int}   \; ,
\end{equation}
where $U_{\rm int}$ is given in Eq.~(\ref{eq:U_int}). 

The pressure estimator is derived from
\begin{equation}
\left\langle \mathcal{P} \right\rangle = 
   \frac{1}{\beta} \, \frac{\partial \ln Z}{\partial V}
\end{equation}
with the result 
\begin{equation}
  {\cal P} = \frac{NL}{\beta V} - \frac{2}{3V} U_{\rm spr} -
    \frac{1}{L} \sum_{i=1}^{L} \frac{\partial U_{i}}{\partial V}  \, .
\label{eq;P_esti_PIMC}
\end{equation}
The volume derivative can be calculated either analytically or numerically
by considering that an increase of the actual volume by $\triangle V$
will change each bead coordinate of the simulation cell by a scaling
factor $s = (1+\Delta V/V)^{1/3}$. 

A final important relation for the isothermal compressibility $\kappa_T$, 
valid in the $NPT$ ensemble, is the following:
\begin{equation}
 \kappa_T = - \frac{1}{\langle V \rangle} \, 
   \frac{\partial \langle V \rangle}{\partial P} = 
   \frac{\beta \, \sigma_V^2}{\langle V \rangle}  \; ,
\label{eq:iso_compressibility}
\end{equation}
where the volume fluctuations are
\begin{equation}
  \sigma_V^2 =  \langle V^2 \rangle - \langle V \rangle ^2  \; .
\end{equation}
A deep analysis on the derivation of compressibilities from PI
simulations was presented by Ses\'e \cite{se03}.

\subsection{Path-integral molecular dynamics
\label{sub:Path-integral-molecular} }

First of all, it is important to make clear that the dynamics considered 
in the PIMD method is artificial. 
It is suitable for efficiently sampling the partition
function in Eq.~(\ref{eq:Z_NVT}) and thus for evaluating static or
equilibrium properties, but it does not represent the real quantum
dynamics of the systems under investigation.

An efficient implementation of classical dynamics to sample the microstates
associated to the discretized partition function in Eq.~(\ref{eq:Z_NVT})
requires the use of numerical techniques adapted to this problem.
The potential term in Eq.~(\ref{eq:U_spr}) implies that we have a
cyclic chain of coupled harmonic oscillators giving rise to a wide
spectrum of different vibrational frequencies. Integration of the
dynamic equations is numerically more stable if all modes oscillated
in the same time scale. Besides, if the number of beads $L$ increases,
then for a given replica the harmonic coupling constant 
in Eq.~(\ref{eq:U_spr}) becomes larger while the effective 
interaction potential on each bead becomes smaller,
because of the factor $L^{-1}$ in Eq.~(\ref{eq:U_int}). Dominance
of harmonic forces makes that ergodicity problems can be present in
the classical bead dynamics. 

The set-up of effective algorithms to perform PIMD simulations has
been described in detail by Martyna~{\em et al.} \cite{ma96,ma99b}
and by Tuckerman \cite{tu02,tu10}. The main steps of a possible
implementation are based on the use of: $i)$ staging variables to
bring harmonic modes to the same time scale; $ii)$ massive thermostatting
techniques to avoid ergodicity problems; $iii)$ reversible integrators
and multiple time steps to integrate the classical equations of motion
by a factorization of the classical Liouville operator. A detailed
account of these algorithms is outside the scope of the present review
so that we limit ourselves to present a brief overview of this PIMD
implementation.

Staging bead coordinates $u_{ij}$ are defined by a linear transformation
that diagonalizes the harmonic energy in Eq.~(\ref{eq:U_spr})
\cite{tu98,tu02}.
For a given atom $j$ the staging mode coordinates are defined as
\begin{equation}
\mathbf{u}_{1j} = \mathbf{R}_{1j}   \; ,
\end{equation}

\begin{equation}
\mathbf{u}_{ij} = \mathbf{R}_{ij} - \frac{i-1}{i} \, \mathbf{R}_{(i+1)j} - 
        \frac{1}{i} \, \mathbf{R}_{1j} \;, \; i=2,\dots,L   \; .
\end{equation}
The masses of the staging modes are defined as 
\begin{equation}
    m_{1} = 0,
\end{equation}
\begin{equation}
  m_{i} = \frac{i}{i-1} \, m  \;,\;  i = 2,\dots,L   \; ,
\end{equation}
where $m$ is the mass of atom $j$. The harmonic term in 
Eq.~(\ref{eq:U_spr}) is transformed to a diagonal form with the staging 
coordinates: 
\begin{equation}
  U_{\rm spr} = \sum_{j=1}^{N}\sum_{i=2}^{L}\frac{m_{i}L}{2\beta
       \text{\texttwosuperior}\hbar^{2}} \, \mathbf{u}_{ij}^{2}   \; .
\end{equation}
Momentum variables for the MD algorithm are introduced through the
substitution of the prefactor in the partition function in 
Eq.~(\ref{eq:Z_NVT}) by a Gaussian integral over momentum coordinates
\begin{equation}
  \left(\frac{mL}{2\pi\beta\hbar^{2}}\right)^{3NL/2}=C\int\prod_{j=1}^{N}
    \prod_{i=1}^{L}d\mathbf{P}_{ij}
    \exp\left(-\frac{\beta \, \mathbf{P}_{ij}^{2}}{2 \, m_{i}'}\right)  \; .
\end{equation}
$\mathbf{P}_{ij}$ is the momentum associated to the $i$'th staging
mode of the $j$'th particle, while $C$ is a constant that depends
on the masses and has no influence in the calculated equilibrium
properties. The masses $m_{i}'$ are chosen so that the staging modes
with $i = 2,\ldots L$ move on the same time scale
\begin{equation}
  m'_{1} = m\;,
\end{equation}
\begin{equation}
  m'_{i} = m_{i}, \; \mathrm{f\mathrm{or}} \; i = 2,\ldots,L  \; .
\end{equation}
With this extension, the sampling of the partition function in 
Eq.~(\ref{eq:Z_NVT}) is made now by a classical MD algorithm in the phase
space expanded by the set of coordinates 
$\left\{ \mathbf{u}_{ij},\mathbf{P}_{ij} \right\}$.
An additional advantage of the staging modes is that both 
the transformation
$\left\{ \mathbf{u}_{ij}\right\} \rightarrow
   \left\{ \mathbf{R}_{ij}\right\}$
and its inverse are easily done by simple recursion relations \cite{tu93}.
An alternative to the staging variables is based on the normal mode
transformation of the bead coordinates \cite{tu98}.

The temperature control in either the $NVT$ or $NPT$ ensembles is
achieved by a massive thermostatting of the system that implies a chain 
of Nos\'e-Hoover thermostats coupled to each of the staging variables
$\mathbf{u}_{ij}$ \cite{ma92,tu98}. The thermostat `mass'
parameter, $Q$, is set to evolve in the scale of the harmonic bead
forces as \cite{tu98} 
\begin{equation}
  Q = \frac{\beta \, \hbar^2}{L}  \; .
\label{eq:mass_ther_Q}
\end{equation}
In the case of the $NPT$ ensemble, the volume itself is a dynamic
variable coupled to a chain of barostats in order to maintain a constant
pressure. The volume and barostat `mass' parameters are set 
as \cite{ma99b}
\begin{equation}
  W = \frac {3 \, (NL+1)} {\beta \, \omega_b^2}  \; ,
\end{equation}

\begin{equation}
  Q_p = \frac {1} {\beta \, \omega_b^2}  \; .
\end{equation}
The frequency $\omega_{b}$ is taken to be 1 ps$^{-1}$.

The dynamic equations in the $NPT$ ensemble are the time derivatives
of both position and momentum coordinates of each dynamic degree of
freedom of the extended system, i.e., the staging modes, the chains
of Nos\'e-Hoover thermostats, the volume, and the chain of 
barostats \cite{ma99b}.
The thermostats and barostats introduce friction terms in the dynamic
equations, which cause that the extended dynamics is not Hamiltonian.
Nevertheless there is a quantity, playing the role of a total energy
in the extended system, that is conserved along the time evolution.
The conservation of this energy is a useful check for the numerical
integration of the equations of motion.

The reversible extended-system multiple time-step integration is well
documented in Refs.~\cite{ma99b,tu98,tu06,tu92}. The
factorization of the classical Liouville propagator, 
$\exp \left(t\frac{\partial}{\partial t}\right)$, appears at first 
sight rather complicated, but it is in fact straightforward and 
relatively easy to implement as a computer code. For this implementation
it is advisable to start programming the factorization of the exponential
propagator in the microcanonical $NVE$ ensemble, as both thermostat
and barostat variables are absent. Having a checked $NVE$ integrator,
one can extend the algorithm to the $NVT$ ensemble, which implies
to add the somewhat lengthy factorization associated to the integration
of the Nos\'e-Hoover chains. The next step would be to include both
volume and the chain of barostats as dynamic variables to allow for
$NPT$ simulations. The last step of this build-up would be to allow
for the full flexibility in the simulation cell of the $NPT$ 
ensemble \cite{ma99b}. 

The estimators in the PIMD method are similar to those presented for
PIMC . The only difference is the term associated to the momentum
variable. For example, the PIMC pressure estimator shown 
in Eq.~(\ref{eq;P_esti_PIMC}) changes to 
\begin{equation}
  {\cal P} = \frac{2}{3V} \sum_{i=1}^{L} \sum_{j=1}^{N}
   \frac{\mathbf{P}_{ij}^{2}}{2 \, m'_{i}} -
   \frac{2}{3V} \, U_{\rm spr} - \frac{1}{L}
   \sum_{i=1}^{L} \frac{\partial U_{i}}{\partial V}  \; .
\label{eq:P_est_PIMD}
\end{equation}
Note that the equipartition principle implies that the ensemble average
of the first term on the right-hand side is
\begin{equation}
  \sum_{i=1}^{L} \sum_{j=1}^{N}
    \left\langle \frac{\mathbf{P}_{ij}^{2}}{2 \, m'_{i}} \right\rangle 
    = \frac{3NL}{2 \, \beta}  \; ,
\end{equation}
so that the PIMC and PIMD pressure estimators in Eqs.~(\ref{eq;P_esti_PIMC})
and (\ref{eq:P_est_PIMD}) should provide identical expectation values
in converged equilibrium simulations.

\subsection{PIMC versus PIMD: solid Ne simulations
\label{sub:PIMC-versus-PIMD}}

PIMC is conceptually simpler than PIMD. First, it requires to
work in the configurational space expanded by the bead coordinates,
while PIMD needs also bead velocities. Second, the algorithm for MC
simulation requires to know the potential energy of the system but
not the atomic forces, contrary to the PIMD method. And third, the
control of external parameters, as the pressure and temperature, is
easier in a MC approach. In addition, an efficient treatment of particle
exchange (fermionic or bosonic degrees of freedom) is only possible
in a MC approach \cite{ce95}.

In spite of the larger simplicity of the PIMC algorithm, the code
of a PIMC program may require more effort than that of a PIMD method.
The reason is that in a MC sampling one needs to calculate the potential
energy change associated to random moves of the system. An efficient
code restricts this calculation only to the energy terms affected
by the move, avoiding to recalculate those that remain invariant.
E.g., for a pairwise potential a random move of a particle requires
to calculate the interaction energy of the moving particle with the
rest of the system, all other pairwise interactions remain invariant
and need not to be recalculated. Thus, an efficient MC implementation
can not be done for an arbitrary potential model, but requires the
development of a specific computer code for both the particular move
type and the model at hand. 

Then, the PIMC approach might be a disadvantage if one needs to use
different potential models, because it requires usually writing and
checking large parts of new computer code. However, such implementation
is straightforward in PIMD. Here the update algorithm (integration
of the equations of motion of both position and momentum coordinates) 
is coupled to the potential model only
by the transfer of an array with the actual value of the atomic forces.
Thus, implementation of a new model is easier as it does not affect
the general structure of a PIMD code. 

Another important difference between both methods is that a typical
MC move implies a \textit{sequential} updating of each bead coordinate.
In contrast, in a MD simulation all bead coordinates are simultaneously 
updated at each time step. 
Then, as the number of atoms increases over certain
boundary, trajectories generated by PIMD may explore the configuration
space much faster than PIMC, which translates into lower requirements
of computational resources. Besides, the global updates of the PIMD
method, as opposed to PIMC, imply an easier implementation of parallelization
strategies in the computer code.

\begin{figure}
\vspace{-1.5cm}
\includegraphics[width= 9cm]{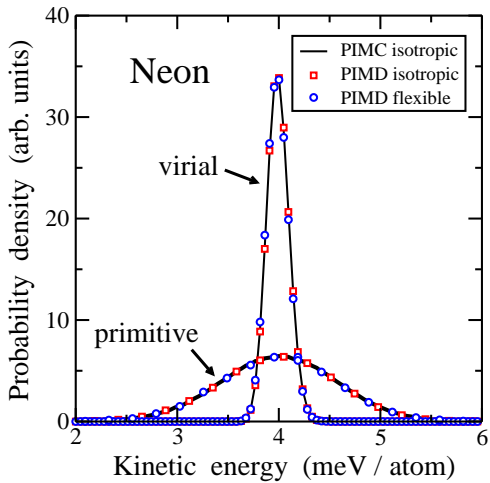}
\vspace{-0.8cm}
\caption{
Probability distribution of the kinetic energy per atom for solid neon
at $T = 20$~K and $P = 0$, as derived by primitive
and virial estimators. The results correspond to three different $NPT$
simulation methods: PIMC and PIMD with isotropic volume fluctuations,
as well as PIMD with full cell flexibility.
}
\label{fig_teo1}
\end{figure}

To compare results obtained from both methods with a simple interatomic
potential, PIMC and PIMD test simulations are presented here for solid Ne. 
The noble-gas atoms are treated as particles of mass $m = 20.18$ amu 
(mean mass for natural isotopic composition)
interacting through a Lennard-Jones 6-12 potential of the form
\begin{equation}
U(r) = 4 \, \epsilon \left[ \left( \frac{\sigma}{r} \right)^{12} -
     \left( \frac{\sigma}{r} \right)^6 \right] \; ,
\label{ljpot}
\end{equation}
with parameters $\epsilon = 3.2135$~meV and $\sigma = 2.782$~\AA. 
This pair potential is truncated at distances $r$ larger than 
$r_c = 2.5 \, \sigma$. The simulation cell is a $3\times3\times3$ supercell 
of the standard fcc unit cell, containing $N = 108$ atoms subject to 
periodic boundary conditions. Standard long-range corrections 
were computed assuming that the pair correlation
function is unity for $r > r_c$, leading to well-known corrections
for the pressure and internal energies \cite{jo93}. The number
of beads was set by the condition $L \, T = 200$~K. A PIMC run
consisted of $10^6$ MC steps with a previous equilibration of $10^{5}$
steps. Each step included: 
$i)$ a sequential trial move of each of the $N \times L$ beads; 
$ii)$ a sequential trial move of each centroid in the $N$ ring-polymers; 
$iii)$ a trial move of the simulation cell volume. 

PIMD simulations were performed with $10^{6}$ MD steps with
an equilibration of $10^{5}$ steps. A chain of four Nos\'e-Hoover
thermostats was coupled to each of the staging variables and a chain
of four barostats was coupled to the volume. The equations of motion
were integrated using a time step of $\triangle t=3$ fs for both the 
Lennard-Jones and barostat interactions, and a smaller one of 
$\triangle t/4$ for both the thermostat and harmonic interactions. 
PIMD simulations in the $NPT$ ensemble were performed by allowing either 
isotropic volume fluctuations or full cell flexibility. In the latter case 
the simulation cell was not allowed to rotate in space. With this 
restriction the number of independent degrees of freedom to describe 
a fully flexible unit cell is reduced from 9 to 6 \cite{cita}.

The equivalence between PIMC and PIMD simulations in the $NPT$ ensemble
is illustrated by the results on solid Ne. In both simulation methods
we allow for isotropic volume fluctuations, and in the case of PIMD
we additionally consider simulations with a full flexibility of the
simulation cell. In Fig.~\ref{fig_teo1} the distribution function
of the kinetic energy of the solid per simulation cell is presented at 
$T=20$~K and $P = 0$, using both the virial and the primitive estimators.
For each estimator the results obtained by both PIMC and PIMD methods
are shown. Notice that the virial estimator produces a significantly
narrower kinetic energy distribution. However, the expected value of
the kinetic energy is, within the statistical noise, identical for
both estimators. The agreement between PIMC and PIMD simulations, as
well as between PIMD data for isotropic volume fluctuations and full
cell flexibility, is a strong evidence that the correct ensemble 
distribution is being explored in the various simulations.

\begin{figure}
\vspace{-1.5cm}
\includegraphics[width= 9cm]{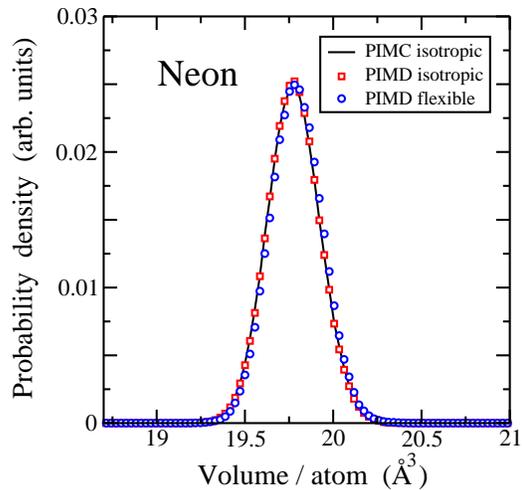}
\vspace{-0.8cm}
\caption{
Probability distribution of the volume for the $NPT$ simulation
of solid Ne at $T = 20$~K and $P = 0.2$~GPa. The data were derived by
PIMC and PIMD with isotropic volume fluctuations, as well as by PIMD
with full cell flexibility.
}
\label{fig_teo2}
\end{figure}

This conclusion is further supported by the probability density 
of the volume of solid Ne at $T=20$~K and $P=0.2$~GPa displayed
in Fig.~\ref{fig_teo2}. An unbiased sampling should provide the expected
value of equilibrium properties and also the fluctuations of 
the thermodynamic
properties according to the employed ensemble. We recall that in the
$NPT$ ensemble the volume fluctuation has an important physical
meaning, as it is related to the compressibility of the solid 
[see Eq.~(\ref{eq:iso_compressibility})].
The volume distributions shown in Fig.~\ref{fig_teo2} provide a non-trivial
test for the pressure control in the PIMD method. This pressure control
is much easier, and less prone to error, using a PIMC algorithm. Thus,
the agreement found between the volume sampled by both methods is
a necessary condition for the correctness of the employed algorithms.
Also relevant is the agreement found in the PIMD simulations using
fully flexible and rigid simulation cells. We stress that a pre-requisite
for this agreement is to set the correct number of degrees of freedom
in the dynamic equation of the barostat coupled to the volume \cite{cita}.

\begin{figure}
\vspace{-1.5cm}
\includegraphics[width= 9cm]{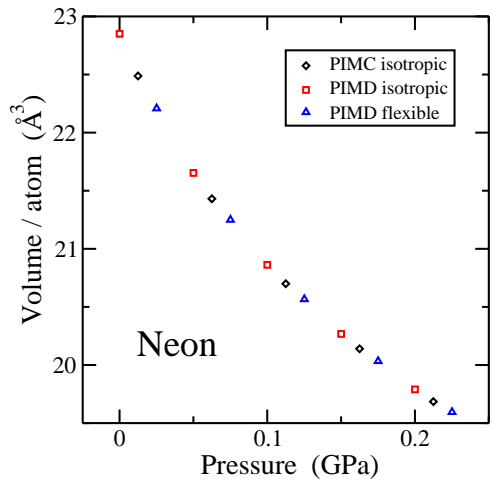}
\vspace{-0.8cm}
\caption{
Equation of state $P$--$V$ of solid Ne at 20~K. The data correspond to
$NPT$ simulations by three different methods: PIMC and PIMD with
isotropic volume fluctuations, as well as PIMD with full cell
flexibility.
}
\label{fig_teo3}
\end{figure}

The equation of state $P$--$V$ of solid Ne at $T = 20$~K is presented
in Fig.~\ref{fig_teo3} for pressures in the range from 0 to 0.23~GPa.
The state points were studied by both PIMC and PIMD simulations, and
in the latter case by sampling either isotropic or fully flexible volume
fluctuations. Note the good agreement obtained between the three sets of
simulations.

\subsection{Quantum transition-state theory
\label{sub:Quantum-Transition-State}}

Classical transition-state theory is a well-established approximation
for the calculation of rate constants associated to the kinetics of
infrequent events. For the activated diffusion of an impurity in a
solid, the key element in this theory is the calculation of a ratio
between the probability for finding the impurity at a barrier region
($B$) and at its stable configuration ($S$), where the barrier region
must be specified for the given kinetic process. A quantum generalization
of classical transition-state theory has been developed based on the 
PI formulation. According to this approach, 
the calculation of jump rates depends on the ratio,
$P_{c}$, between the probabilities of finding the diffusing
particle at the saddle-point and at the stable site. Thus, the jump
rate constant is given by
\begin{equation}
k = \frac{1}{2} \, \overline{v} \, \frac{P_{c}}{l}  \; ,
\end{equation}
where $\overline{v}$ is a factor weakly dependent on temperature,
taken to be the thermal velocity $\overline{v}=\sqrt{2/(\pi\beta m)}$
of the jumping particle, and $l$ is the jumping distance.

The probability ratio $P_{c}$ is proportional to the Boltzmann factor
$\exp(-\beta\triangle F)$, where $\triangle F$ is the free energy
barrier for the activated process. Thus, the jump rate (a dynamical
quantity) is closely related to the free energy difference $\triangle F$
(a time-independent quantity), so that its temperature dependence
can be calculated from equilibrium simulations without any dynamical
information.

A direct estimation of the probability $P_{c}$ by a simulation is
difficult in the case that the diffusion event is infrequent. A convenient
approach to this calculation was first proposed by Gillan \cite{gi87b}
as the reversible work done on the system when the centroid of the
diffusing particle ($\mathbf{X}_{j}$) moves along the diffusion path
$S\rightarrow B$
\begin{equation}
P_c = \exp \left( \beta \int_S^B  \mathbf{f}(\mathbf{X}_{j}) \, 
      d\mathbf{X}_{j} \right)
\label{eq:delta_f}
\end{equation}
where $\mathbf{f}(\mathbf{X}_{j})$ is the mean force acting on the
jumping impurity with its centroid fixed on $\mathbf{X}_{j}$
\begin{equation}
  \mathbf{f}(\mathbf{X}_{j}) = - \left\langle \frac{1}{L} \sum_{i=1}^{L}
   \nabla_{\mathbf{R}_{ij}}U_{i}\right\rangle _{\mathbf{X}_{j}}
\label{eq:mean_force}
\end{equation}
The meaning of $\nabla_{\mathbf{R}_{ij}}U_{i}$ was explained after
Eq.~(\ref{eq:kin_vir}). The average value in Eq.~(\ref{eq:mean_force})
is taken over a sampling of configurations where the ring-polymer
of the jumping impurity is constrained to be fixed at $\mathbf{X}_{j}$.
Note that by fixing the centroid of the impurity, we are suppressing
its classical spatial delocalization. The integration 
in Eq.~(\ref{eq:delta_f})
is performed along a path $S\rightarrow B$ typically discretized
into about ten equidistant $\mathbf{X}_{j}$ points. Quantum effects
that may give rise to substantial deviations from the classical jump
rate are taken into account within the QTST. In this way, jump
rates for kinetic processes can be readily approximated for realistic,
highly nonlinear many-body problems, even at quite low temperatures.
Thus, this technique provides a methodology to study the influence of
vibrational mode quantization and quantum tunneling on impurity
jump rates.

In contrast to the case of equilibrium properties, where PI methods are
well established and controllable (at least for distinguishable and
bosonic particles), the PI simulation of time-dependent properties still
remains as an open numerical problem. 
Several approximations, such as centroid MD \cite{ca94,ca94b,ja99} and 
ring-polymer MD \cite{cr04}, have been formulated to tackle this problem, 
but there is at present no widespread consensus on their general reliability. 
We refer to the recent literature for those readers interested in this 
topic \cite{pe09,ha13}.

\subsection{Linear-response approach for vibrational frequencies
\label{sub:Linear-response}}

An interesting application of path-integral simulations to solids
is the calculation of vibrational frequencies by using the static
isothermal susceptibility tensor $\chi^{T}$ \cite{ra05}.
This tensor gives the linear response (LR) of a system (e.g., solid or
molecule) in thermal equilibrium to vanishingly small forces applied 
on the atomic nuclei.
For solids, in particular, this method offers a practical approach 
to derive phonon energies
by a non-perturbative method, and the representation of the response 
function within the path-integral formulation offers a simple way for 
its numerical calculation \cite{ra01,lo03}.
In fact, $\chi^{T}$ can be readily derived
from PI simulations of a solid at equilibrium, without having to
explicitly impose any external forces during the simulation.
This LR approach is able to realistically reproduce
vibrational properties that are strongly affected by anharmonicity, and
thus represents a significant improvement as compared to
the standard harmonic approximation.
A sketch of the method is given in the following.

Let us denote the set of $N$ centroid positions 
$\left\{ \mathbf{X}_{1},\ldots,\mathbf{X}_{N}\right\} $
of the atoms in the simulation cell as a vector with components $X_{j}$
($j=1,\dots,3N$). See Eq.~(\ref{eq:centroid}) for the centroid definition.
The susceptibility tensor ${\chi}^{T}$, with dimensions $3N\times3N,$
is defined in terms of the centroid coordinates as \cite{ra01} 
\begin{equation}
  \chi_{ij}^{T} =  \beta \, \sqrt{m_{i}m_{j}} \; \mu_{ij}  \; ,
\label{eq:chi_3d}
\end{equation}
where $m_{i}$ is the mass of the atom associated to component $i$, 
$\mu_{ij} = \langle X_{i}X_{j}\rangle - 
            \langle X_{i}\rangle\langle X_{j}\rangle$
is the covariance of the centroid coordinates $X_{i}$ and $X_{j}$,
and $\langle\dots\rangle$ indicates an ensemble average.

The tensor ${\chi}^{T}$ allows us to derive a LR approximation to
the excitation energies of the vibrational system, that is applicable 
even to highly anharmonic situations. The LR approximation
for the vibrational frequencies reads 
\begin{equation}
  \omega_{n,{\rm LR}} =  \frac{1}{\sqrt{\Delta_n}}   \; ,
\end{equation}
where $\Delta_{n}$ ($n=1,\dots,3N$) are the eigenvalues of $\chi^{T}$,
and the LR approximation to the first excitation energy of vibrational
mode $n$ is given by $\hbar \, \omega_{n,{\rm LR}}$. More details on 
the method and illustrations of its ability for predicting vibrational 
frequencies of solids and molecules can be found elsewhere 
\cite{ra01,ra02,lo03,ra05}.

\subsection{Free-energy calculation}

The simulation of the phase diagram and phase coexistence properties
of a given material requires the calculation of its free energy. 
Thermodynamic integration (TI) is one of the most widely employed methods 
to compute free energies \cite{ki35,al87,fr96}. It is based on
the construction of a thermodynamic path defined by making the canonical 
partition function $Z_{\lambda}$ to depend upon a control parameter
$\lambda$ that is varied in the range [0,1]. The Helmholtz free
energy is defined as
\begin{equation}
  F_{\lambda} = -k_{B} T \ln Z_{\lambda}  \; .
\label{eq:F_lambda}
\end{equation}
The free energy difference between the final ($\lambda=1$) and initial
states ($\lambda=0$) may be calculated as
\begin{equation}
  \triangle F = \int_0^1 d\lambda 
  \left( \frac{\partial F_{\lambda}}{\partial\lambda}\right)  \; .
\label{eq:delta_F_def}
\end{equation}
Here $\triangle F$ is the reversible work performed on the system along
the considered thermodynamic path, where $\lambda$ behaves as a generalized 
displacement and $\partial F_{\lambda}/\partial\lambda$ is a generalized 
force. If the initial state ($\lambda = 0$) is a reference model of known
free energy, then Eq.~(\ref{eq:delta_F_def}) allows us to obtain the 
free energy of the system for $\lambda = 1$. Specializing to the particular 
case of quantum systems described with the PI formulation, a thermodynamic
path may be defined by coupling the potential energy $U_{\rm int}$ of
Eq.~(\ref{eq:U_int}) with that of a reference system, $U_{\rm int,ref}$, as
\begin{equation}
  U_{\rm int,\lambda} = \lambda \, U_{\rm int} + 
           (1 - \lambda) \, U_{\rm int,ref}  \;.
\label{eq:U_int_lambda}
\end{equation}
Then $\lambda = 0$ corresponds to the reference system, while $\lambda=1$
is associated to the system of interest. Intermediate values of $\lambda$
correspond to a fictitious system resulting from the coupling between both
limiting cases. If $Z_{\lambda}$ is defined by Eq.~(\ref{eq:Z_NVT})
using $U_{\rm int,\lambda}$ as interaction potential, then the Helmholtz
free energy is derived from Eqs.~(\ref{eq:F_lambda}) and 
(\ref{eq:delta_F_def}), giving the result
\begin{equation}
\triangle F = \int_0^1 d\lambda \, \left< \frac{\partial U_{\rm int,\lambda}}
      {\partial\lambda} \right>_{\lambda} = \int_0^1 d\lambda \,
      \langle U_{\rm int}-U_{\rm int,ref} \rangle _{\lambda}  \; ,
\label{eq:rev_work}
\end{equation}
where the brackets $\left\langle \cdots\right\rangle _{\lambda}$
designate an ensemble average for a given value of the coupling parameter
$\lambda$. If the averages were calculated in the $NPT$ ensemble,
then the result on the rhs of the last equation would be the Gibbs
free energy difference $\triangle G$.

A typical reference system for solid phases is the Einstein crystal,
whose free energy is analytic \cite{fr84b,po00}. Here the
atoms are assumed to be fixed to their equilibrium positions, 
$\mathbf{R}_{i}^{eq}$,
by means of harmonic springs and the interaction energy term is
\begin{equation}
  U_{\rm int,ref} = \frac{1}{L}  \sum_{j=1}^{N} \sum_{i=1}^{L} 
        \frac{1}{2} \, m \, \omega^2
        \left| \mathbf{R}_{ij}-\mathbf{R}_{i}^{eq} \right|^{2}  \; ,
\end{equation}
where $\omega$ is the frequency of an Einstein oscillator of mass
$m$. Another useful reference system is defined as
\begin{equation}
  U_{\rm int,ref} = U(\mathbf{X}_{1},\mathbf{X}_{2},\ldots,\mathbf{X}_{N})
    \; ,
\end{equation}
where the interaction energy is a function of the centroid positions.
Morales and Singer have shown that a TI with this reference potential can 
be used to calculate the excess free energy of the quantum system with 
respect to the classical limit \cite{mo73b}.

Interestingly one can use the harmonic interaction term $U_{\rm spr}$ in 
Eq.~(\ref{eq:U_spr}) to define the thermodynamic path. One possibility 
is to use a changing atomic mass 
\begin{equation}
  m_{\lambda} = \lambda \, m + (1-\lambda) \, m_0\;.
\end{equation}
where $m$ and $m_{0}$ are the final (actual) and the reference atomic 
masses. Using the mass $m_{\lambda}$ to define $Z_{\lambda}$, the change
in free energy along the considered thermodynamic path is given by
\begin{multline}
 \triangle F = (m-m_{0}) \int_{0}^{1} d\lambda
     \left(- \, \frac{\langle K \rangle _{\lambda}}{m_{\lambda}} \right)
 = \\ - \int_{m_{0}}^{m} dm_{\lambda} \, \frac{\langle K \rangle _{\lambda}}
     {m_{\lambda}}   \; ,
\end{multline}
where $K$ is the kinetic energy [see Eq.~(\ref{meank})].
The last expression was derived by a simple change of variable, 
and is useful to study isotope effects in phase coexistence properties 
by PI simulations. Moreover, extending the TI to the limit of high 
atomic mass ($m_0 \to \infty$) provides 
an alternative to the Morales-Singer method for the calculation of
quantum excess free energies \cite{ra10}.

An interesting alternative to TI is the adiabatic switching
method \cite{wa90}, in which the coupling parameter $\lambda$ is 
varied on the fly during a simulation, so that the work along the 
thermodynamic path is computed using the `instantaneous' generalized 
force $\partial U_{\rm int,\lambda}/\partial\lambda$, and one has
\begin{equation}
  \triangle F \leq  \int_{0}^{1} d\lambda \, 
  \frac{\partial U_{\rm int, \lambda}}{\partial \lambda}  \; ,
\label{eq:irrev_work}
\end{equation}
instead of the canonical average 
$\langle \partial U_{\rm int,\lambda}/\partial \lambda \rangle _{\lambda}$
in Eq.~(\ref{eq:rev_work}). Here the parameter $\lambda$ is changed
at a uniform rate along a single simulation run, so that the system
is no longer in equilibrium. This leads to dissipation of energy, and
therefore the computed work is not reversible, hence the inequality
in Eq.~(\ref{eq:irrev_work}). The idea here is that, in order to
minimize the effects of dissipation, the switching should be 
quasi-adiabatic, or in other words, the simulation run should 
be long. An advantage of this method with respect 
to standard TI is that the free energy difference can be determined 
from a single simulation run.

\section{Noble-gas solids}

Noble-gas solids provide us with systems allowing fruitful 
comparisons between experiment and theory.
The simplicity of these solids makes them specially interesting for 
detailed studies of structural, vibrational, and thermodynamic properties. 
The interatomic forces are weak, short ranged, and fairly well
understood, so that one can rather easily check the ability of 
theories to predict properties of noble-gas crystals.
In particular,
the thermodynamic properties of these weakly bound solids are
interesting due to the large anharmonicity of their lattice vibrations.
 Solid helium is an extreme case where short-range quantum correlation
effects are important. For heavier elements, quantum effects are
less significant, but some of them can still be observable at low
temperatures, even for solid xenon, due to the anharmonicity 
of the lattice dynamics.

Several effective interatomic potentials have been employed to study 
noble-gas solids. The most popular is the Lennard-Jones pair potential given 
in Eq.~(\ref{ljpot}), with parameters $\epsilon$ and $\sigma$ slightly
differing in several works \cite{va98,cu97,cu93,mu95,th84,he02,he03a}.
Other model potentials have been employed for the effective interaction 
between noble-gas atoms, such as Aziz-type pair potentials
\cite{az95,ti96,ce95} and three-body interactions 
\cite{lo87,bo94b,ch01,he06c}.

\subsection{Helium}

Among the most known applications of path integrals in condensed
matter are the different phases of helium, in particular superfluid
$^4$He.  One can accurately calculate the properties of helium
using path-integral methods, since the interatomic potential is
well known \cite{ba89,bo94b}.
Moreover, for bosonic and distinguishable particle systems, these
methods provide one  with equilibrium properties directly from
an assumed Hamiltonian, without significant approximation.
As mentioned in the Introduction, nuclear exchange is not relevant in
general for the solids considered here, and
even in the case of solid $^3$He, effects of Fermi statistics may be
neglected for $T >$~0.1~K and densities slightly away
from melting \cite{dr00}.

Anharmonic effects in solid helium are expected to be appreciable,
due to the low atomic mass and weak interatomic forces, which cause
large vibrational amplitudes.
PIMC simulations were carried out by Draeger and Ceperley \cite{dr00}
for solid $^3$He and $^4$He at temperatures between 5 and 35~K, and
a wide rage of densities.  These authors found that the 
mean-squared displacement from lattice sites
exhibits finite-size scaling consistent with a crossover between
quantum and classical limits of $N^{-2/3}$ and $N^{-1/3}$,
respectively, $N$ being the number of atoms in the simulation cell.
They computed the static structure factor $S(k)$ and obtained for
the Debye-Waller factor an anisotropic $k^4$ term, which
indicates the presence of non-Gaussian corrections to the density
distribution around lattice sites. These results, extrapolated to
the thermodynamic limit, were found to agree with those of
scattering experiments.

Other physical observables that can be calculated from
path-integral simulations are the momentum distribution $n(k)$ and
kinetic energy $K$ of the atomic nuclei in the considered solid.
Thus, $n(k)$ in solid $^4$He has been
calculated by means of PIMC methods \cite{ro11}.
For perfect crystals, $n(k)$ was found to be nearly independent of
temperature and different from the classical Gaussian
shape of the Maxwell-Boltzmann distribution, even though such
discrepancies decrease for increasing density.  In crystals
including vacancies, it was found that for
$T \geq$ 0.75~K, $n(k)$ displays the same behavior as in the
perfect crystal, but it presents a peak at lower temperature
for $k \to 0$.

Path-integral calculations and measurements of the kinetic energy
of condensed $^4$He were reported in Ref.~\cite{ce96}.
An overall dependence of $K$ on temperature
for densities less than 70 atoms nm$^{-3}$ was constructed.
In the solid phase $K$ was fount to be nearly temperature
independent and smaller than that corresponding to the
fluid near freezing at the same density.

Comparison between properties of solid $^3$He and $^4$He in the 
hcp and fcc phases has been carried out from PIMC calculations
\cite{he06c}.
Simulations in the isothermal-isobaric ensemble up to pressures of
about 50~GPa allowed to analyze the temperature and pressure
dependence of isotopic effects upon the crystal volume and vibrational 
energy on a large region of the phase diagram.
Due to anharmonicity, the kinetic energy $K$ of solid helium
turns out to be larger than the vibrational potential energy 
$U_{\rm vib}$, and the ratio $K/U_{\rm vib}$
decreases for rising pressure, converging to the harmonic limit
($K/U_{\rm vib} = 1$) at high pressures.

\begin{figure}
\vspace{-2.0cm}
\includegraphics[width= 9cm]{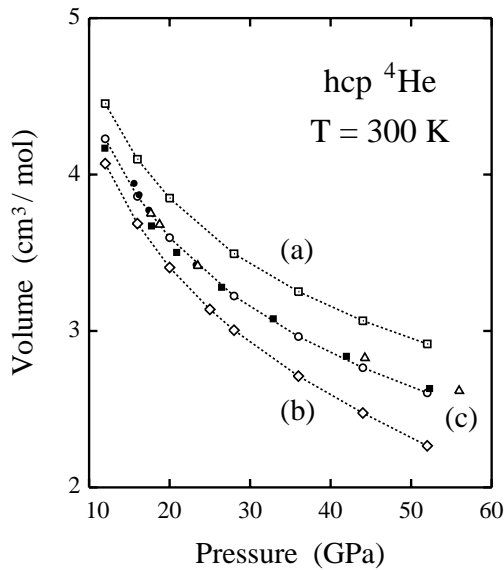}
\vspace{-2.3cm}
\caption{
Equation of state $P$--$V$ of hcp $^4$He at 300~K, as derived
from PIMC simulations for different interatomic potentials
\cite{he06c}.
(a) Open squares: only two-body interactions with an
Aziz-type potential \cite{az95}; (b) open diamonds: two-body terms as
in Ref. \cite{az95} and three-body interactions, as defined in
Ref. \cite{lo87};
(c) open circles: same potential as (b), but with the exchange
three-body interaction rescaled by 2/3.
Error bars of the simulation results are less than the symbol size.
 Dotted lines are guides to the eye.
Open triangles show results obtained from PIMC simulations in
the $NVT$ ensemble \cite{ch01}, with the attractive exchange interaction
rescaled as in (c).  Black symbols show experimental data obtained by
Mao~{\em et al.} \cite{ma88} (filled circles) and Loubeyre~{\em et al.}
\cite{lo93} (filled squares).
}
\label{fig_hel1.1}
\end{figure}

A question raised in this kind of computational studies has been the
availability of effective potentials to describe helium at high
pressures. This has been checked by analyzing the pressure-volume
equation of state.
The pressure dependence of the molar volume for hcp $^4$He is
shown in Fig.~\ref{fig_hel1.1} at 300~K \cite{he06c}.
Open symbols represent results of PIMC simulations
with different interatomic potentials: (a) only two-body interaction
with an Aziz-type potential \cite{az95} (squares); (b) two-body
interaction as in Ref.~\cite{az95} plus three-body terms as in
Ref.~\cite{lo87} (diamonds);
(c) the same two-body potential and three-body interaction with the
effective exchange part rescaled by 2/3, as proposed in Ref.~\cite{bo94b} 
(circles).  For comparison, we also display earlier results yielded 
by PIMC simulations
in the $NVT$ ensemble \cite{ch01}, with the exchange interaction
rescaled by the same factor 2/3 (triangles).
Filled symbols indicate experimental results obtained by Mao~{\em et al.}
\cite{ma88} (filled circles) and Loubeyre~{\em et al.} \cite{lo93}
(filled squares).  The interatomic potential (c) yields results for
the equation of state of hcp $^4$He near the experimental data.
For a given pressure, the only consideration of two-body terms predicts
a molar volume larger than the experimental one. On the contrary,
including both two- and three-body terms derived from
{\em ab initio} calculations underestimates the volume of solid helium.
This agrees with results obtained from PIMC simulations
in the $NVT$ ensemble in Refs.~\cite{bo94b,ch01}.
These results indicate that three-body terms are necessary to reproduce
the high-pressure results, and are suitable for the pressure range
nowadays experimentally available.  However, this kind of effective
interatomic potentials will probably yield a poor description of
solid helium at very high pressures ($\gtrsim$ 60~GPa) \cite{ch01}.

The compressibility of solid $^3$He and $^4$He in the hcp and
fcc phases was studied by PIMC in Ref.~\cite{he08}.
Simulations were carried out in the canonical ($NVT$) and
isothermal-isobaric ($NPT$) ensembles at temperatures between
10 and 300~K, showing consistent results in both ensembles.
At a given pressure, the bulk modulus $B$ decreases as temperature 
rises. For pressures between 4 and 10~GPa, the change in $B$ was found
to be in the order of 10\%, when temperature increases from the low-$T$
limit to the melting temperature.
Solid $^3$He is more compressible than $^4$He. At a given $T$,
the difference between bulk moduli of both solids increases as
pressure rises, but the relative difference between them decreases.

The coexistence between the hcp and bcc phases of solid $^4$He at
fixed pressure was studied by Rota and Boronat \cite{ro11b} using
PIMC simulations.  They reported microscopic results for the
energetic and structural properties of both phases.
Differences between them were found to be small, with
the exception of the static structure factor. When crossing
the phase transition line, most appreciable changes are observed
in the kinetic energy per particle and in the Lindemann ratio, both
suggesting a less correlated quantum solid for the bcc crystal.

Turning now to the magnetic properties of condensed helium,
solid $^3$He has been studied
over the last four decades because at millikelvin temperatures,
it is an almost pure spin-1/2 fermion system with a simple crystal
structure. To study these properties, Candido~{\em et al.} \cite{ca11}
used PIMC simulations, and calculated ring exchange frequencies in
the bcc phase of solid $^3$He, for densities ranging from melting to
the highest stable density.
Exchange frequencies were evaluated for two atoms and for long
cycles including up to eight atoms.
Using a fit to these frequencies, the contribution to the
Curie-Weiss temperature, $\Theta_{CW}$, was calculated as well as
an upper critical magnetic field, $B_{c2}$, for even
longer exchanges, using a lattice Monte Carlo procedure.
It was found that contributions from seven- and eight-particle
exchanges make a significant contribution to $\Theta_{CW}$ and
$B_{c2}$ at melting density.

Other phases of $^4$He, including amorphous solids, were studied
by Boninsegni~{\em et al.} \cite{bo06c}, who employed PIMC
simulations based on a worm algorithm.
This study included simulations that started from a
high-temperature gas phase, which was subsequently `quenched'
down to $T$ = 0.2~K. The low-temperature properties of
the system were found to crucially depend on the initial state, so 
that the disordered system was found to freeze into a superglass, i.e., 
a metastable amorphous solid displaying off-diagonal long-range order
(ODLRO) and superfluidity.

As a final point in this brief survey of solid helium, we will
consider the possibility of a supersolid phase, exhibiting
nondissipative flow.
The apparent discovery of a nonclassical moment of inertia in solid
$^4$He by Kim and Chan \cite{ki04} provided a possible
experimental evidence for a supersolid, although the interpretation
in terms of supersolidity of the ideal crystal phase has been
questioned \cite{ki12}.  This launched a series of path-integral 
studies, whose results in general
were not compatible with the existence of a supersolid helium phase.

 The possibility of superfluid behavior in bulk hcp $^4$He was
investigated in Ref.~\cite{ce04} by using PIMC simulations.
Frequencies of ring exchange were calculated for the bosonic atoms
of $^4$He.
The obtained frequencies were fitted to a lattice model in order to
examine whether such atoms could become a supersolid.
It was found that the scaling with
respect to the number of exchanging atoms is such that superfluid
behavior is not expected to be observed in a perfect $^4$He crystal.

To study the order in solid $^4$He, Clark and Ceperley \cite{cl06}
carried out PIMC simulations to calculate the ODLRO, which for this 
purpose is equivalent to Bose-Einstein condensation.
They did not find ODLRO in a defect-free hcp crystal at the melting
density, and  discussed their results in relation to proposed quantum
solid trial functions, concluding that the solid $^4$He wave function
has correlations which suppress both vacancies and Bose-Einstein
condensation.

  To estimate the onset temperature $T_0$ of Bose-Einstein
condensation in $^4$He crystals presenting vacancies, the temperature
dependence of the one-body density matrix was calculated by PIMC
simulations in Ref.~\cite{ro12}.
The temperature $T_0$ was found to depend on the vacancy
concentration $X_v$, but did not follow the law  $T_0 \sim X_v^{2/3}$,
expected for noninteracting vacancies.
For $X_v = 1/256$, it was obtained $T_0$ =  0.15 $\pm$ 0.05~K.
Below $T_0$, vacancies did not behave as classical
point defects, but became completely delocalized entities.

\subsection{Heavier elements}

Path-integral simulations have been used to study structural,
thermodynamic, and vibrational properties of heavier noble-gas solids
\cite{cu97,mu95,ne00,ch02,ti96,ne02}.
This technique has turned out to be well-suited to analyze isotopic
effects in different properties of these solids \cite{mu95}.
For example, the isotopic-mass dependence of the molar volume 
can be described well from this kind of simulations \cite{he99,he02}. 
Moreover, in this context, several authors
developed effective (temperature-dependent) classical potentials
that reproduce accurately various properties of quantum
solids \cite{gi86,fe86,cu92}.
Thus, Acocella~{\em et al.} \cite{ac00} applied an improved
effective-potential Monte Carlo theory \cite{ac95a}
to study thermal and elastic properties of noble-gas solids.

     The capability of path-integral simulations to
describe thermodynamic and structural properties of solids at low 
temperatures was studied in detail by M\"user~{\em et al.} \cite{mu95}, 
considering noble-gas crystals as examples.  They investigated 
solid Ar at constant volume, as well as isotope effects in the lattice 
parameter of $^{20}$Ne and $^{22}$Ne at zero pressure, with special
emphasis on the convergence of their results at low temperatures.
To reduce the systematic 
limitations due to finite Trotter and particle number, 
these authors proposed a combined Trotter and finite-size scaling. 
  At very low temperatures much effort is necessary to
avoid discretization effects in the phonon spectra, resulting in an
artificial fast decrease in the specific heat as temperature is lowered.
Better approximants than the primitive algorithm may be necessary
to compute the exponent describing the vanishing of the heat
capacity as $T \to 0$ (e.g., a Debye $T^3$ law for insulators).
This calculation turns out to be difficult, because
the exponent can only be measured at temperatures much lower than
the Debye temperature of the solid. 
Taking all this into account, M\"user~{\em et al.} \cite{mu95}
concluded that the lattice parameter and associated structural
properties can clearly be resolved, as well as the isotope shift in
the molar volume.

Given an interatomic potential, for volume $V$ and temperature $T$
the internal energy of a solid, $E(V,T)$, can be written in our
context as \cite{he02}:
\begin{equation}
  E(V,T) = U_0 + U_S(V) + E_{\rm vib}(V,T)   \, ,
\label{evt}
\end{equation}
where $U_0$ is the minimum potential energy for the (classical) crystal
at $T$ = 0~K, $U_S(V)$ is the elastic energy, and $E_{\rm vib}(V,T)$ is
the vibrational energy: $E_{\rm vib}(V,T) = K(V,T) + U_{\rm vib}(V,T)$.
For a given volume $V$, the classical energy at $T = 0$ increases by
an amount $U_S(V)$ with respect to the minimum energy $U_0$.
This elastic energy $U_S$ depends only on the volume, but at finite
temperatures and for the quantum solid, it depends implicitly
on $T$ because of the temperature dependence of $V$ (thermal
expansion).  The elastic energy at low temperatures is basically due to
the `zero-point' lattice expansion.

\begin{figure}
\vspace{-2.0cm}
\includegraphics[width= 9cm]{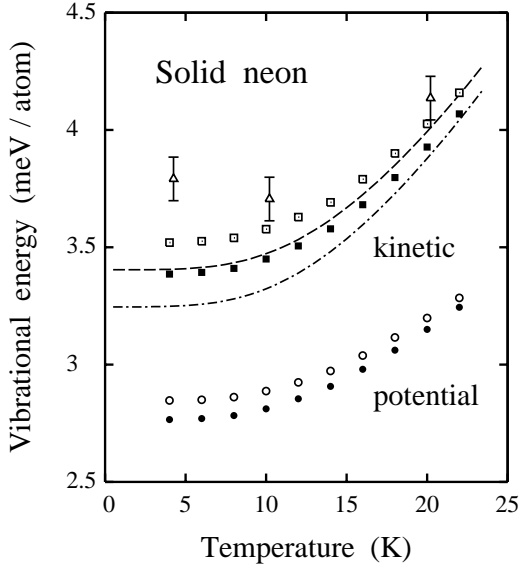}
\vspace{-2.1cm}
\caption{
Temperature dependence of the vibrational energy of solid neon.
Squares and circles correspond to kinetic and potential energy,
respectively, as derived from PIMC simulations \cite{he02}.
Open symbols, $^{20}$Ne; black symbols, $^{22}$Ne.
Error bars of the simulation results are less than the symbol size.
The dashed and dashed-dotted lines correspond to a Debye model for
$^{20}$Ne and $^{22}$Ne, respectively, with
$\Theta_D(^{20}{\rm Ne})$ = 70~K.
Triangles are results for the kinetic energy of $^{20}$Ne, obtained by
Timms~{\em et al.} \cite{ti96} from inelastic neutron scattering
in solid neon with natural isotopic composition
($\overline{m}$ = 20.18 amu).
}
\label{fig_nobl1.2}
\end{figure}

The vibrational energy, $E_{\rm vib}(V,T)$, depends on both $V$ and $T$,
and can be obtained by subtracting the elastic energy from the
internal energy. Path-integral simulations allow one to obtain
separately the kinetic, $K$,  and potential energy, $U_{\rm vib}$,
associated to the lattice vibrations \cite{gi88}. Both energies are
shown in Fig.~\ref{fig_nobl1.2} for solid $^{20}$Ne (open symbols)
and $^{22}$Ne (black symbols). Squares and circles correspond to
$K$ and $U_{\rm vib}$ obtained in Ref.~\cite{he02}.
These results for the kinetic energy are close to those previously
obtained from PIMC simulations with Lennard-Jones \cite{cu93,ti96,cu97}
and Aziz \cite{ti96} interatomic potentials.
Triangles in Fig.~\ref{fig_nobl1.2} indicate the kinetic energy
of $^{20}$Ne, found by Timms~{\em et al.} \cite{ti96} from inelastic 
neutron scattering in solid neon with natural isotopic composition.
The results of PIMC simulations indicate that the vibrational
potential energy is smaller than the kinetic energy for both
neon isotopes.
Dashed and dashed-dotted lines in Fig.~\ref{fig_nobl1.2} represent
$K$ and $U_{\rm vib}$ for $^{20}$Ne and $^{22}$Ne obtained
from a harmonic Debye model for the lattice vibrations \cite{he02}.

 Neumann and Zoppi \cite{ne02} performed PIMC simulations of liquid 
and solid Ne, in order to derive the kinetic energy as well
as the single-particle and pair distribution functions of Ne atoms in
condensed phases.  The simulations were carried out using Aziz-type
and Lennard-Jones potentials.
 The single-particle distribution function $n(r)$ was employed to
derive the momentum distribution and to obtain an estimate of $K$.
Differences between the considered potentials, as measured by the 
properties investigated, turned out to be not very large, especially
when compared with the precision of the available experimental data.

\begin{figure}
\vspace{-2.0cm}
\includegraphics[width= 9cm]{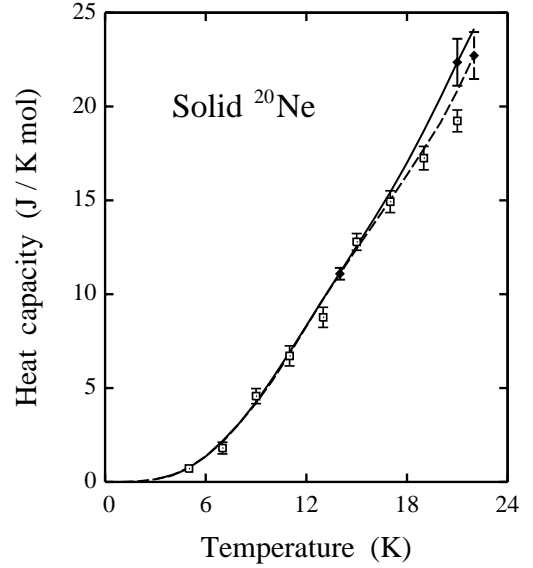}
\vspace{-2.1cm}
\caption{
Heat capacity, $C_P$, of solid neon as a function of temperature.
Squares indicate results of PIMC simulations for $^{20}$Ne \cite{he02}.
The continuous and dashed lines are experimental results for solid
$^{20}$Ne \cite{so71} and for solid neon with natural isotopic
composition \cite{fe66}, respectively.
Error bars are given for three experimental points,
indicated by black diamonds.
}
\label{fig_nobl1.5}
\end{figure}

Path-integral simulations in the isothermal-isobaric ensemble 
have been also used to calculate the heat capacity $C_P$ of noble-gas
solids. In Fig.~\ref{fig_nobl1.5} we present simulation results 
\cite{he02} for $^{20}$Ne at $P$ = 1~bar (open squares), to be compared 
with experimental data \cite{so71} (solid line). For comparison,
experimental results for natural neon \cite{fe66} are shown as a 
dashed line. 
Results of the PIMC simulations follow closely the experimental data
up to $T$ = 18~K, and at $T \ge$ 18~K they seem to be slightly lower.
However, some experimental uncertainties have been reported for
$T \gtrsim$ 18~K \cite{so71}.

\begin{figure}
\vspace{-2.0cm}
\includegraphics[width= 9cm]{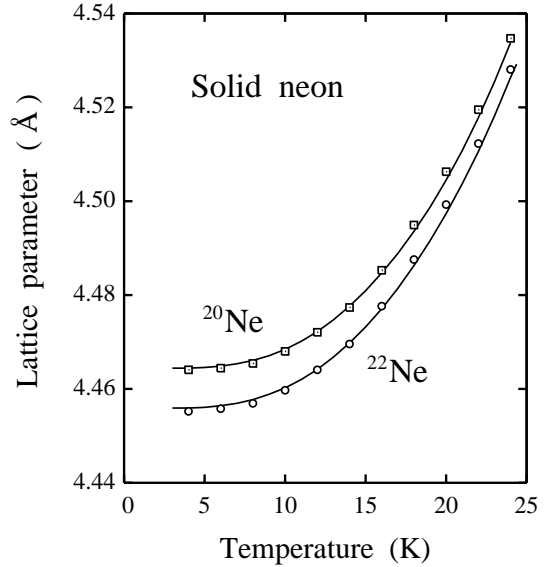}
\vspace{-2.1cm}
\caption{
Temperature dependence of the lattice parameter of isotopically-pure
crystals of solid neon, as derived from PIMC simulations for $^{20}$Ne
(squares) and  $^{22}$Ne (circles) \cite{he02}. Error bars are less
than the symbol size.  Solid lines represent results derived from x-ray
diffraction experiments by Batchelder~{\em et al.} \cite{ba68}.
Error bars of the experimental data are less than the line width.
}
\label{fig_nobl1.7}
\end{figure}

As mentioned above, the thermal lattice expansion is an anharmonic
effect that is well captured by path-integral simulations.
The lattice parameter $a$ derived by the PIMC method in Ref.~\cite{he02}
was found to follow closely the experimental data for $^{20}$Ne and
$^{22}$Ne up to 24~K, as displayed in Fig.~\ref{fig_nobl1.7}.
Classical simulations yield a nearly linear temperature dependence
for the lattice parameter \cite{mu95}, which converges at
low $T$ to the value corresponding to the minimum potential
energy of the solid.
One finds for the quantum solid $^{20}$Ne a zero-temperature lattice
parameter 4\% larger than the classical limit \cite{he02}.
This increase, due to anharmonicity of the zero-point motion, 
amounts to about twice the change in $a$ caused by thermal expansion 
between 0~K and the melting temperature of neon.

\begin{figure}
\vspace{-1.1cm}
\includegraphics[width= 9cm]{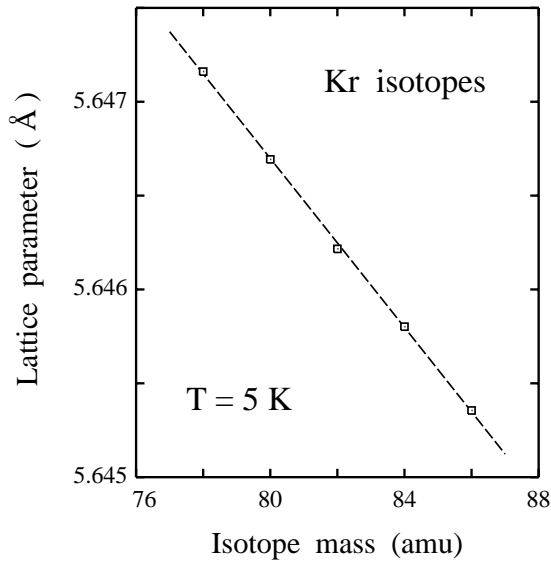}
\vspace{-2.4cm}
\caption{
Lattice parameter of isotopically-pure Kr crystals as a function of
isotopic mass at $T$ = 5~K and $P$ = 1~bar. Symbols indicate results
of PIMC simulations \cite{he03a}.
The dashed line is a least-square fit to the simulation results.
Error bars are less than the symbol size.
}
\label{fig_nobl2.4}
\end{figure}

The isotopic effect on the lattice parameter $a$ of other noble-gas 
solids (Ar, Kr, Xe) was studied in Ref.~\cite{he03a} 
as a function of temperature and pressure.
A linear dependence of $a$ on the isotopic mass $m$ was found in all
cases. As an example we show in Fig.~\ref{fig_nobl2.4} the parameter 
$a$ for some stable krypton isotopes at 5~K, as derived from PIMC 
simulations with a Lennard-Jones-type potential.

\begin{figure}
\vspace{-1.1cm}
\includegraphics[width= 9cm]{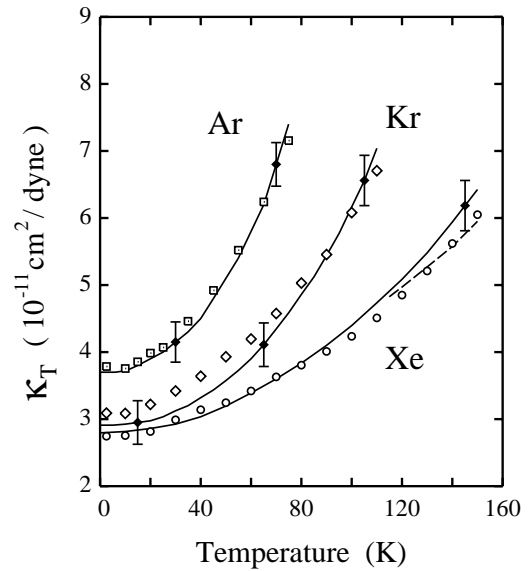}
\vspace{-2.3cm}
\caption{
Isothermal compressibility $\kappa_T$ at $P$ = 1 bar as a function
of temperature. Symbols indicate results derived from PIMC simulations
for noble-gas solids with natural isotopic composition: Squares, Ar;
diamonds, Kr; circles, Xe \cite{he03a}. Error bars of the calculated
compressibilities are smaller than the symbol size. Lines represent
experimental data obtained by different authors: Ar \cite{pe66},
Kr \cite{lo68}.
For Xe, the solid line is from \cite{po64}, whereas the dashed
line represents more recent results from \cite{gr81}.
Error bars are given for some experimental points, indicated by black
diamonds.
}
\label{fig_nobl2.7}
\end{figure}

The isothermal compressibility $\kappa_T$ can be derived 
from the volume fluctuations obtained in PI simulations in the
isothermal-isobaric ensemble, as given by
Eq.~(\ref{eq:iso_compressibility}).
In this way, $\kappa_T$ was derived for noble-gas solids in
Ref.~\cite{he03a}, and is presented here in Fig.~\ref{fig_nobl2.7}
(symbols) as a function of temperature. Lines represent experimental
data for crystals with natural isotopic composition
\cite{pe66,po64,gr81,lo68}.
The overall agreement between calculated and experimental results
is good, given the uncertainty in the measurements \cite{he03a}.
Comparison with the classical expectancy indicates that
quantum effects give rise to an appreciable increase
in the low-temperature compressibility: 19\% for Ar, 9\% for Kr, and
5\% for Xe (for Ne, it is about 70\% and depends on
the isotope) \cite{he02,he03a}.

\begin{figure}
\vspace{-1.1cm}
\includegraphics[width= 9cm]{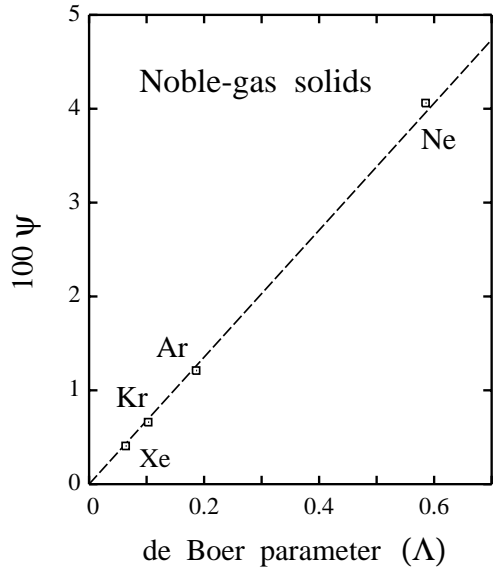}
\vspace{-2.3cm}
\caption{
Low-temperature relative change in the lattice parameter,
$\psi = (a_{\rm q} - a_{\rm cl}) / a_{\rm cl}$,
vs the de Boer parameter $\Lambda$ for noble-gas solids with
fcc structure and natural isotopic composition.
$a_{\rm q}$ and $a_{\rm cl}$ denote the quantum and classical lattice
parameter for $T \to 0$.
Symbols represent results derived from PIMC simulations \cite{he03a}.
The line is a guide to the eye.
}
\label{fig_nobl2.2}
\end{figure}

In connection with noble-gas solids,
Chakravarty \cite{ch02} performed path-integral simulations to study 
structural and thermodynamic properties of quantum Lennard-Jones
solids, as a function of the de Boer parameter 
$\Lambda = h / \sigma \sqrt{m \epsilon}$, which measures 
the quantum delocalization of the considered particles. 
These simulations revealed a strong dependence of the 
density on the parameter $\Lambda$. 
The lattice expansion of the quantum solids, with respect to their 
classical counterparts, is accompanied by an appreciable decrease 
in the binding energy.
The relation between zero-temperature lattice expansion and the
parameter $\Lambda$ for noble-gas solids was presented in
Ref.~\cite{he03a}, and is shown here in Fig.~\ref{fig_nobl2.2}.
The degree of solid-like order and the average coordination number
were also found to depend markedly on $\Lambda$.
Moreover, the calculated kinetic energy per particle indicates that a 
Lennard-Jones solid is far from the classical equipartition regime at 
temperatures as high as 70\% of the melting temperature \cite{ch02}. 
To assess the nature of the quantum-corrected energy landscape, effective
pair potentials have been defined using the pair correlation function
$g(r)$ of the quantum system \cite{ch11b}. 
For rising $\Lambda$, these effective potentials become increasingly 
softer, shallower, and of longer range, with the potential 
minimum shifted to larger distances.  

Given the importance of anharmonic effects in noble-gas solids,
PIMC simulations have been employed to assess the accuracy of
harmonic or quasi-harmonic approximations for vibrational modes 
in these solids under pressure \cite{he05b}.
This allowed to quantify the overall anharmonicity of the lattice 
vibrations and its influence on several structural and thermodynamic 
properties. The vibrational energy $E_{\rm vib}$ 
increases with pressure, but this increase is slower than that 
of the elastic energy $U_S$ [see Eq.~(\ref{evt})], which dominates at 
high pressures. 
Results of these PIMC simulations indicated that the accuracy of the 
QHA to describe noble-gas solids increases as pressure is 
raised \cite{he05b}. This is mainly a consequence of the relative 
importance of elastic and vibrational energy, as the latter becomes 
comparatively irrelevant as pressure rises. 
For large pressures, even a classical description of the vibrational  
modes can be precise enough to predict structural and thermodynamic 
properties of these solids.
However, vibrational properties usually require the full quantum
treatment, with the consideration of zero-point anharmonic effects. 

\begin{figure}
\vspace{-3.5cm}
~\hspace{0.8cm}
\includegraphics[width= 9.5cm]{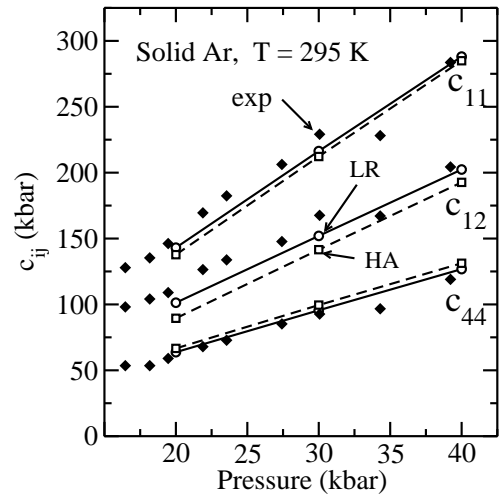}
\vspace{-0.9cm}
\caption{
Pressure dependence of the elastic constants of solid Ar at 295~K.
Results of the LR (circles) and harmonic approximation (HA, squares)
procedures \cite{ra05} are compared to the adiabatic elastic constants
determined by Brillouin spectroscopy \cite{sh01} (black diamonds).
The lines are guides to the eye.
}
\label{fig_nobl5.9}
\end{figure}

Another interesting application of path-integral simulations is the 
calculation of vibrational frequencies by using the static susceptibility 
tensor $\chi^T$ in the LR approach discussed in Sect.~II.H.
This method has been applied to study several properties of solid 
Ne and Ar as functions of pressure and temperature \cite{ra05}.
The LR approach predicts anharmonic shifts in the phonon frequencies 
in reasonable agreement to experimental data. 
This procedure has been also applied to calculate elastic constants 
of noble-gas solids from the propagation velocity of low-energy
acoustic phonons. 
The temperature dependence of the elastic constants of solid
Ne compares well with those derived from Brillouin scattering
experiments \cite{en75,mc75}.
For Ar, the pressure dependence of the elastic constants predicted 
by the LR method shows an overall agreement to experiment
\cite{sh01}, as presented in Fig.~\ref{fig_nobl5.9}.
Adiabatic elastic constants derived from Brillouin spectroscopy
measurements are displayed by solid diamonds \cite{sh01}.
The values of $c_{11}$ and $c_{12}$ are slightly underestimated by the 
LR approach, maybe as a consequence of neglecting three-body forces
\cite{ra05}.
The elastic constants obtained from a pure harmonic approximation 
are also plotted in Fig.~\ref{fig_nobl5.9}.
Elastic constants of solid Ar and $^3$He (hcp, fcc, and bcc) were also
calculated by Sch\"offel and M\"user \cite{sc01} from PI
simulations in the $NVT$ and $NPT$ ensembles. In particular, in the
isothermal-isobaric ensemble they exploited the relationship between 
strain fluctuations and elastic constants, and found good agreement
with available experimental data.

\begin{figure}
\vspace{-1.3cm}
\includegraphics[width= 9cm]{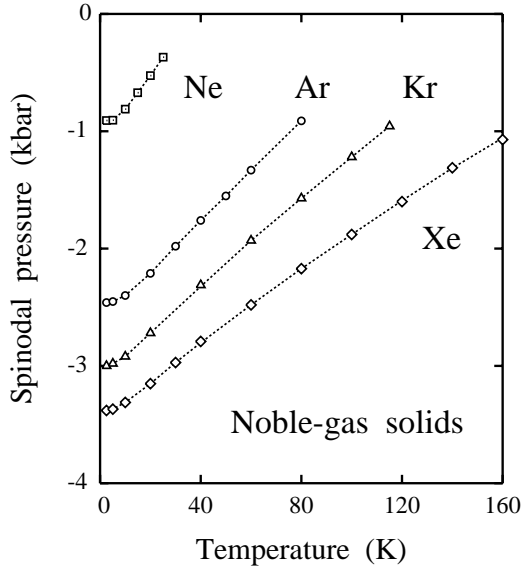}
\vspace{-2.3cm}
\caption{
Calculated spinodal pressure for noble-gas crystals as a function of
temperature \cite{he03b}. Squares, neon; circles, argon;
triangles, krypton; diamonds, xenon.  Error bars are less than the
symbol
size, and dotted lines are guides to the eye.
}
\label{fig_nobl3.3}
\end{figure}

Turning to the thermodynamic properties of noble-gas solids, an
interesting application of path-integral simulations consists in the
assessment of quantum effects in their limits of mechanical
stability. To this end, PIMC simulations in the isothermal-isobaric
ensemble were carried out at negative pressure, which allowed to
determine the solid-gas spinodal line \cite{he03b}.
This line (spinodal pressure $P_s$ vs temperature) can be determined
as the locus of points where the bulk modulus $B$ vanishes
(the compressibility $\kappa_T$ diverges).
The resulting data for the spinodal pressure are plotted in
Fig.~\ref{fig_nobl3.3} as a function of temperature.
In all cases, $P_s < 0$, becoming more negative as the atomic
mass of the noble gas increases.
Quantum effects were found to affect appreciably the spinodal line
at low temperatures, and make $P_s$ less negative than the classical
expectancy.  For $T \to 0$, this change in $P_s$ ranges
from 43\% for neon to 6\% for xenon \cite{he03b}.

\begin{figure}
\vspace{-0.1cm}
\includegraphics[width= 9cm]{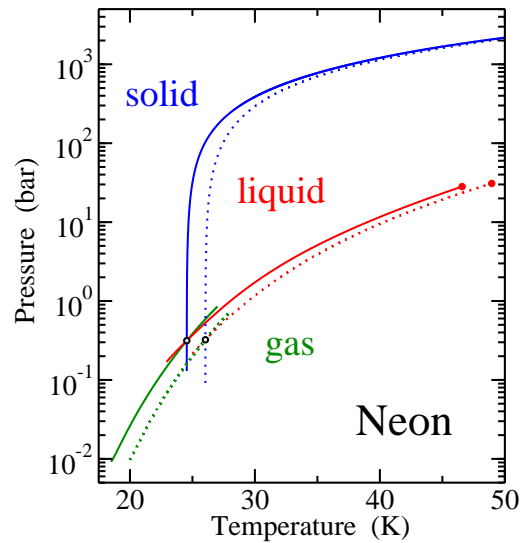}
\vspace{-0.9cm}
\caption{
Phase diagram of natural neon in the pressure-temperature domain.
Results are shown for the three coexistence lines between solid,
liquid,
and gas phases \cite{ra08c}. Continuous and dotted lines were derived
from PIMC and classical MC simulations, respectively.
Open circles represent the triple point and filled ones the critical
point.
\label{fig_nobl7.8}
}
\end{figure}

Free energy calculations based on PI simulations have
been also employed to study the phase diagram of noble gases.
For neon, in particular, this kind of simulations have demonstrated
that free energy techniques, previously used in classical
simulations, can be adapted and generalized to cover the case of quantum
systems \cite{ra08c,ra08b,br12}.
This includes the adiabatic switching \citep{wa90} and 
reversible scaling \citep{ko99} methods, 
which are based on algorithms where either the Hamiltonian,
a state variable (pressure, temperature) or even an atomic mass are
adiabatically changed along a simulation run (see Sect.~II.I). 
Significant quantum effects were found in the phase diagram of neon 
at pressures below 2~kbar, where the solid-gas
and liquid-gas coexistence lines are located \cite{ra08c}. The main 
quantum effect found for these two lines in the $P$--$T$ diagram 
is a shift of about 1.5~K towards lower temperatures. For the 
solid-liquid coexistence, the temperature shift decreases
with pressure, from a value of 1.5~K at triple point conditions to
0.6~K at 2~kbar (see Fig.~\ref{fig_nobl7.8}). 
This means that including quantum effects in the atomic
dynamics of Ne in the solid and liquid phases lowers the melting
temperature by 1.5~K compared to the classical result, i.e., 
a 6\% of its actual value.
A shift of 0.14~K in the triple-point temperature was found 
between $^{20}$Ne and $^{22}$Ne, in good agreement
with experimental results \cite{ra08c}. 

In connection with solid models including simple interatomic interactions, 
some path-integral simulations have been carried out for quantum 
hard-sphere solids.
In particular, fcc, hcp, and bcc structures were recently studied
in Ref.~\cite{se13} by PIMC, along with the fluid-solid 
coexistence lines. 
No significant differences between the relative stabilities of
fcc and hcp lattices were found within the attained accuracy.
Starting from a bcc solid, the simulations yielded either
irregular lattices keeping some traces of the initial one,
or spontaneous transitions to hcp-like lattices. 
It was discussed the relation between these transitions for
quantum hard-sphere solids and solid-solid equilibria at low
temperatures in real systems \cite{se13}.

\section{Group-IV materials}

Another important group of materials, for which path-integral
simulations have been carried out, are those formed by elements of the 
group IV of the periodic table, in particular those with a cubic 
structure. This includes diamond and the well known semiconductors 
silicon, germanium, and silicon carbide (type 3$C$).
Contrary to the noble-gas solids discussed above, where interatomic
interactions are of van der Waals type, in group-IV materials the
atoms are four-fold coordinated and connected by covalent bonds.

\subsection{Silicon}

PIMC simulations of crystalline silicon have been performed in both 
the canonical ($NVT$) and isothermal-isobaric ($NPT$) 
ensembles \cite{ra93,no96}, using the empirical Stillinger-Weber 
potential.  This allowed to study several finite-temperature properties 
of the material, such as potential energy, radial distribution function 
(RDF), and atomic delocalization.
The simulations in the $NPT$ ensemble \cite{no96} allowed to study
properties of silicon such as lattice parameter, thermal expansion 
coefficient, heat capacity $C_P$, and bulk modulus. The calculated 
quantities showed an overall agreement with experimental data.
These quantum simulations led to a good description of quantum 
effects like zero-point vibrations (mean-square displacements),
and the results were found to converge to the classical
limit at high temperatures. There appeared some deviations of
the resulting vibrational energies when compared with the experimental 
ones, as the potential model seems to overestimate them.

\begin{figure}
\vspace{-1.4cm}
\includegraphics[width= 9cm]{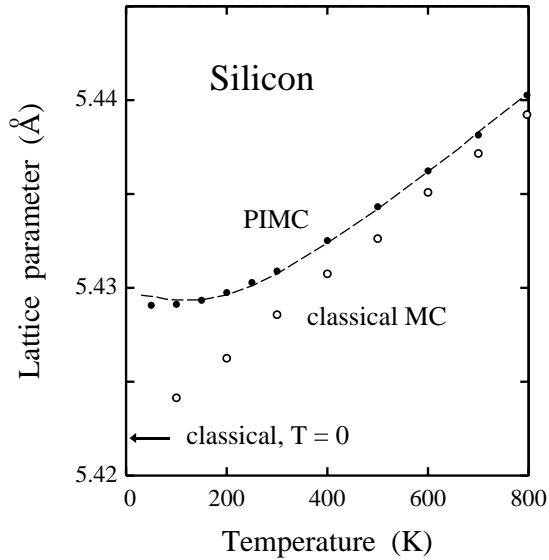}
\vspace{-2.3cm}
\caption{
Temperature dependence of the lattice parameter $a$ for natural
silicon.
Data points were derived from PIMC simulations (filled squares)
and in the classical limit (open squares) \cite{he99}.
Experimental results \cite{la82} are given as a dashed line.
An arrow indicates the lattice parameter corresponding to the minimum
potential energy of the crystal.
}
\label{fig_si1}
\end{figure}

The relevance of anharmonicity and quantum effects on the properties 
derived from the PIMC simulations was addressed by comparison with the
results of QHA calculations and classical simulations.
A comparison of results derived in both 
quantum and classical approaches for several structural and
thermodynamic properties of silicon was presented in Ref.~\cite{no96}.
In particular, the low-temperature lattice expansion due to 
anharmonicity of the zero-point vibration was found to be nonnegligible. 
In Fig.~\ref{fig_si1} we show the temperature dependence of the
lattice parameter $a$ for $T$ up to 800~K. Filled and open circles are
results of PIMC and classical MC simulations, respectively, whereas  
the dashed line corresponds to experimental 
measurements \cite{la82,ha61,sh72}.
 The results of the PIMC simulations follow closely the experimental 
data, but they do not reproduce the negative thermal
expansion observed for silicon at $T <$ 100~K.  
 This seems to be a drawback of empirical interatomic potentials, 
such as that employed in Refs.~\cite{no96,he99}.
Classical MC simulations with the same effective potential yield 
an almost linear dependence of $a$ vs the temperature, 
with the lattice parameter converging at low $T$ to the value
corresponding to the minimum potential energy of the silicon crystal
(indicated by an arrow in Fig.~\ref{fig_si1}).
We note that the lattice expansion due to anharmonicity of the zero-point
motion ($\Delta a$ = 0.007 \AA) is in the order of the thermal expansion 
between 0~K and the Debye temperature of silicon ($\Theta_D \sim$ 650~K).

\begin{figure}
\vspace{-3.2cm}
\includegraphics[width= 9cm]{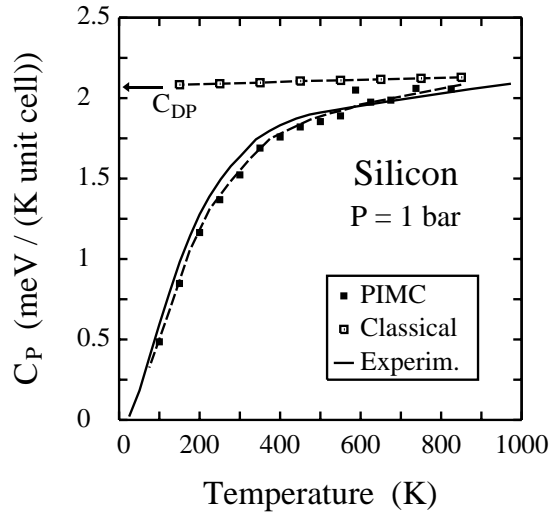}
\vspace{-0.7cm}
\caption{
Temperature dependence of the heat capacity $C_P$ of crystalline
silicon at $P = 1$~bar. Filled and open squares are results derived from
PIMC and classical MC simulations by numerical differentiation of the
enthalpy values \cite{no96}. The dashed lines are smooth functions
through the simulation data.
The solid line shows the experimental temperature dependence of
$C_P$ \cite{ge65,sh63}. An arrow indicates the $C_P$ value
corresponding to the Dulong-Petit's law, $C_{\text{DP}} = 3 N k_B$,
for $N = 8$ atoms (the fcc unit cell of silicon).
}
\label{fig_si2}
\end{figure}

In this context, a point of interest has been the 
capability of effective interatomic potentials to reproduce several
properties of crystalline solids, and group-IV materials in particular. 
In general, this kind of potentials were designed to
carry out classical simulations, so that its validity for quantum
simulations is in principle not guaranteed.
However, quantum path-integral results compared with experimental
data much better than those derived from classical simulations,
e.g., lattice parameter, bulk modulus and its pressure derivative, 
temperature dependence of the Gr\"uneisen constant, and specific heat.
At low temperatures, this could be expected for some quantities
such as thermal expansion coefficient or specific heat, which should
vanish for $T \to 0$, whereas usually they take finite (positive) 
values in a classical approximation.
  In Fig.~\ref{fig_si2} we display the heat capacity $C_P$ of silicon
at $P$ = 1~bar, as derived from PIMC (black symbols) and classical
MC (open squares) simulations \cite{no96}.
Experimental results \cite{ge65,sh63} are represented by a solid line. 
The agreement between values obtained in the PIMC simulations with the
Stillinger-Weber potential and experimental data was satisfactory.

As shown above for noble-gas solids,
an interesting application of path-integral simulations in the study
of solids is the assessment of various isotopic effects.
Maybe the most direct of these applications for PI simulations in
the $NPT$ ensemble is the calculation of the isotopic effect in the
crystal volume, which can be carried out in a nonperturbative way.
Thus, the dependence of the lattice parameter $a$ of silicon on isotopic 
mass was studied by PIMC simulations in the isothermal-isobaric
ensemble at ambient pressure \cite{he99}.
At 300~K, the isotopic effect leads to a decrease of 0.8 and 
$1.6 \times 10^{-4}$ \AA\ in the lattice parameter of
isotopically pure crystals of $^{29}$Si and $^{30}$Si, respectively, as
compared to the solid with natural isotopic composition.
At lower temperatures ($T \approx$ 50~K), this isotopic effect is 
about 50\% larger than at room temperature.
The fractional change in the lattice parameter of
$^{30}$Si with respect to natural silicon was found to
be $\Delta a / a = -4.4$ and $-2.9 \times 10^{-5}$
at 100 and 300~K, respectively. These results compare well with 
those derived from calculations using density-functional perturbation 
theory \cite{pa94,ba01}.  We emphasize that this dependence 
on isotopic mass results from a combination 
of the anharmonicity of the interatomic potentials and the quantum nature
of atomic nuclei, which manifests itself in the zero-point vibrations
(this isotopic effect does not appear in a classical model).

Another application of path-integral simulations in solids has
been the study of amorphous materials.
For amorphous silicon, in particular, quantum effects on the atomic 
delocalization were studied by PIMC simulations in the temperature
range from 30 to 800~K \cite{he98}.
In this material, the quantum delocalization was found to be appreciable 
vs. topological disorder, as seen from structural observables such as 
the RDF. At $T \approx$ 50~K, the width 
of the first peak in the RDF increases by a factor of 1.5 because of
quantum effects.
The overall anharmonicity of the solid vibrations at finite
temperatures in amorphous silicon resulted to be significantly larger
than in the crystalline material. Low-energy vibrational modes
were found to be mainly located on coordination defects in the amorphous 
solid.

In connection with PI simulations of silicon, we mention that
similar simulations have been carried out to study several structural
and dynamical properties of silica polymorphs, in particular
quartz \cite{mu01,ri01} and $\beta$-cristobalite \cite{ri01}.

\subsection{Germanium}

As in the case of silicon, an application of path-integral
simulations for germanium consisted in studying isotopic effects in
this material. 
 Thus, the dependence of the lattice parameter upon isotopic mass 
for isotopically pure Ge crystals was studied by PIMC
simulations \cite{no97}. 
At 50~K, the isotopic effect was found to yield an increase of 
$2.3 \times 10^{-4}$ \AA\ in the lattice parameter of $^{70}$Ge with 
respect to $^{76}$Ge, i.e., $\Delta a/a = 4.1 \times 10^{-5}$. 
Similar to silicon,
a comparison of the simulation results with experimental data 
for $^{74}$Ge indicated that effective interatomic potentials 
combined with path-integral simulations give 
a realistic description of this anharmonic effect in germanium. 

The path-integral results were also compared to those derived from 
QHA calculations of the crystal vibrations. 
Within this approximation, the 
fractional change of the lattice parameter of $^{74}$Ge respect 
to a crystal with atoms having the average mass of natural Ge amounts 
to $\Delta a/a = -9.2 \times 10^{-6}$ in the limit $T \to 0$, which
coincides within error bars with the results of the simulations
\cite{no97}.
Data derived from the QHA slowly depart from those of PIMC simulations
as temperature rises, since that approximation does not include the 
whole anharmonicity of the system, whose effects appear more
clearly at high temperatures.

\subsection{Diamond}

For diamond, path-integral simulations have been performed using
different types of interatomic potentials, in particular effective 
Tersoff-type potentials \cite{te88a} and tight-binding Hamiltonians.
In Ref.~\cite{he01} several structural and thermodynamic properties 
of diamond were studied by PIMC simulations in the isothermal-isobaric 
ensemble, using a Tersoff-type potential.
Such properties were investigated as functions of both temperature and 
hydrostatic pressure. 
The resulting lattice parameter, heat capacity, thermal expansion 
coefficient, and bulk modulus showed an overall agreement with 
experimental data.
The relevance of quantum effects and anharmonicity on the properties 
derived from the path-integral simulations was estimated by comparison 
with results obtained from classical simulations with the same
interatomic potential, as well as with those yielded by QHA calculations.

\begin{figure}
\vspace{-1.5cm}
\includegraphics[width= 9cm]{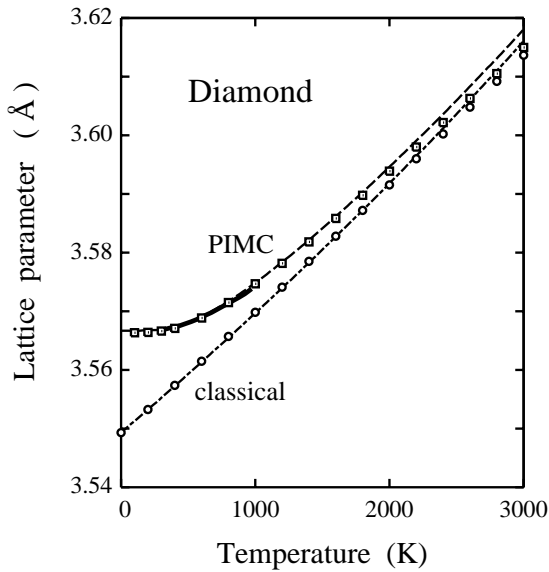}
\vspace{-2.3cm}
\caption{
Temperature dependence of the lattice parameter $a$ of diamond
at $P = 1$~bar.  The data shown were derived from PIMC (squares) and
classical MC simulations (circles) \cite{he01}.
Dashed and dashed-dotted lines were obtained in a quantum and classical
QHA, respectively.
Experimental results \cite{la82} are displayed as a bold line.
}
\label{fig_diam1.8}
\end{figure}

In Fig.~\ref{fig_diam1.8} we present the temperature dependence of
the lattice parameter $a$ obtained in PIMC simulations (open squares),
along with results of classical MC simulations (open circles) and 
experimental data (bold line). For comparison,
a dashed line shows the lattice parameter obtained in a QHA with 
the Tersoff-type potential employed in Ref.~\cite{he01}.
This approximation yielded for the parameter $a$  values
close to those obtained from the PIMC simulations, but at $T >$ 2000~K 
the $a$ values derived from the QHA gradually deviate from the 
PIMC data.
From the results of these simulations, we emphasize that zero-point
motion of carbon atoms causes an increase in the lattice parameter of
$1.7 \times 10^{-2}$ \AA\ (a relative change of 0.5\%) and a decrease
in the bulk modulus by about 5\%.
These relative values are on the order of the precision obtained from  
{\em ab initio} electronic structure methods. 
This means that, at least for solids with light atoms,
further improvements in electronic structure calculations cannot reduce 
the error bars in the computed values of structural observables, 
unless the effects associated to the quantum nature of atomic nuclei 
are taken into account.

The influence of quantum effects on the properties of solids is
nontrivial in the presence of anharmonicity.  For situations in which 
anharmonic effects can be treated perturbatively 
(i.e., low temperatures), the QHA gives a good description of physical 
observables, but this approximation becomes less accurate as temperature
rises, as mentioned above in the case of heavier atoms (Si and Ge). 
Results of QHA calculations for several properties of diamond at 
room temperature are close to those yielded by PIMC simulations with 
a Tersoff-type potential \cite{he01}. This agreement is still good
at pressures in the order of those attained in diamond anvil cells.
Nevertheless, appreciable differences between QHA and PIMC results are 
found for diamond at temperatures higher than 1000~K,
especially for the bulk modulus, thermal expansion coefficient,
and heat capacity. 
At such high temperatures, anharmonic effects are enhanced and the
QHA can only yield qualitative trends of the material properties.

 Despite the limitations associated to employing empirical
potentials originally optimized for classical simulations,
a reasonable agreement was found between PIMC and experimental data
for various finite-temperature properties of diamond, such as 
heat capacity and thermal expansion coefficient.
For the heat capacity $C_P$, values slightly lower than the experimental 
ones were obtained, due to the rigidity of the Tersoff potential,
which causes a hardening of vibrational modes.
Nevertheless, this rigidity does not seem to affect the lattice 
parameter and thermal expansion. One expects that these shortages
of effective potentials could be surmounted by using  {\em ab initio} 
PI simulations, where the interatomic interactions
are obtained from first-principles calculations.

\begin{figure}
\vspace{-0.2cm}
~\hspace{0.3cm}
\includegraphics[width= 9cm]{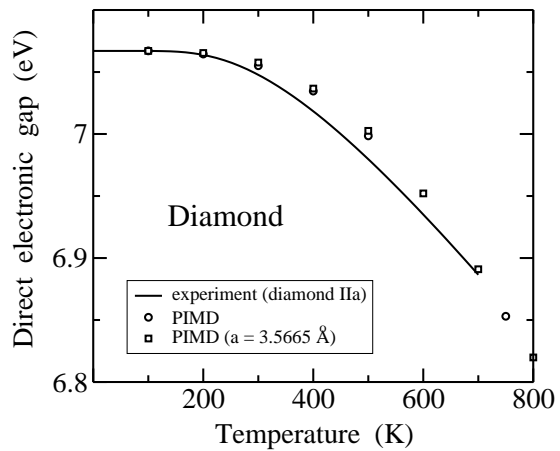}
\vspace{-5.1cm}
\caption{
Temperature dependence of the direct electronic gap of diamond,
obtained from PIMD simulations \cite{ra06}. Open circles were derived
from simulations that took into account the thermal expansion of the
lattice. Open squares correspond to simulations where the cell volume
was fixed to the equilibrium value at 100~K.
The continuous line is a fit to the experimental data given in
Ref.~\cite{lo92} for diamond IIa.
}
\label{fig_diam4.5}
\end{figure}

 Diamond has been studied more recently by PIMD simulations using 
a tight-binding Hamiltonian to describe
the electronic structure and total energy \cite{ra06}.
This kind of simulations are based on handling electrons and atomic 
nuclei as quantum particles, in the framework of the Born-Oppenheimer 
approximation.  In this context,
using the path-integral formalism for the nuclei allows one to
assess the effect of zero-point vibrations and finite temperatures 
on the vibrational and electronic properties of solids.
For diamond, this method predicted a reduction of the direct electronic 
gap by 0.7~eV (i.e., a 10\%), as a consequence of the electron-phonon
coupling mediated by zero-point vibrations.
The direct gap was also predicted to decrease with temperature, in good 
agreement with experimental data available up to 700~K \cite{ra06}
(see Fig.~\ref{fig_diam4.5}), as well as with theoretical results 
based on perturbation theory \cite{zo92}.

Anharmonic vibrational frequencies of diamond were obtained
from the linear-response approach described in Sect.~II.H.
In particular, the temperature dependence of the zone-center optical
phonon (at the $\Gamma$ point) was derived from PIMD simulations
with a tight-binding Hamiltonian.
A redshift of this phonon frequency was found, which turned out
to be overestimated in comparison to experimental data at
$T >$ 500~K \cite{ra06,he91}.
Better agreement was found in the comparison of the elastic constants
obtained from the simulations with experimental values derived 
from Brillouin scattering experiments \cite{zo98}.

Diamond has been also employed as benchmark for checking new
simulation techniques related to path integrals.
Thus, Ceriotti~{\em et al.} \cite{ce09} presented a method, based on a 
non-Markovian Langevin equation, aiming at including quantum corrections 
to the classical dynamics of ions in quasi-harmonic systems. 
These authors showed that fitting the correlation function of
the noise, one can change the fluctuations in positions and momenta as
functions of the vibrational frequency, and match them to reproduce 
the quantum-mechanical behavior. 
They applied this procedure to diamond and ice Ih, and found 
results in agreement with full PI simulations, but
appreciably reducing the computational effort.

\subsection{Silicon carbide}

Tight-binding Hamiltonians, similar to that employed for diamond,
 have been used to carry out PIMD simulations of cubic (3$C$) silicon 
carbide, as a function of pressure and temperature \cite{ra08}.
These simulations were able to reproduce several experimental 
properties of 3$C$-SiC.
Moreover, a comparison to classical simulations allowed to assess 
quantum effects in structural and thermodynamic properties,
as well as to evaluate the separate contribution of each element 
(either C or Si) to various properties of the material.

The zero-point renormalization of the cell parameter $a$ (lattice
expansion with respect to the classical value) derived from the 
PIMD simulations was found to be $\Delta a/a=2.5\times10^{-3}$. 
This value is close to the sum of zero-point renormalizations
obtained when only one type of atomic nuclei (either Si or C) is
treated quantum mechanically \cite{ra08}.
Moreover, the tight-binding potential employed in those simulations 
predicted a temperature dependence of the cell parameter, $a(T)$, in 
good agreement with experimental data. The main discrepancy was found
for $T >$ 600~K, where the linear expansion coefficient was
predicted to be about 8\% larger than the experimental one.
Other properties of 3$C$-SiC, such as the bulk modulus and its 
pressure derivative at room temperature displayed good agreement 
with the available experimental data.  

As shown above for diamond, an interesting effect of the electron-phonon 
interaction is the decrease in the direct electronic gap of 3$C$-SiC 
for rising temperature.
Results of the PIMD simulations for the gap showed satisfactory
agreement with experimental data, although one observes that the 
tight-binding model employed in Ref.~\cite{ra08} tends to overestimate 
the magnitude of the electron-phonon interaction.
At low temperatures, zero-point motion led to a gap renormalization 
$\Delta E_0 = -0.69$ eV, which means a relative value 
$\Delta E_0 / E_0 = -0.10$.

PIMD simulations in the isothermal-isobaric ensemble 
have been also performed to investigate the dependence of the
lattice parameter $a$ of 3$C$-SiC on isotopic mass \cite{he09c}.
In these simulations, atomic nuclei were also treated as quantum 
particles interacting via a tight-binding-type potential.
The isotopic effect was studied by considering separately changes in
the C or Si mass, and assuming in each case for the other element
the average mass of the natural isotopic composition.
Considering changes in the C mass,
the difference $\Delta a$ between lattice parameters of
crystals with $^{12}$C and $^{13}$C at 300~K was found to amount to
$2.1 \times 10^{-4}$~\AA. The effect due to Si isotopes is much
smaller, and amounts to $3.5 \times 10^{-5}$~\AA\ when replacing
$^{28}$Si by $^{29}$Si.
These results of PIMD simulations can be interpreted in terms of
a QHA for the lattice vibrations \cite{he09c}.

An important conclusion of this kind of simulations in group-IV
materials is that they are well-suited to describe anharmonic effects
related to the phonon-phonon interaction, as well as for the
treatment of the electron-phonon coupling at finite temperatures.
Thus, this type of simulations is an alternative to perturbational
calculations, with the advantage of being also applicable in cases
where a perturbational series might converge rather slowly.
A prerequisite to account for phonon-phonon and electron-phonon
interactions is a quantum description that includes both electrons
and atomic nuclei.

\section{Molecular solids}

\subsection{Solid hydrogen}

Solid hydrogen, a simple system consisting only of protons and
electrons, exhibits remarkable properties under extreme conditions
of pressure and temperature. Thus, a variety of structural phase 
transitions at high pressures has been found by various experimental
and theoretical methods.
Due to the light mass of hydrogen, solid H$_2$ and its isotopes 
solid deuterium (D$_2$) and tritium (T$_2$) appear as natural
candidates to show significant nuclear quantum effects. In fact, 
in this case atomic nuclei are protons, deuterons, or tritons.
The properties of hydrogen under extreme conditions have been recently
reviewed by McMahon~{\em et al.} \cite{mc12b}.

Experimental studies based on static compression 
revealed three relevant phases of solid molecular hydrogen: 
phase I (high-temperature, relatively low-pressure phase), 
phase II (low-temperature, low-pressure phase), and 
phase III (high-pressure phase).
These H$_2$ phases have been studied in recent years with
path-integral simulations, using effective \cite{zo91,cu97b,cu01,cu02} 
and {\em ab initio} potentials \cite{bi98,ki00,ts01,ge12,li13}.
A new phase IV has been recently proposed \cite{ho12}.

Kitamura~{\em et al.} \cite{ki00,ts01} reported quantum distributions
of protons in phases I, II, and III.
These H$_2$ phases were investigated by first-principles
PIMD simulations, where the interatomic forces were given by
DFT calculations.
The resulting crystal structures were found to have different
symmetries from those predicted by simulations where protons were 
treated classically.
For phase II, an interesting result was that molecular rotation is 
hindered by quantum fluctuations of protons.
This kind of quantum localization was rationalized by analyzing the
potential energy surface for the molecular rotation, derived from
DFT calculations \cite{ts01}.

The orientational order of H$_2$ molecules in these solid phases has 
been also studied from PIMC simulations in Ref.~\cite{cu02}, employing 
an empirical potential.
The results of these simulations allowed to propose
the following kinds of distribution of molecules:
(I) orientationally disordered hcp;
(II) orientationally ordered hcp with $Pa3$-type local orientational
order; and
(III) orientationally ordered orthorhombic structure of $Cmca$
symmetry.
The transition between phases I and III 
was found to be essentially induced by thermal fluctuations, but
quantum fluctuations were reported to be important in determining
the transition temperature, by effectively hardening the
intermolecular interaction \cite{cu01}.
From their results, these authors estimated the melting line for
phase III.
Other possible ordered structures at pressures $P \sim$ 150~GPa
were analyzed in Ref.~\cite{su97}.

The phase transition associated to orientational ordering in hcp 
H$_2$ was studied by Kaxiras and Guo \cite{ka94} using PIMC 
simulations with an effective potential obtained from DFT calculations.
These authors analyzed short- and long-range angular correlations
as well as the behavior of the corresponding order parameters, and
found this transition to be of first order.  

Classical and quantum ordering of protons in cold high-pressure
solid H$_2$ have been recently investigated by Li~{\em et al.}
\cite{li13}, at several points on the phase diagram, using 
{\em ab initio} PIMD simulations.
These calculations revealed that the transition between phases
I and II is strongly quantum in nature,
which results from a competition between thermal plus quantum
fluctuations that enhance molecular rotation, and anisotropic
intermolecular interactions that hinder it.
In this work, no evidence for the `quantum localization' reported
earlier \cite{ki00} was found (see above). 
The transition from phase II to III is more classical, since nuclear 
quantum motion plays a secondary role, and thus the transition
is mainly controlled by the underlying potential energy surface.
A $C2/c$ structure was proposed for phase III, which is now 
considered as the most likely candidate for this H$_2$ 
phase \cite{mc12b}.
Concerning the interatomic interactions, it was found that
taking into account van der Waals forces improved the agreement
between experiment and theory \cite{li13}.

Biermann~{\em et al.} \cite{bi98} presented
{\em ab initio} path-integral simulations of solid H$_2$ in
the pressure range from 150 to 700~GPa.
These authors compared results of the quantum-mechanical simulations
with classical ones, which allowed them to assess nuclear quantum
effects on the solid structure at different pressures.
They showed that a QHA is insufficient to correctly describe the
zero-point motion at megabar pressures.
These simulations suggested the possibility that the structure of
solid hydrogen at very high compression may be diffuse due to quantum
effects, showing characteristics similar to fluid phases.

Turning to isotopic effects in these solid phases of hydrogen,
PIMC simulations were carried out in Ref.~ \cite{ru92} to study
pressure-induced orientational ordering transitions in solid
para-H$_2$, as well as in ortho-D$_2$ and para-T$_2$
(all of them with even angular momentum).
These simulations were performed utilizing a potential energy
derived from a local density approximation in the framework of DFT.
At $T$ = 0, the transition pressure was found to change strongly
as a function of the considered hydrogen isotope. In fact, pressures 
of 75, 26, and 12~GPa were obtained for H$_2$, D$_2$, and T$_2$,
respectively. 
This procedure yielded phase diagrams for H$_2$ and D$_2$ in good
agreement with experiment for orientational ordering up to
pressures of about 100~GPa.

A strong isotope effect in phase II of solid H$_2$ and D$_2$
was reported by Geneste~{\em et al.} \cite{ge12}.
These authors investigated nuclear quantum motion at 80~K and 
pressures up to 160~GPa by first-principles PIMD calculations.
An important result is that molecular orientations are well
defined in phase II of D$_2$, but solid H$_2$ displays large and
asymmetric angular quantum fluctuations, so that it is difficult
to associate a single
classical structure to this H$_2$ phase. The mechanism for the
transition to phase III was also discussed in this contribution.

\subsection{Ice}

Condensed phases of water have been studied in recent years by means of
PI simulations.
For the different ice polymorphs, this kind of simulations have been
mainly addressed to understand how the mass of the lightest atom,
hydrogen, can influence the structural and thermodynamic properties 
of these solids. This includes consideration of isotopic effects in
these properties, which can be nontrivial in some cases.

Computer simulation of water in condensed phases has a long
history, that dates back to about 1970 \cite{ba69,ra71}.
Studies of ice using path-integral simulations have been
performed later by using mainly effective potentials, and were
focused on structural and dynamic properties of several ice 
phases \cite{ga96b,he05,he06b,pa08,mc09}.
Different types of empirical potentials have been employed in these
simulations \cite{ko04,jo05,ab05,pa06,mc09}.  
Many of them assume a rigid geometry for the water 
molecule (such as TIP4P and SCP models), and
some others include molecular flexibility either with harmonic
or anharmonic OH stretches (as the q-TIP4P/F model \cite{ha09}). 
Also, in some cases the polarizability of the water molecule has 
been taken into account in the considered model potentials \cite{ma01}.
Moreover, some simulations of water using
{\em ab initio} calculations for the electronic structure have 
appeared in the literature \cite{fe06,mo08,li11,ko12}.
Nevertheless, the hydrogen bonds present in condensed phases of water
seem to be difficult to describe with available energy
functionals, which causes that some properties are poorly reproduced
by DFT simulations \cite{yo09}.
This scenario is now changing by the development of new functionals that
account for van der Waals interactions \cite{co13}.

Several properties of ice Ih have been investigated by using 
PI simulations with effective rigid-water potentials.
Gai~{\em et al.} \cite{ga96b} employed the SCP model to study ice Ih
near the melting temperature, and concluded that nuclear quantum 
effects are appreciable at temperatures $T \lesssim$ 280~K, so
that hydrogen bonds become effectively weaker.
The TIP4P model was later employed to study hexagonal ice
by centroid molecular dynamics \cite{he05}.
The effect of quantization on the local structure, as measured 
by the radial and spatial distribution functions, as
well as on the energy and lattice vibrational modes was presented. 
Comparison of results from quantum and classical simulations indicated 
that the increase in potential energy due to quantum motion is 
similar to that obtained by rising the temperature around 80~K. 
This study was extended to other rigid-water models available in
2006, and the results obtained for equilibrium and dynamical
properties of ice showed no important discrepancies between the 
different potential models \cite{he06b}.

Later versions of the TIP4P potential, such as the so-called TIP4PQ/2005
model, have been used to investigate the influence of nuclear quantum
delocalization on several ice phases. 
Path-integral simulations with this effective potential
were found to reproduce well a number of physical properties of ice
polymorphs, such as density, structure, heat capacity, and relative 
stability \cite{mc09,ve10}.
The TIP4PQ/2005 model was proposed in order to correct some deviations
of the properties predicted by earlier TIP4P-type potentials with
respect to experimental values. 
This question raised for rigid models of water is a problem generally 
appearing when performing path-integral simulations with effective 
potentials tailored to fit results of classical simulations to 
experimental data.
In such cases, quantum effects may worsen the agreement with
experimental results, because the effective potential already 
incorporates the effects of quantum zero-point motion 
by fitting its predictions to the actual experimental data.
Thus, path-integral simulations may incur in a `double counting' 
of these quantum effects, hence over-correcting the classical simulation 
data.

\begin{figure}
\vspace{-0.0cm}
~\hspace{-0.5cm}
\includegraphics[angle=-90,width= 9cm]{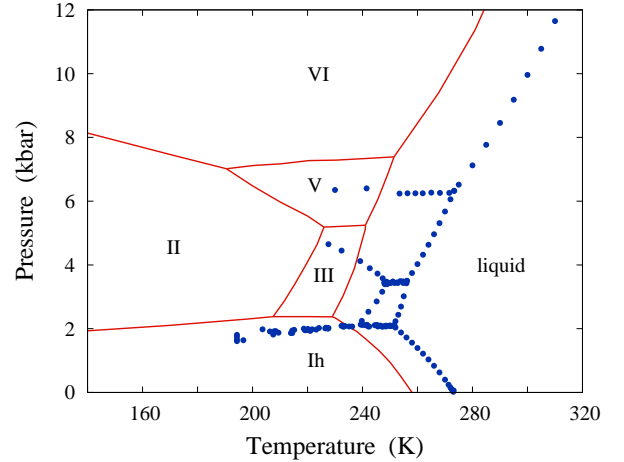}
\vspace{-0.0cm}
\caption{
Phase diagram of water from PI simulations with the TIP4PQ/2005 model.
Continuous lines correspond to the coexistence between phases, as
derived from the simulations. Labels indicate liquid water and
various ice polymorphs.
Experimental results (dots) are also shown for comparison
(from Ref.~\cite{mc12}, with permission of The Royal Society
of Chemistry).
}
\label{fig_ice_phdiag}
\end{figure}

An interesting application of the TIP4PQ/2005 model consisted in
calculating the phase diagram of water using PIMC \cite{mc12}. 
The coexistence lines between different phases were traced out using 
thermodynamic integration, after calculating the free energies
of liquid water and solid polymorphs. 
The resulting phase diagram was found to agree reasonably well with 
the experimental one, the former being displaced to lower temperatures 
by 15-20~K (see Fig.~\ref{fig_ice_phdiag}). 
In this work, the relevance of nuclear quantum effects on phase 
transitions was found to be significant. These quantum effects were
rationalized in terms of the degree of tetrahedral ordering in 
the different ice polymorphs and the volume variation associated 
to each phase transition \cite{mc12}. 

An extension to other phases of water consisted in studying
hydrate solid structures \cite{co10}.
A classical description of hydrates is able to predict correctly the
densities at $T >$ 150~K and the relative stabilities between the
hydrates and ice Ih.
Below 150~K, however, the inclusion of nuclear quantum effects was
found to be necessary for a reliable description of the density,
sublimation energy, heat capacity, and radial distribution
function of these water phases.

\begin{figure}
\vspace{-0.9cm}
\includegraphics[width= 8.0cm]{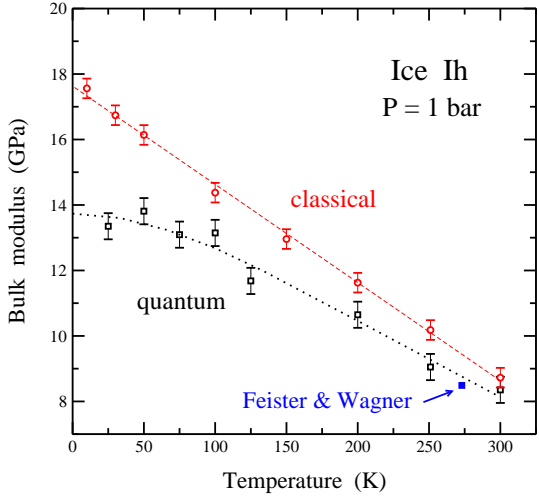}
\vspace{-0.5cm}
\caption{
Bulk modulus of ice Ih as obtained from PIMD (open squares)
and classical MD (open circles) simulations \cite{he11b}.
Error bars show the statistical uncertainty in the values of $B$
found from the simulations.
A filled square represents the value obtained at 273~K from the
equation of state derived by Feistel and Wagner \cite{fe06b} from
experimental data. Lines are guides to the eye.
}
\label{fig_ice_iso8}
\end{figure}

Among the effective potentials available for flexible water molecules, 
one of the most employed to carry out path-integral simulations of 
liquid water and ice has been the q-TIP4P/F potential, developed by 
Habershon~{\em et al.} in 2009 \cite{ha09}.
This potential was used in Ref.~\cite{he11} to analyze 
finite-temperature quantum effects in ice Ih from 25 to 300~K at 
atmospheric pressure.
It was shown that this kind of quantum simulations are necessary to 
reproduce some properties of ice, that are not captured by classical
simulations.
Among these properties, PIMD simulations were able to reproduce 
the negative thermal expansion of hexagonal ice Ih at low
temperatures. Also, these simulations gave the apparently
anomalous decrease of the intramolecular O--H distance for increasing
temperature. A good check of this interatomic potential was
the calculation of macroscopic observables such as the bulk modulus $B$
shown in Fig.~\ref{fig_ice_iso8}, where results from
classical and PIMD simulations are presented \cite{he11}.
In the classical simulations, $B$ is found to decrease approximately 
linearly as temperature is raised, and PIMD indicates that quantum 
ice Ih is `softer' ($B$ is smaller).
In the low-temperature limit a value $B = 13.8 \pm 0.4$~GPa was found, 
about 3.8~GPa smaller than the classical value, which means that 
quantum motion reduces the bulk modulus by more than 25\%.

Isotope effects in ice Ih were analyzed from PIMD simulations by 
considering normal, heavy, and tritiated water \cite{he11}. 
In particular, it was studied the effect of changing
the isotopic mass of hydrogen upon the kinetic energy and atomic
delocalization in the crystal, as well as on structural properties such
as interatomic distances and molar volume. 
For D$_2$O ice Ih at 100~K, it was found a
decrease in the crystal volume and intramolecular O--H distance of 
0.6\% and 0.4\%, respectively, as compared to H$_2$O ice.
These simulations, however, did not reproduce the inverse isotopic effect
in the crystal volume, observed in Ref.~\cite{ro94}, i.e. the molar
volume of D$_2$O ice derived from synchrotron x-ray diffraction is larger 
than that of H$_2$O ice (contrary to simpler solids, where heavier
isotopes have smaller volume; see Fig.~\ref{fig_nobl1.7}).
This effect in ice has been later reproduced quantitatively by using DFT 
calculations in combination with a QHA for the solid vibrations 
\cite{pa12}.

The isotopic effect in the melting temperature $T_m$ of ice Ih
has been also calculated using path-integral simulations with the
q-TIP4P/F potential \cite{ra10}.
At ambient pressure, these calculations predicted for D$_2$O and
T$_2$O ice an increase in $T_m$ of $6.5 \pm$ 0.5~K and $8.2 \pm$ 0.5~K,
respectively, vs H$_2$O ice. These temperature shifts show the same
trend as experimental results, although the theoretical ones are larger
in magnitude.
This isotopic effect was rationalized by considering the coupling
between intermolecular interactions and molecular flexibility in
liquid water and ice \cite{ra10}.

The kinetic energy of H and O nuclei in ice Ih at $P$ = 1~bar was 
studied as a function of temperature in Ref.~\cite{ra11}.
Results obtained for the proton kinetic energy in hexagonal ice 
compared well with data derived from deep inelastic neutron scattering
experiments, although some discrepancies were observed in the case of
liquid water.  
(The data analysis of some of these experiments in water has been
controversial \cite{so09,pi09}.)
The temperature dependence of $K_{\rm H}$
predicted by the q-TIP4P/F model for ice Ih was found in good
agreement with results of path-integral simulations using 
{\em ab initio} DFT \cite{ra11}.
Comparing calculated values of $K_{\rm O}$ and $K_{\rm H}$ in water 
and ice at isothermal conditions, it was shown that the larger results
always correspond to the solid phase.
This fact determines the sign of the isotopic shift in the melting
temperature of ice upon isotopic substitution of either H or O atoms.
In both cases the model predictions agreed to experimental data
\cite{ra11}.

\begin{figure}
\vspace{-0.9cm}
\includegraphics[width= 8.0cm]{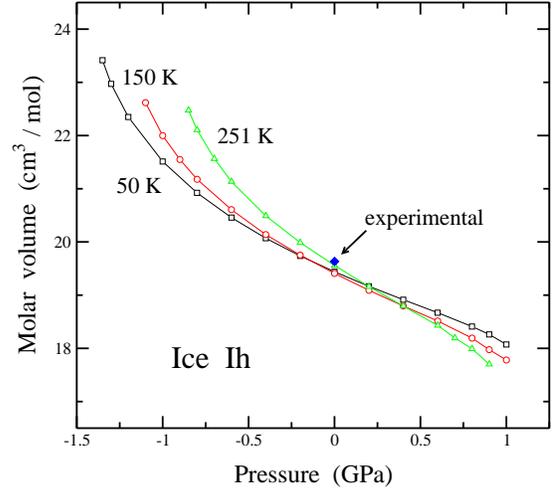}
\vspace{-0.5cm}
\caption{
Molar volume of ice Ih as a function of pressure, as derived
from PIMD simulations at various temperatures: 50~K (squares),
150~K (circles), and 251~K (triangles) \cite{he11b}.
Error bars are in the order of the symbol size.  Lines are guides
to the eye.  A solid diamond represents experimental data from
Refs.~\cite{gi47,da68} at atmospheric pressure and 273~K.
}
\label{fig_ice_pres1}
\end{figure}

Another application of path-integral simulations with a flexible-water
model (q-TIP4P/F) has been an study of ice Ih under pressure \cite{he11b}.
Positive (compression) and negative (tension) pressures were considered,
which allowed to approach the limits for the mechanical stability of
hexagonal ice. Several structural and thermodynamic quantities were 
studied as functions of pressure: crystal volume, bulk modulus, 
interatomic distances, atomic delocalization, and kinetic energy.
The overall agreement with available experimental data was good.
In Fig.~\ref{fig_ice_pres1} we show the pressure dependence of the 
molar volume at three different temperatures: 50, 100, and 251~K.
At a given temperature, one observes the common volume decrease for
increasing pressure ($d V/d P < 0$), and in most of the considered 
pressure region one has $d^2 V/d P^2 > 0$. Nevertheless, around 
$P$ = 0.5~GPa it was found a change in the trend of the first derivative, 
as indicated by the existence of an inflection point with $d^2 V/d P^2 = 0$.
This change in the trend of the $P$--$V$ curve has been argued to be
related with to the proximity of ice amorphization \cite{mi84,sc95,st04}.
It is also remarkable that the volume-pressure curves cross at
$P \sim$ 0.2--0.3~GPa, as discussed in \cite{he11b}.
This anomalous behavior of the crystal volume as a function of
temperature seems to be due to the negative sign of the Gr\"uneisen 
parameter for TA vibrational modes in ice Ih \cite{st04}.

\begin{figure}
\vspace{-0.9cm}
\includegraphics[width= 8.0cm]{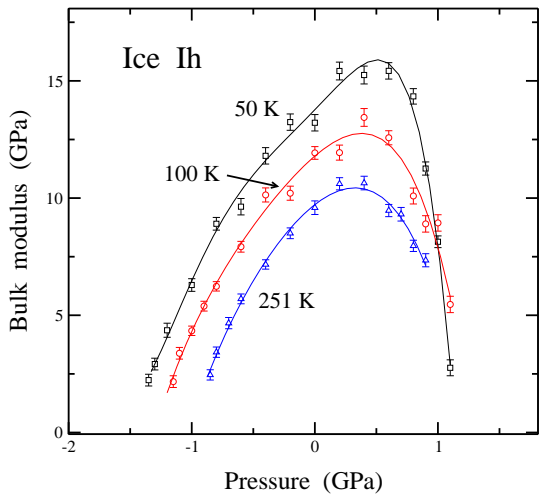}
\vspace{-0.5cm}
\caption{
Pressure dependence of the bulk modulus of ice Ih as obtained from
PIMD simulations at several temperatures: 50~K (squares), 150~K
(circles),
and 251~K (triangles) \cite{he11b}.
Error bars show the statistical uncertainty in the values of $B$
found from the simulations.  Lines are guides to the eye.
}
\label{fig_ice_pres2}
\end{figure}

In connection with the pressure dependence of the crystal volume, the
bulk modulus $B$ is also expected to show some related anomalies. 
Thus, at positive and negative pressures, $B$ was found to 
extrapolate to zero (the compressibility diverges) at the corresponding 
spinodal pressure $P_s$,
which depends on temperature (see Fig.~\ref{fig_ice_pres2}).
In this respect, it was found that nuclear quantum effects reduce 
the metastability region of ice Ih, for both negative and positive 
pressures \cite{he11b}.

\begin{figure}
\vspace{-0.9cm}
\includegraphics[width= 8.0cm]{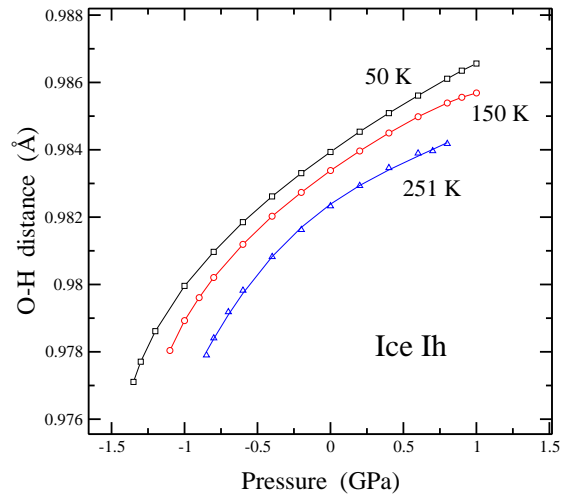}
\vspace{-0.5cm}
\caption{
Pressure dependence of the mean intramolecular O--H distance, as
derived
from simulations of ice Ih \cite{he11b}. Open symbols represent results
of PIMD simulations at different temperatures:
50~K (squares), 150~K (circles), and 251~K (triangles).
Error bars are in the order of the symbol size.
Lines are guides to the eye.
}
\label{fig_ice_pres5}
\end{figure}

Notwithstanding the volume reduction for increasing pressure, the
interatomic distances in ice were found to change in a peculiar way.  
Thus, the distance between oxygen atoms in adjacent molecules decreases 
as pressure is raised, but the intramolecular O--H distance becomes 
larger (see Fig.~\ref{fig_ice_pres5}).
This is due to the hydrogen bonds between contiguous molecules,
which become stronger as the volume (or O--O distance) is reduced,
causing a weakening of the intramolecular O--H bonds \cite{he11b}.

Amorphous ice has been likewise investigated by path-integral
simulations. Thus, structural and thermodynamic properties of
high-density amorphous (HDA) ice were studied by PIMD simulations
in the isothermal-isobaric ensemble, using the q-TIP4P/F potential
model \cite{he12}.
Quantum nuclear motion was found to affect several observable
properties of the amorphous solid.
At $T$ = 50~K, the molar volume of HDA ice was found
to increase by 6\%, and the intramolecular O--H distance rose by
1.4\% due to quantum motion.
Moreover, PIMD simulations showed a broadening of the peaks
in the RDFs, as compared with classical MD simulations. 
For different isotopes, changes in the RDFs of HDA ice
were observed. In particular, the width of the peaks in the
O--H RDF was found to depend on the considered hydrogen isotope,
but the general features of this RDF were substantially the
same for H$_2$O and D$_2$O ice.

These conclusions contrast with those of earlier PIMC simulations in
the $NVT$ ensemble, using the SPC/E rigid-water model \cite{ga96}.
In these simulations, the structure of H$_2$O amorphous ice was
found to be significantly different from that in D$_2$O ice, and this
difference was attributed to the larger quantum delocalization of the
protons, as compared to the deuterons.
These authors argued that the hydrogen bonding network that is present 
in D$_2$O amorphous ice either disappears
or is totally mixed with the second nearest-neighbor shell.
Apart from the constant volume employed by Gai~{\em et al.} \cite{ga96}
in their simulations vs the constant pressure employed in PIMD
simulations with the q-TIP4P/F potential, the main difference between
both kinds of calculations was the interatomic potential:
rigid molecules in Ref.~\cite{ga96} vs flexible molecules in
Ref.~\cite{he12}.
Nevertheless, it does not seem straightforward to find a convincing
explanation why rigid molecules should yield very different
structures for H$_2$O and D$_2$O amorphous ice, as suggested by the
results of Gai~{\em et al.}

Some measurements of the momentum distribution in water and ice
revealed that protons are in a considerably softer potential in ice Ih
than in water or the free monomer, which is compatible with the large 
redshift observed in the vibrational spectrum of ice.
Burnham~{\em et al.} \cite{bu06} showed that many water models,
which consider the intramolecular potential as unchanged by the hydrogen
bonding cannot reproduce the momentum distribution.
Moreover, although they can substantially reproduce the redshift,
they are unable to explain the large rise in intensity observed in
the infrared spectrum when going from the water monomer to ice Ih.
To correct these limitations, an empirical flexible and polarizable 
water model (TTM4-F) has been developed \cite{bu08}.
Using PIMD simulations, this model was found to give an improved
description of the momentum and position distributions, as well as
the infrared spectrum.
It was shown that the over-rigidity of the OH stretch and the
underestimation of infrared intensity, observed for other empirical
models, are connected to the failure to reproduce the correct
monomer polarizability surface.

 More recently, Lin~{\em et al.} \cite{li11b} analyzed the effective
potential experienced by protons in ice Ih from PI and
phonon calculations. They found an anisotropic quasi-harmonic effective
potential with three distinct principal frequencies corresponding
to different molecular orientations.  Anharmonicity in the
ground-state dynamics of protons turned out to be appreciable.
From these calculations, it was concluded that the full momentum
distribution is required for a reliable comparison with experimental
results.

Among other applications of PI simulations to condensed
phases of water, quantum motion of H$_2$O molecules has been shown to
appreciably affect the surface premelting of ice Ih \cite{pa08}.
This behavior was revealed by calculating structural and dynamical
properties of the ice surface below the melting point.
Results of PIMD and centroid molecular dynamics simulations
with a polarizable and flexible water model predicted 
the existence of a quasi-liquid phase extending about 10 \AA\ into 
the bulk, in line with results of optical ellipsometry measurements.  
Surface disorder in the topmost layers of the quantum ice was found to 
persist at temperatures of about 10--20~K lower than its classical 
counterpart.

Effective potential models for condensed phases of water usually treat the 
H$_2$O molecules as well-defined unbreakable entities, connected by 
H-bonds. However, high-pressure phases such as ice X, include symmetric 
O$-$H$-$O bonds, thus preventing their modeling with those empirical
potentials. This drawback does not affect {\em ab initio} potentials,
where in principle any geometry or bonding configuration can be
considered. 
Thus, Benoit~{\em et al.} \cite{be98b} investigated the role of quantum 
effects in proton ordering and hydrogen-bond symmetrization in
high-pressure phases of ice, by using PIMD simulations with an
internuclear interaction given by {\em ab initio} DFT. 
They found a sequence of transitions from antiferroelectric ice VIII 
to paraelectric (proton disordered) ice VII, and finally to symmetric 
ice X.
Nuclear quantum effects were found to influence these transitions: 
the antiferroelectric disordering transition is affected by proton
tunneling, whereas the symmetrization transition is influenced by
zero-point motion. The coordination of H in ice X, with a unimodal 
proton distribution centered at the O--O midpoint, was explained on the 
basis of an effective potential for hydrogen. 
This potential evolves from a double-well shape at lower pressures to a 
single-well shape with a stronger proton localization at higher pressures.
Thus, this kind of simulations have allowed to study the shape of the
proton-transfer potential between a donor and an acceptor molecule in
ice \cite{be99}.

More recently, the momentum distribution of protons in high-pressure 
ice phases was studied by Morrone~{\em et al.} using {\em ab initio} 
PIMD \cite{mo09}.
They found that the symmetric H-bonded phase (ice X) is characterized 
by a momentum distribution narrower than phases containing the usual
intramolecular covalent OH bonds, 
such as ice VIII, in agreement with experimental findings. 
The transition to the tunneling regime displays similarities
to some aspects observed in experiments on confined water. 
The results of these simulations were rationalized with 
a one-dimensional double-well potential model, that captures
some of the effects observed in the simulations. This allowed to 
distinguish between ground state and mixed state tunneling effects
in the proton dynamics \cite{mo09}.

\subsection{Other molecular solids}

Path-integral simulations have been also carried out to study other
molecular solids. We present here a few of them.
 Solid nitrogen was investigated by Presber~{\em et al.} \cite{pr98} 
using PIMC simulations in the isothermal-isobaric ensemble. 
They assumed an N$_2$ crystal composed of rigid molecules interacting
via Lennard-Jones and electrostatic potentials.
The effect of quantum fluctuations on molar volume, energy, and 
temperature for the orientational order/disorder transition was studied 
as a function of pressure.
The low-temperature molar volume and energy derived from the 
PIMC simulations were close to the corresponding experimental values, 
in contrast to classical simulations.
At $P = 0$, the transition temperature from a high-temperature 
orientationally disordered cubic phase to a low-temperature phase with 
$Pa3$ structure was found to be reduced by about 11\% as a consequence 
of quantum motion. 
For increasing pressure, the transition temperature rises and the 
difference between classical and quantum values is reduced. 

The quantum rotator phase transition in solid methane was analyzed in
Ref.~\cite{mu96} by using PIMC simulations.
With this purpose, an efficient quantum propagator for asymmetric tops 
was developed.
Quantum fluctuations were found to reduce the transition temperature
from a high-temperature plastic phase to a low-temperature  
orientationally ordered state.

The temperature dependence of several physical properties of
crystalline polyethylene at $P$~=~0 was studied by 
using self-consistent quasi-harmonic lattice dynamics \cite{ru98}, 
as well as classical and PIMC simulations \cite{ma98,ru98}.
Special emphasis was laid upon the classical approximation, the onset 
of anharmonicities in the atomic dynamics, not captured by the QHA, 
and finite size effects. 
It was shown that quantum effects are significant for 
$T \lesssim$ 300~K, and above 250~K (about 2/3 of the melting 
temperature) the QHA becomes increasingly inadequate.
Taking into account the computational costs, it was concluded that
quasi-harmonic lattice dynamics may be a very efficient alternative to 
study solid polyethylene at $T <$ 250~K.

\section{Point defects in solids}

Point defects in solids have been investigated in the last decades
using different experimental and theoretical techniques.
Over the last 20 years, path-integral simulations have been applied 
to study the structure (geometry) and dynamics of this kind of
defects in a variety of materials. Several works have been devoted to
hydrogen-like impurities, since they are expected to show relevant
nuclear quantum effects. Other applications have included heavier
impurities, such as oxygen in semiconductors \cite{ra97} and group-III 
elements in solid hydrogen \cite{kr00,mi02}.

\subsection{Structure}

Several path-integral simulations have dealt with atomic hydrogen in
silicon and diamond.
Thus, finite-temperature properties of isolated hydrogen and deuterium 
impurities in crystalline silicon were studied by the PIMC method,
in the temperature range from 50 to 600~K \cite{he95}.
Interactions between Si atoms were modeled by an effective
Stillinger-Weber potential, and
the Si--H interaction was parameterized by following the
 results of pseudopotential-DFT calculations. 
Hydrogen and deuterium were found to be stable at the bond-center
(BC) site, midway between two adjacent Si atoms.  
Average values of the kinetic, $\langle K \rangle$,
and potential, $\langle U \rangle$, energy
of the defect were compared with those expected for the impurity
within a harmonic approximation. At low temperatures,
$\langle K \rangle$ is larger than $\langle U \rangle$, as a
consequence of the strong anharmonicity of the potential surface
for the impurity around the BC site.

Similar simulations were carried out for muonium in crystalline Si,
which behaves for this purpose as a light pseudo-isotope of hydrogen.
In contrast to hydrogen, the same interatomic potential model predicted 
muonium to be delocalized in the interstitial region. This mass-induced
effect was argued to be a consequence of the uncertainty principle,
that prevents localization of muonium at the BC site due to an increase
in its kinetic energy \cite{ra94}.
These hydrogenic impurities in silicon were later studied by
the PIMD method using DFT-based interatomic potentials \cite{mi98}.
It was shown that muonium has a high density at the center of the silicon
cage (the so-called T site) due to quantum effects, i.e., far from the
BC site found for hydrogen.  

A similar PIMC study was carried out for hydrogenic impurities in 
boron-doped silicon, using empirical potentials for the Si--Si and
Si--B interactions, and a potential for hydrogen parameterized from DFT
calculations \cite{no97c}.
Isotopic effects on the properties of B--H complexes were studied by
considering also B--D pairs. Thus, it was analyzed the influence of 
the isotope mass on the impurity energy, atomic delocalization and 
lattice relaxation. 
The reorientation rate of these complexes was obtained from quantum 
transition-state theory, based on the PI centroid formalism \cite{no97b}.
A change in the Arrhenius plot for the jump rate of hydrogen was 
obtained at $T \sim$ 60~K, indicating a crossover from thermally activated
quasi-classical motion over a barrier to thermally assisted quantum
tunneling, in line with experimental data.
For deuterium, a break of the Arrhenius law was found at $T \sim$ 35~K.  
These PI simulations showed that the defect complex 
undergoing quantum tunneling consists of hydrogen, boron and 
the nearest silicon atoms, rather than the hydrogen atom alone.

With respect to hydrogen-like impurities in diamond, PIMD simulations
in the canonical ensemble have been performed to study
isolated hydrogen, deuterium, and muonium in this material \cite{he06}.
Finite-temperature properties of these point defects were
analyzed in the range from 100 to 800~K. 
The interatomic interactions were modeled by a tight-binding potential
fitted to DFT calculations. 
The most stable position for these hydrogenic impurities was found to
be a BC site between carbon atoms, in agreement with earlier
theoretical calculations.
Vibrational frequencies associated to the defect complexes were 
calculated within the LR approach discussed in Sect.~II.H. 
The resulting frequencies displayed a large anharmonic effect in the
impurity vibrations at the BC site, causing a hardening of the
vibrational modes with respect to a harmonic approximation. 
Quantum zero-point motion was found to produce an appreciable shift of 
the defect level in the electronic gap, as a consequence of 
electron-phonon interaction \cite{he06}. 

Turning to molecular hydrogen as an impurity in solids, 
PIMD simulations have been carried out to analyze isolated H$_2$
molecules in silicon and graphite \cite{he09b,he10b}.
In these works, tight-binding-type potentials were used to model
the interatomic interactions.
For H$_2$ in silicon, the most stable position was found at the 
interstitial T site with the hydrogen molecule rotating freely in 
the Si cage \cite{he09b}.
The H--H stretching frequency derived from the LR approach
showed a large anharmonic effect, resulting to be softened with respect
to a harmonic approximation by about 300 cm$^{-1}$.
Moreover, the coupling between rotation and vibration was found to give
rise to an important decrease in this frequency for rising temperature.

Using the same procedure, finite-temperature properties of H$_2$ molecules
adsorbed between graphite layers were analyzed in the temperature range
from 300 to 900~K \cite{he10b}.
In its lowest-energy position, the H$_2$ molecule was found to be disposed
parallel to the sheets plane. At finite temperatures, the molecule
explored other orientations, but its rotation turned out to be 
partially hindered by the adjacent graphite layers. 
For the H--H stretching vibration, a frequency of 3916 cm$^{-1}$ was found
at 300~K, more than 100 cm$^{-1}$ lower than the frequency corresponding to
an isolated H$_2$ molecule (in vacuo). 
For D$_2$ molecules in graphite, a stretching frequency of 2816 cm$^{-1}$
was obtained at room temperature, which compared with the H$_2$
frequency gives an isotopic ratio of 1.39, close to $\sqrt{2}$.

Going to heavier impurities in solids, a well-known case is that of
oxygen in semiconductors, and silicon in particular.
This is an archetypal example of isolated impurities in solids showing
low-energy highly-anharmonic quantum dynamics, a question for which
path-integral simulations have turned out to be a well-suited tool.
In particular, interstitial oxygen in silicon was studied by
PIMC simulations in Ref.~\cite{ra97}.
An interesting problem arises since the definition of a geometry for 
this defect is conditioned by the presence of a large anharmonicity. 
This means that
the configuration with minimum potential energy may not correspond
to a maximum in the probability density of the impurity atom.
In fact, the minimum energy corresponds to a bent Si-O-Si configuration,
whereas the PIMC simulations yielded the maximum density for oxygen at
a BC site, midway between silicon atoms.
In contrast, for the oxygen impurity in germanium, a different geometry is
found and the Ge-O-Ge structure turns out to be puckered, with O
rotating around the Ge-Ge axis \cite{ar97}.

We now turn to atomic impurities in molecular solids.
First, we consider B and Al impurities in solid para-H$_2$.
The equilibrium properties of a boron atom trapped in solid
para-H$_2$ were investigated by Krumrine~{\em et al.} \cite{kr00} 
using PIMD simulations.
Due to its singly filled $2p$ orbital, the B atom interacts
anisotropically with the hydrogen molecules in the solid.
To assess the effect of this electronic anisotropy, these authors
performed simulations in which an orientation-averaged B-H$_2$
potential was used.
It was found that small distortions of the lattice allow for an 
energetically favorable orientation of the $2p$ orbital, even in
the absence of a vacancy.
When the B impurity is located near the surface, the
spherically-averaged potential gives a very different description
from the case of the anisotropic potential \cite{kr00}.

For Al impurities in para-H$_2$, it was particularly considered the
interaction between these impurities and vacancies in the solid \cite{mi02}.
In the presence of a vacancy, the orientation-dependent Al-H$_2$
potential causes the Al atom to displace to the midpoint between 
its substitutional site and the vacancy.
Thermodynamic results indicated that the presence of a neighboring
vacancy stabilizes the Al impurity much more than in the case of the B
impurity \cite{kr00}.

Apart from B and Al, isolated Li impurities in solid para-H$_2$ and 
ortho-D$_2$ have been studied by Scharf~{\em et al.} \cite{sc93} using 
PIMC simulations in the isothermal-isobaric ensemble.
These calculations predicted the impurities to occupy preferentially
a three-vacancy trapping site in para-H$_2$, whereas in
ortho-D$_2$ a four-vacancy trapping site seemed to be favored.
For comparison, a variational Einstein model predicted that the
four-vacancy trapping site is favorable in both para-H$_2$ and
ortho-D$_2$ \cite{sc93}.
The effect of pressure on the trapping site structures and adsorption
spectra of Li in solid H$_2$ has been studied later by PIMC simulations
\cite{ma01b}. The applied pressure was found to cause a reduction in
the Li--H$_2$ and H$_2$--H$_2$ distances, but no appreciable
rearrangement of the local structure around the Li impurity was
observed for pressures up to 4.8~GPa. 

Other point defects, such as vacancies in solid $^4$He have been also
studied using PIMC simulations \cite{cl08}.
A relevant question in this case is the energy necessary to
create a single vacancy, which was found to be 11.5~K (0.99~meV).   
As expected, the number of atom exchanges in the system increases with
the vacancy concentration. 
As a result, it was observed in the simulations that the solid
becomes unstable to melting whenever more than a few percent of
the crystal sites become vacant.
It was shown that vacancies behave in an attractive way, and the 
simulations allowed to quantitatively study the inter-vacancy
attraction.

\subsection{Diffusion}

Concerning impurity diffusion in solids, two methods based on 
path integrals have been mainly applied to calculate jump rates and 
diffusion coefficients.
One of them is based on quantum transition-state theory, as described
above in Sect.~II.G. Other methods rely on real-time MD simulations, 
where the ring-polymers describing the quantum
particles move according to the forces derived from interatomic
interactions, giving an approximation to the actual quantum 
dynamics \cite{ca94,ca94b,ja99,cr04,pe09,ha13}.

One of the earliest applications of atomistic PI simulations 
of solids was the work by Gillan \cite{gi87,gi88}, devoted to study 
hydrogen-like impurities in metals.
In particular, he considered hydrogen, deuterium and muonium in Pd and Nb, 
based on empirical interaction models, and carried out PIMC 
simulations to study the influence of quantum effects on the properties
of these isolated impurities. 
 Results for the density distribution of D, H, and the positive muon 
over the unit cell showed an important increase of quantum effects along 
this series. 
Gillan calculated effective diffusion barriers for these point defects,
and found that they change appreciably with the impurity mass.
The calculated diffusion coefficient for H in Nb displayed an 
experimentally observed break in the Arrhenius slope at 250~K, 
which was interpreted as a cross-over from classical (high $T$) to 
quantum (low $T$) behavior. 
Moreover, calculations of the activation energy for diffusion showed
the relevance of excited states at high temperature,
suggesting that the hydrogen diffusion is approximately classical
in this regime \cite{gi87,gi88}.

For hydrogen-like impurities in silicon, QTST was applied 
to study the jump rate of atomic hydrogen and deuterium \cite{he97}. 
In this case the interatomic interactions were modeled with
potentials similar to those described for this system in Sect.~IV.A.
The hydrogen jump rate was found to follow an Arrhenius law, describable 
with classical transition-state theory at $T >$ 100~K.
At $T \sim$ 80~K, a change in the slope of the Arrhenius plot was
obtained for hydrogen, as expected for the
onset of a diffusion regime controlled by phonon-assisted tunneling
of the impurity. In these calculations, no change of slope was observed 
for deuterium in the studied temperature range down to 40~K.

\begin{figure}
\vspace{-1.5cm}
\includegraphics[width= 9cm]{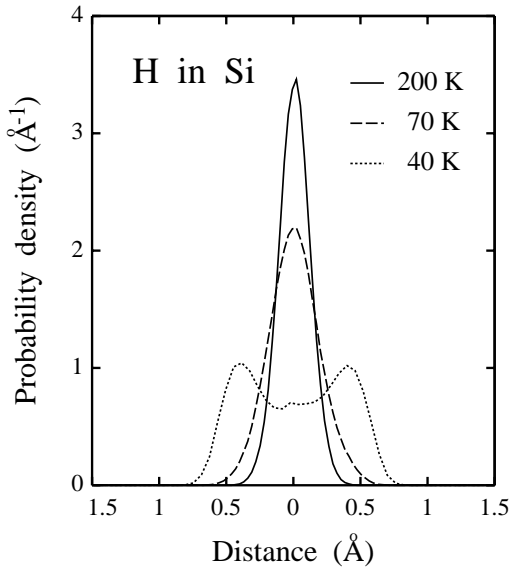}
\vspace{-2.3cm}
\caption{
Probability density for hydrogen in silicon, as derived from PIMC
simulations with the path-centroid of the impurity fixed at the saddle
point C$^*$ \cite{he97}.
Shown is the projection of the density along the [110] crystal
direction.
Continuous line, 200~K; dashed line, 70~K; dotted line, 40~K.
The appearance of two well-defined maxima in the probability
density at $T$ = 40~K is an indication of impurity tunneling.
}
\label{fig_sih1}
\end{figure}

\begin{figure}
\vspace{-1.5cm}
\includegraphics[width= 9cm]{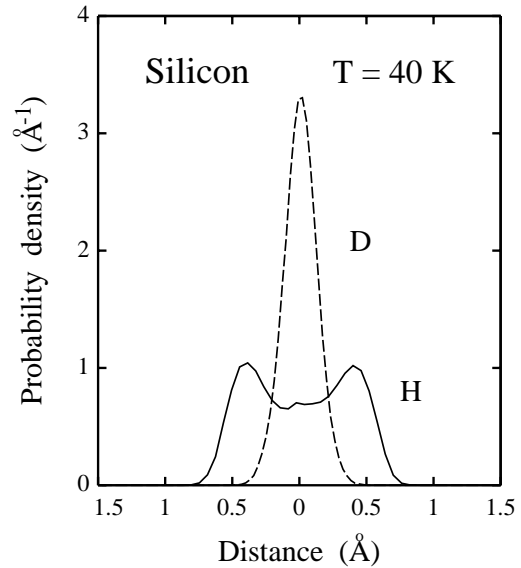}
\vspace{-2.3cm}
 \caption{
Probability density for hydrogen and deuterium in silicon, found in
PIMC
simulations at 40~K with the path-centroid of the impurity fixed at the
saddle point C$^*$ \cite{he97}.
Shown is a projection of the density as in Fig.~\ref{fig_sih1}.
Continuous line, hydrogen; dashed line, deuterium. Note the different
shape of the probability density for both isotopes at this temperature.
}
\label{fig_sih2}
\end{figure}

In this context it is interesting to analyze the probability distribution 
of the hydrogenic impurities with the path-centroid fixed 
at the saddle point (the so-called C$^*$) in the diffusion barrier. 
This provides information on the delocalization of the considered particle 
in the two potential wells, corresponding to two adjacent energy minima 
(BC sites).
A projection of the probability density of hydrogen in silicon
along the [110] crystal direction is displayed in Fig.~\ref{fig_sih1}.
The spatial distribution of the impurity broadens as temperature is 
lowered, and eventually its shape changes from a single maximum around 
the saddle point to a bimodal distribution, which is a fingerprint of 
quantum tunneling \cite{he97,ma93d}.
This change in the distribution of the quantum paths as temperature 
decreases, is associated to a transition from semiclassical motion
over the effective barrier to quantum tunneling through the barrier.
The effect of the impurity mass on the quantum delocalization is 
important at low temperatures, as shown in Fig.~\ref{fig_sih2}, 
where dashed and continuous lines correspond to projections of the 
impurity density on the [110] silicon axis for D and H, respectively, 
at 40~K. The deuterium distribution has the usual shape for particles 
undergoing semiclassical motion, in line with the fact that at 
$T \geq$ 40~K a linear Arrhenius plot was found for this 
impurity \cite{he97}.

The diffusion of H and D in crystalline silicon was also studied by 
path-integral methods in Ref.~\cite{fo98}.
In this work, the site-to-site hopping rate of the hydrogenic  impurities 
was calculated over a wide range of temperatures,
and the hopping rate was derived from the time integral of a flux 
correlation function. 
It was found that quantum tunneling is relevant in the diffusion
process even above room temperature.

Similar studies were carried out for the diffusion of hydrogen-like 
impurities in diamond. In this case, jump rates of hydrogen and muonium
were calculated by QTST, using interatomic
potentials derived from a tight-binding model \cite{he07}.
This technique allowed to analyze the effect of vibrational mode 
quantization on the effective free-energy barriers $\Delta F$ for 
impurity diffusion, which are renormalized with respect to 
the zero-temperature classical calculation. 
For the transition of the impurity from a tetrahedral T site to 
a BC position, $\Delta F$ results to be larger for hydrogen 
than for muonium, and the opposite occurs for the transition from a BC
to a T site. The effective barriers decrease for rising
temperature, except for the muonium transition from T to BC sites.
In the case of muonium, the calculated jump rates could be compared
with muon spin rotation data, giving good agreement with the
experimental results \cite{he07}.

For the diffusion of impurities in molecular solids, particular
attention has been focused on the case of Li in para-H$_2$.
In particular, the recombination of two Li atoms trapped in 
one-vacancy defect sites in this molecular solid has been studied 
at $T$ = 4~K by QTST and PIMD simulations \cite{ja98}.
The results of this investigation indicated that
the two Li atoms begin to recombine at a distance of about twice 
the lattice spacing, and the height of the effective free-energy barrier 
relative to the metastable well resulted to be about 80~K (6.9 meV). 
The recombination step at 4~K takes place on the millisecond time
scale, and given that the defects are metastable for
longer times, these results suggested that the recombination process
may be diffusion limited.
These quantum simulations showed that the Li nuclei display appreciable 
tunneling behavior through their classical barrier. 

Another vacancy-related problem in solids treated with path-integral
simulations has been the question of vacancy-induced superflow in 
solid $^4$He. This was
investigated by Boninsegni~{\em et al.} \cite{bo06b} using
a worm algorithm based on PIMC in the grand-canonical ensemble. 
Their results indicated that vacancies are unstable in a $^4$He 
crystal, i.e., they form clusters and the system separates into 
a vacancy-rich phase and a perfect, insulating crystal. 
They also explored the possibility that nonequilibrium vacancies
relaxing on defects of polycrystalline samples could provide an 
explanation for the experimental results suggesting the presence of 
the superflow in solid $^4$He.

As an example of ring-polymer MD simulations in solids,
we will mention a study of the diffusion of H, D, and muonium in 
hexagonal ice Ih as well as in liquid water over a wide temperature
range (from 8 to 361~K) \cite{ma08}.
Quantum effects were found to reduce the diffusion of muonium in water
relative to that predicted by classical simulations.
This is due to an increase in the effective radii of the quantum
particles, which can be considered as a `swelling' of classical
particles caused by finite-temperature quantum fluctuations, and can
be quantified in terms of the `radius of gyration' of the corresponding
ring-polymers \cite{ma08}.
Results of those simulations indicated that the diffusion mechanism in 
the liquid phase is analogous to the intercavity hopping observed in ice, 
with an additional diffusion of the cavities in the liquid.
An observed crossover in the $c$-axis diffusion coefficients of H and D
in ice Ih at about 200~K was satisfactorily reproduced by these
simulations.
Good agreement with experimental data was also found for the diffusion
of muonium in hexagonal ice at 8~K, a temperature at which the process is
purely quantum mechanical.

\section{Surfaces and adsorbates}

Path-integral simulations have been applied to the study of solid
surfaces and adsorbates on different types of materials.
Several works have been devoted to the physisorption of diatomic molecules 
(N$_2$, H$_2$, CO, ...) on graphite, as well as to the chemisorption of 
hydrogen on this material. 
Adsorbates on other types of materials have been also studied with this
kind of techniques. Most of these works were carried out using empirical
interatomic potentials, but in recent years several papers dealing with 
{\em ab initio} PI simulations have appeared in the literature. 

Questions related to the adsorption of molecules on solid surfaces are 
the adsorption sites and two-dimensional (2D) structure of the adsorbate, 
as well as the diffusive and rotational molecular motions. 
In particular, at low temperatures it is expected that molecular rotation 
will be affected by quantum effects.
A good example of this is the case of nitrogen adsorbed on
graphite, which forms a layer with `herringbone' ordering.
 The effects of quantum fluctuations on this ordering pattern
of N$_2$ molecules have been studied via PIMC simulations,
using a procedure optimized for the simulation of rotational motion 
in many-body systems \cite{ma93}.  It was found that
quasi-classical and quasi-harmonic calculations agreed with PIMC
simulations for high and low temperatures, respectively, but only 
the quantum simulations gave satisfactory results over the whole
temperature range. 
An important conclusion of this work was that the inclusion of quantum 
fluctuations results in a 10\% lowering of the orientational-ordering 
transition temperature of the molecular monolayer.
Moreover, a decrease in the ground-state ordering parameter of
about 10\% was also observed, which turned out to be due to quantum 
librations of the N$_2$ molecules in the potential wells \cite{ma93}. 

That work was extended by considering
the rotational motion of homonuclear diatomic molecules confined to
two dimensions at finite temperatures in Ref.~\cite{ma93c}.
For single rotators, the symmetry restriction on the total wave
function (including nuclear spin and rotations) was carried out for 
fermionic and bosonic diatomic molecules in the PIMC simulations.
This method was applied to single N$_2$ and H$_2$ rotators adsorbed on
graphite in the frozen-in crystal field due to a commensurate
herringbone phase.
Contrary to N$_2$, exchange effects were found to be appreciable for
H$_2$ in the relevant temperature range \cite{ma93c}.

Other phases of molecular hydrogen on graphite have been also studied
by PIMC simulations \cite{nh02}.
For an ideal flat substrate, the first molecular layer was found to
have a solid-gas coexistence phase at low densities, and a
triangular solid phase above some equilibrium density.
For full H$_2$-graphite interaction (including the substrate corrugations),
a variety of solid, fluid, and gas phases were found, depending on
the coverage and temperature.
In particular, at low coverages and low enough temperatures, the
adsorbate consists of solid clusters surrounded by vapor. 
At coverages below a critical density, defining a tricritical point, such 
clusters melt into a uniform fluid phase as the system is heated up.
 In connection with this, adsorption of ortho-D$_2$ and para-H$_2$ films 
on a graphite substrate, preplated with a single atomic layer of krypton,
was studied by PIMC simulations at low temperature \cite{tu07}.
The thermodynamically stable adsorbed films were found to be
solid, with no clear evidence of any liquid-like phase.
Quantum exchanges of ortho-D$_2$ and para-H$_2$ were essentially
absent down to zero temperature, thus concluding that the 
system displays no superfluidity in this limit.

Apart from homonuclear diatomic molecules, quantum effects upon
heteronuclear molecules on solid surfaces have been considered.
Among these molecules, one of the most studied 
has been carbon monoxide. In this case, head-tail ordering in
CO monolayers on graphite was studied by performing classical and PIMC
simulations with an {\em ab initio} potential along with 
finite-size scaling \cite{ma94}.
The order/disorder transition was found to belong to the 2D Ising
universality class. 
This assignment as well as the critical amplitude of the
heat capacity agreed with experimental data.
Quantum fluctuations were found to appreciably modify the critical
temperature obtained in the classical simulations, and in fact a
renormalization of this temperature by about 10\% was found.
This study also showed that ordering of molecular adsorbates is a
very sensitive phenomenon to check the accuracy of intermolecular
potentials \cite{ma94}.

Likewise, more complex molecules adsorbed on solid surfaces have been
investigated by path-integral methods. Thus,
the influence of nuclear quantum effects upon water-hydroxyl
overlayers on transition metal surfaces was studied by
using {\em ab initio} PIMD \cite{li10}.
The metal substrates cause a reduction in the classical barriers
for proton transfer within the overlayers, and
a shortening of the intermolecular hydrogen bonds.
It was argued that, depending on the substrate and the
intermolecular distances, the distinction between covalent and
hydrogen bonds may be partially lost, as happens for Pt(111) and Ru(0001),
or almost entirely, as occurs for Ni(111).

\begin{figure}
\vspace{-1.7cm}
\includegraphics[width= 9cm]{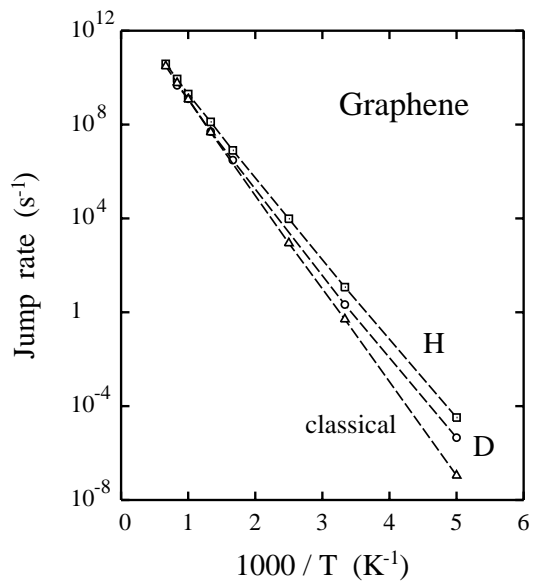}
\vspace{-2.3cm}
\caption{
Rate for impurity jumps on a graphene sheet as a function of the
inverse temperature. Symbols represent results derived from
QTST for hydrogen (squares), deuterium (circles), and
classical limit (triangles) \cite{he09a}.
 Dashed lines are guides to the eye.
}
\label{fig_grafe1}
\end{figure}

In recent years, some simulations of adsorbed species on a single
graphene layer have appeared in the literature.
The chemisorption of atomic H and D on graphene was
studied by PIMD simulations \cite{he09a}.
Finite-temperature properties of these point defects were obtained
by using a tight-binding potential fitted to DFT calculations. 
On one side, vibrational properties of the adatoms were studied at their
equilibrium positions, linked to C atoms. The
vibrations displayed an important anharmonicity, as derived from a
comparison between the vibrational energy of hydrogen and deuterium.
On the other side, motion of H and D was studied by QTST.
For hydrogen, the diffusivity on the graphene
sheet at 300~K turned out to be enhanced by quantum effects 
by a factor $\sim 20$, as can be seen in the Arrhenius-type plot
of Fig.~\ref{fig_grafe1}.
QTST based on PIMC simulations has been also employed to study
the diffusion of hydrogen on the Ni(100) surface
\cite{ma93d,ma95b,ma97}.

More recently, the motion of hydrogen on the Ru(0001) surface between 
75 and 250~K was studied in Ref.~\cite{mc13} by a combination of 
{\em ab initio} PIMD simulations, QTST, and helium 
spin-echo measurements.
Quantum effects were found to be significant at temperatures as high
as 200~K, while for $T <$ 120~K a tunneling-dominated
temperature-independent motion was observed, with a large jump rate of
$\sim 10^9$ s$^{-1}$.
The crossover to tunneling-dominated diffusion at low temperatures
was well reproduced by the QTST calculations, but 
the tunneling rate turned out to be underestimated when compared with
experimental data \cite{mc13}.

Even though superfluidity in bulk molecular hydrogen has not been observed, 
reduction of dimensionality has been regarded as a possible way
for the detection of a superfluid phase at low temperatures.
This has motivated experimental and theoretical investigation of 
adsorbed films of para-H$_2$ on several substrates. 
With this question in mind,
the low-temperature phase diagram of an ideal 2D sheet of para-H$_2$
was studied by Boninsegni \cite{bo04} from PIMC simulations.
For $T \to 0$, the system was found to be a crystal at equilibrium,
with a triangular lattice structure.
No metastable liquid phase has been observed, and the system remained
as a solid down to its spinodal density, breaking into solid clusters
at lower densities. The equilibrium crystal was found to melt at
$T \sim$ 7~K.
A more realistic system was later studied by considering a 2D
sheet of para-H$_2$ embedded in a crystalline matrix of alkali
atoms \cite{bo05}.
For this system, the thermodynamically stable phase at $T \lesssim$ 5~K
was a solid, commensurate with the underlying lattice.
For simulated systems of small size, a nonzero superfluid signal
for para-H$_2$ was observed, but results for larger system sizes
were consistent with the absence of superfluidity in the thermodynamic
limit.

In connection with the behavior of 2D H$_2$ films, and turning to 
three-dimensional solid hydrogen, an interesting question is the 
melting of the H$_2$ layers close to the surface, which is expected to
happen at temperatures lower than the bulk melting.
This was studied by Wagner and Ceperley \cite{wa96} using PIMC,
who obtained density profiles perpendicular and parallel to the 
surface, and showed the formation of a liquid adlayer at 6~K,
less than half the bulk melting temperature of para-H$_2$.
The onset temperature and depth of H$_2$ surface melting was
estimated by calculating the static structure factor for the
individual H$_2$ layers, and no crystalline order was found down to
3~K in a partially filled H$_2$ adlayer at the free surface.
Zero-point motion of H$_2$ molecules at the surface turned out
to be larger than in the bulk, so that quantum effects
largely amplify the melting point depression at the surface.

Adsorption of noble gases on a variety of solid surfaces and confined
geometries has been also studied by using path-integral methods. 
Several works have been focused on $^4$He.
In this respect, helium adsorption on graphene was studied by PIMC
calculations using effective interatomic potentials \cite{kw12}.
The first $^4$He layer was found to be commensurate with the substrate,
displaying a $C_{1/3}$ structure. Vacancy states in this structure
were found to behave differently depending on the employed interaction
potential. This means that they may appear as a cluster of localized
vacancies or as mobile vacancies inducing finite superfluid fractions.
For the second helium layer, exchange among $^4$He adatoms
resulted in quantum melting of a $C_{7/12}$ commensurate structure.
In a similar work,
the adsorption of $^4$He on a lithium substrate at low temperature was
studied by Boninsegni~{\em et al.} \cite{bo99}. These authors employed 
PIMC simulations with an effective interatomic potential, as well as DFT
calculations.
Evidence for continuous growth of a $^4$He film was presented,
with the appearance of a superfluid monolayer at low coverage and no
well-defined layering at higher coverages.
The coverage dependence of the estimated superfluid transition
temperatures was found to be similar to the experimental one.

 PIMC simulations have been also employed to determine the phase diagram 
of an isolated $^4$He film in a range of temperatures
and coverages where it undergoes solidification, superfluidity, and a
liquid-gas transition \cite{go98}.
The phase-transition densities were calculated, as well as
the coefficients for a functional form of the free energy for
the different phases.
The resulting phase diagram turned out to be similar to that derived
from experimental measurements of a second layer of helium on graphite.
Furthermore, the superfluidity of grain boundaries in solid $^4$He 
has been studied by PIMC simulations based on a continuous-space worm 
algorithm \cite{po07}.
These boundaries are generically superfluid at low $T$, with a
transition temperature of $\sim$ 0.5~K at the melting pressure.
Superfluid grain boundaries were found far from the melting line,
and a grain boundary in contact with the superfluid liquid at
the melting curve was found to be mechanically stable.

In addition to these contributions to the study of $^4$He in 
low-dimensional geometries, several PI works have been 
devoted to heavier noble gases in confined spaces.
Thus, neon adsorbed in the interstitial channels of a bundle of
carbon nanotubes was studied using PIMC calculations \cite{go04b}.
The chemical potential of Ne atoms could be evaluated, allowing 
for an estimation of the temperature and pressure conditions for 
neon adsorption to take place in those channels.
In connection with this, phase transitions and quantum effects in 
condensates of Lennard-Jones particles inside cylindrical pores were 
also discussed from PIMC results \cite{ho03}.
Several properties of this system were analyzed, such as the
effects of pore diameter and adsorbate--particle interaction
strength on the emerging structures and location of phase boundaries.
For strong wall--particle interactions, qualitative agreement was
found with experimental results for the freezing of Ar--pore
condensates, while for lighter Ne-like particles, 
quantum delocalization effects gave rise to unexpected condensate 
structures.  As a result of these simulations, it was concluded that 
nuclear quantum effects can change 
the freezing temperature by an amount as large as a 10\% \cite{ho03}.

In the context of surface science, PIMC techniques have been also 
applied to study model adsorbates with internal quantum states. 
Marx~{\em et al.} \cite{ma93b} considered an adsorbate in the 
strong-binding and small-corrugation limit.
In this case, the resulting 2D fluid was treated in 
the adiabatic approximation: particle translations were treated 
classically, whereas the internal degrees of freedom were
modeled by interacting two-state tunneling systems.
With these conditions, the temperature-coverage phase diagram was 
obtained by combining PIMC simulations with finite-size scaling.  
In spite of its simplicity, this adsorbate model presents a 
surprisingly complex phase diagram, including gas, liquid, and solid 
phases, as well as first- and second-order transitions \cite{ma93b}. 
It is remarkable the detection of a square-lattice solid phase in
coexistence with a gas phase at low temperatures.

Alfe and Gillan have recently employed path-integral 
simulations in combination with {\em ab initio} DFT calculations to 
construct practical techniques for computing the chemical potential of
molecules adsorbed on surfaces, with full inclusion of nuclear
quantum effects \cite{al10}.
These authors presented practical calculations for the case of H$_2$O 
on the MgO (001) surface at low coverage. 
It was shown that the number of time slices required in the PI
discretization for the higher vibrational frequencies of 
the H$_2$O molecule can be strongly reduced by employing a 
thermodynamic integration with respect to the number of beads.
The correctness of their path-integral results was demonstrated 
by calculations on a set of simple model systems, for which 
quantum contributions to the free energy can be exactly obtained 
from analytic arguments.

\section{Summary and outlook}

The path-integral formulation of quantum mechanics has rendered it possible
to theoretically tackle many problems that require the use of statistical 
mechanics with full consideration of quantum effects. 
The quantum-classical isomorphism derived for the statistical partition 
function allows the application of efficient classical simulation
techniques, such as MC or MD,  to the quantum domain. 
PI simulations reproduce quantum effects 
related to tunneling and anharmonic zero-point motion
of atoms in solids, as well as other properties associated
to the indistinguishability of the atomic nuclei. In this respect,
the treatment of bosonic degrees of freedom has been proven to be
feasible, while the consideration of fermionic systems remains in
general as a challenging task. Also, the path-integral simulation of 
dynamic properties, such as time correlation functions, poses a challenge 
that at present can only be treated by approximate algorithms.

The fast increase in computing facilities, and in particular the use
of parallel computers, opens new routes for PI simulations, both in
terms of the kind of system and properties to be studied, as well as
in terms of the computational methods. Limitations associated to the use 
of empirical potentials are currently being removed by the application of
\textit{ab initio} electronic structure methods. The appealing 
`all-quantum' approach, combining the PI simulation of atomic nuclei 
with an \textit{ab initio} DFT description of the electrons 
is, however, not free of shortcomings. 
For example, the DFT treatment of condensed phases of water has revealed 
to be elusive in reproducing certain thermodynamic properties,
so that the development of new improved functionals is an active area
of research. 

This kind of quantum simulations are necessary to reproduce some
observable properties of solids, that are not correctly described by 
classical simulations.
Thus, PI simulation is an efficient tool to understand various
isotope effects in solids, that at low temperature are a consequence of
the anharmonicity of atomic zero-point vibrations.
These effects are readily explored, since the atomic masses 
appear as input parameters in the PI calculations.
This type of simulations enable one to obtain separately kinetic and
potential energies at finite temperatures, taking into account the
quantization of nuclear motion.
This includes consideration of zero-point motion of the atoms in the
solid, which can be hard to study by analytical approaches in the
presence of light atoms and large anharmonicities.
In this context, a good check of interatomic potentials is the calculation 
of macroscopic observables such as the molar volume, thermal expansion, 
bulk modulus, or heat capacity, which can be directly compared
with experimental data.
In addition, differences between PI and classical calculations can be 
used to assess the importance of quantum effects in the equilibrium
properties of a given substance. 

In this topical review, we have presented PI simulations of different 
types of solids.
Noble-gas solids have allowed to check the methods and their
application to obtain structural and thermodynamic properties, due to
the relative simplicity of the interatomic potentials, as compared to
solids including covalent or hydrogen bonds.
Further applications of PI simulations have been concentrated on
group-IV materials (mainly diamond, and the semiconductors Si, Ge, and
SiC), for which the relevance of quantum effects in the presence of
anharmonicity has been shown.
Molecular solids have been also reviewed, with especial emphasis on
hydrogen and ice, which have revealed important nuclear
quantum effects.
Another field of interest for PI simulations has been
the diffusion of impurities in the bulk and adsorbates on the
surface of various types of solids, a problem for which QTST has 
turned out to be very efficient.

PI simulations of amorphous helium, silicon, and ice have been also
discussed. In a similar manner,
this kind of simulations can yield information on the atom
delocalization and anharmonic effects in other amorphous materials.
In particular, this method may provide relevant information on 
H-containing amorphous solids, whose macroscopic properties can be 
affected by the quantum delocalization of hydrogen.
Moreover, simulations in the context of QTST can give relevant
information on the low-energy tunneling states that control the
acoustic, thermal, and dielectric properties of amorphous materials
at low temperatures.

Hydrogen is of particular interest for PI simulations, as has been 
shown along this topical review. The lowest atomic mass implies 
the prevalence of quantum effects related to it. The presence
of hydrogen in molecules essential for biological systems means
that classical simulations may be not accurate enough
to reproduce some of their properties.  Thus, quantum tunneling 
of protons is supposed to play a relevant role in 
some biological processes, affecting for example the mechanisms for 
enzyme reactivity and oxidation of monoamine neurotransmitters. 
In this line, we expect that the PI approach will reveal in the near
future its capability in the investigation of quantum effects 
in biological systems, as has happened in the past for inorganic 
materials.

\begin{acknowledgments}
This work was supported by Direcci\'on General de Investigaci\'on 
(Spain) through Grant FIS2012-31713,
and by Comunidad Aut\'onoma de Madrid
through Program MODELICO-CM/S2009ESP-1691.
\end{acknowledgments}

\appendix

\section{Free particle density operator
\label{sec:Appendix}}

Computer simulations based on the PIMD and PIMC methods rely on a
discretized algorithm to approximate the temperature dependence of
the exponential density operator, $e^{-\beta\widehat{H}}$. In this
section we rationalize the basic steps of this algorithm by considering
first an arbitrary function of the inverse temperature $f(\beta)$.
One wants to approximate the evolution of $f$ starting at some initial
value $\beta_{1}$. The Taylor expansion of $f$ can be written as
\begin{equation}
f(\beta_{1}+\beta) = {\displaystyle \sum_{n=0}^{\infty} 
\frac{1}{n!}\left(\beta\frac{d}{d\beta}\right)}^{n}f(\beta_{1}) \;.
\label{eq:taylor}
\end{equation}
Here we use the convention of expressing the $n$'th order differential
operator as a power
\begin{equation}
\left(\frac{d}{d\beta}\right)^{n}f(\beta_{1}) \equiv 
\left[\frac{d^{n}f(\beta)}{d\beta^{n}}\right]_{\beta=\beta_{1}}.
\end{equation}
By considering a series expansion of the exponential function
\begin{equation}
  e^{\beta} = \sum_{n=0}^{\infty}\frac{\beta^{n}}{n!}  \; ,
\end{equation}
one obtains from Eq.~(\ref{eq:taylor}) an alternative operator expression
for the Taylor series
\begin{equation}
f(\beta_{1}+\beta) = \exp\left(\beta\frac{d}{d\beta}\right)f(\beta_{1}) \;.
\label{eq:time_evolution}
\end{equation}
We refer to this exponential operator, 
$\exp\left(\beta\frac{d}{d\beta}\right)$,
as an evolution operator. In our context, the first step of a discretized
algorithm is the \textit{exact factorization} of the evolution operator
into $L$ factors 
\begin{equation}
  f(\beta_{1}+\beta) = 
  \left[\exp\left(\frac{\beta}{L}\:\frac{d}{d\beta}\right)\right]^{L}
  f(\beta_{1})  \;.
\end{equation}
This factorization implies the discretization of $\beta$ into $L+1$
points separated by a small interval $\epsilon=\beta/L$. The next and 
most important step is the formulation of a \textit{high-temperature
approximation} for the evolution along the interval $\epsilon$. Note
that a small $\epsilon$ is equivalent to a high temperature $TL$,
as $\epsilon^{-1}=k_B T L$. The HTA is in general dependent on the
problem under consideration. One simple alternative is the first term
of the Taylor series
\begin{equation}
 f(\beta_{1}+\epsilon) \equiv 
  \left[\exp\left(\epsilon\:\frac{d}{d\beta}\right)\right]f(\beta_{1})
  \approx  f(\beta_{1}) + \epsilon 
  \left[ \frac{d \, f(\beta)}{d\beta} \right]_{\beta=\beta_{1}}  \; .
\label{eq:short_time_aprox}
\end{equation}
Once a HTA is computer coded, the whole temperature dependence of
$f(\beta)$ is obtained by a simple \textit{repeating loop} that recalculates
Eq.~(\ref{eq:short_time_aprox} ) for the successive $\epsilon$ intervals.

As a simple example let us apply a discretized algorithm to approximate
the density matrix of a free particle. We denote the canonical density
matrix for a free particle in one-dimension as $\rho(x,x';\beta)$.
Knowing that the high-temperature limit $(\beta\rightarrow0)$ of
$\rho$ is
\begin{equation}
  \rho(x,x';0) = \delta(x-x')  \;,
\label{eq:rho_high_T}
\end{equation}
we aim to derive $\rho(x,x';\beta)$ by a discretization of $\beta$
into steps $\epsilon=\beta/L$. We employ the HTA of 
Eq.~(\ref{eq:short_time_aprox}),
\begin{equation}
  \rho(x,x';\epsilon)\approx\rho(x,x';0)+
    \epsilon \left(\frac{\partial\rho(x,x';\beta)}{\partial\beta}
    \right)_{\beta=0}  \; .
\label{eq:rho_eps}
\end{equation}
The $\beta$ derivative is evaluated with the Bloch equation
\begin{equation}
  \frac{\partial\rho(x,x';\beta)}{\partial\beta} =
  - \widehat{H}\rho(x,x';\beta)\;,
   \quad\mathrm{with}\quad\widehat{H} = 
   - \frac{\hbar^2}{2m} \, \frac{\partial^2}{\partial x^2}  \; .
\label{eq:bloch}
\end{equation}
Using the following representation of the delta function 
\begin{equation}
  \delta(x,x') = \frac{1}{2\pi\hbar}
   \intop_{-\infty}^{\infty} dp\: e^{-ip(x-x')/\hbar}  \; ,
\end{equation}
the first derivative of $\rho(x,x';\beta)$ at $\beta=0$ is obtained
from Eqs.~(\ref{eq:rho_high_T}) and (\ref{eq:bloch}) as
\begin{equation}
  -\widehat{H}\rho(x,x';0) = \frac{1}{2\pi\hbar}
   \intop_{-\infty}^{\infty} dp\: 
   \left(-\frac{p\text{\texttwosuperior}}{2m}\right) 
   e^{-ip(x-x')/\hbar}  \; .
\end{equation}
With the help of the previous equation, the HTA of $\rho(x,x';\epsilon)$
in Eq.~(\ref{eq:rho_eps}) becomes 
\begin{equation}
  \rho(x,x';\epsilon) \approx \frac{1}{2\pi\hbar} 
  \intop_{-\infty}^{\infty} dp\: \left(1-\frac{\epsilon p^{2}}{2m}\right)
   e^{-ip(x-x')/\hbar}  \; .
\end{equation}
A straightforward repeating loop of the previous step gives the approximate
density matrix for the discretized inverse temperatures 
$2\epsilon,3\epsilon,\ldots$.
The final result for $L \epsilon = \beta$ is obtained by induction as 
\begin{equation}
  \rho(x,x';\beta) \approx \frac{1}{2\pi\hbar} 
      \intop_{-\infty}^{\infty}dp\:\left(1-\frac{\epsilon p^{2}}{2m}
      \right)^{L}e^{-ip(x-x')/\hbar}  \; .
\end{equation}
The error of this discretized approximation vanishes in the limit
$L\rightarrow\infty$. One can check it by substitution of 
$\epsilon = \beta/L$, and by considering the definition of the exponential 
function
\begin{equation}
  e^{-x} = \lim_{L\rightarrow\infty} \left(1-\frac{x}{L}\right)^{L}  \;.
\end{equation}
Thus, in the limit $L \rightarrow \infty$
\begin{equation}
  \rho(x,x';\beta) = \frac{1}{2\pi\hbar}
  \intop_{-\infty}^{\infty} dp\: e^{-\beta p^{2}/2m} \, e^{-ip(x-x')/\hbar}
   \; .
\end{equation}
This Gaussian integration is analytic, providing the final exact free
particle result \cite{fe72}
\begin{equation}
  \rho(x,x';\beta) = \left(\frac{m}{2\pi\beta\hbar^2}\right)^{1/2}
   e^{-m(x-x')^{2}/2\beta\hbar^{2}}  \; .
\end{equation}



\begin{thebibliography}{252}
\bibitem[{cit()}]{cita}
\bibinfo{note}{The number of cell degrees of freedom is important because it
  appears as a force factor in the equation of motion related to the first
  barostat that is coupled to the volume. In the original literature this force
  term is given as ${-d^2k_BT}$, where $d=3$ is the spatial dimension [see Eq.
  (D1) in Ref. \onlinecite{ma99b}]. We note that this term should be modified
  in the case that the rotation of the unit cell is eliminated, because the
  number of cell degrees of freedom is reduced from $d^2=9$ to $d(d+1)/2=6$. In
  this case the factor $d^2$ in the force of the first barostat should be
  changed to $d(d+1)/2$ in order to generate the correct ensemble
  distribution.}
\end{thebibliography}
\end{document}